\documentclass[submission, Proceedings]{SciPost}  	
\usepackage{pstricks,pst-coil,pst-fill,pst-plot}
\usepackage{amssymb}
\usepackage{amsmath}
\usepackage{mathrsfs}   
\usepackage{empheq}

 \makeatletter
 \@addtoreset{equation}{section}
 \makeatother

\newcommand{\f}[2]{{\ensuremath{%
    \mathchoice%
    {\dfrac{#1}{#2}}
    {\dfrac{#1}{#2}}
    {\frac{#1}{#2}}
    {\frac{#1}{#2}}
}}}

\let\ua=\uparrow
\let\da=\downarrow

\newcommand{\tf}[2]{\ensuremath{#1/#2}}

\newcommand{\R}{\ensuremath{\mathbb{R}}}
\newcommand{\Cx}{\ensuremath{\mathbb{C}}}
\newcommand{\norm}[1]{\ensuremath{  || #1 || }}

\newcommand{\mc}[1]{\ensuremath{\mathcal{#1}}}
\newcommand{\mf}[1]{\ensuremath{\mathfrak{#1}}}
\newcommand{\msc}[1]{\ensuremath{\mathscr{#1}}}

\newcommand{\bs}[1]{\ensuremath{\boldsymbol{#1}}}

\newcommand{\sul}[2]{\ensuremath{\sum\limits_{#1}^{#2}}}
\newcommand{\pl}[2]{\ensuremath{\prod\limits_{#1}^{#2}}}

\newcommand{\op}[1]{ \boldsymbol{ \texttt{#1} } }
\newcommand{\wt}[1]{\ensuremath{\widetilde{#1}}}

\newcommand{\dd}{\mathrm{d}}
\newcommand{\e}[1]{\ensuremath{\mathrm{#1}}}

\newcommand{\ex}[1]{\ensuremath{\e{e}^{#1}}}
\def \i{ \mathrm i}

\def\a{\alpha}

\def\ga{\gamma}
\def\Ga{\Gamma}

\def\de{\delta}

\def\eps{\epsilon}

\def\la{\lambda}

\def\sg{\sigma}

\def\th{\theta}

\def\Om{\Omega}
\def\om{\omega}
\def\vp{\varphi}

\newcommand\beq{\begin{equation}}
\newcommand\enq{\end{equation}}
\newcommand\bem{\begin{multline}}
\newcommand\enm{\end{multline}}

\def\ba{\begin{array}}
\def\ea{\end{array}}

\newcommand{\Int}[2]{\ensuremath{\int\limits_{#1}^{#2}}}

\def\cadremath#1{\vbox{\hrule\hbox{\vrule\kern8pt\vbox{\kern8pt
			\hbox{ {$\displaystyle #1 $ } }\kern8pt} 
			\kern8pt\vrule}\hrule}}
			
\definecolor{myblue}{rgb}{.8, .8, 1}

\begin{document}

\begin{center}{\Large \textbf{
Solution of Baxter equation for the $q$-Toda and Toda$_2$ chains by NLIE
}}\end{center}

\begin{center}
O. Babelon \textsuperscript{1},
K. K. Kozlowski \textsuperscript{2},
V. Pasquier \textsuperscript{3}
\end{center}

\begin{center}
{\bf 1} 
Sorbonne Universit\'es, UPMC Univ Paris 06, CNRS, UMR 7589, LPTHE, 75005 Paris, France
\\
{\bf 2}Univ Lyon, ENS de Lyon, Univ Claude Bernard Lyon 1, CNRS, Laboratoire de Physique,
 F-69342 Lyon, France
\\
{\bf 3} Univ. Paris Saclay, CNRS, CEA, IPhT, F-91191 Gif-sur-Yvette, France.
\\
\end{center}

\begin{center}
\today
\end{center}
 
\section*{Abstract}
{\bf
We construct a basis of solutions of the scalar $\op{t}-\op{Q}$ equation describing the spectrum of the $q$-Toda and Toda$_2$ chains by using auxiliary non-linear integral
equations. Our construction allows us to provide quantisation conditions for the spectra of these models in the form of 
thermodynamic Bethe Ansatz-like equations.
}

\tableofcontents 

\section{Introduction}

 The Ruijsenaars-Schneider $q$-Toda chain introduced in \cite{Rui90}, appears as a natural $q$ deformation of the quantum mechanical Toda chain. 
As show in \cite{BaKoPa18Td2}, the $q$-Toda chain can be embedded in a larger family of models which can be interpreted as stemming from  the 
quantisation of the classical Toda chain endowed with the second Hamilton structure. 
The quantisation of the spectrum of the $q$-Toda chain was derived in \cite{KT94,KarLebSem02} in the form of a scalar $\op{t}-\op{Q}$ equations by implementing 
the separation of variables transform for the chain. This form of the quantisation conditions is however non-efficient for any practical study of the spectrum
and needs to be recast in a more convenient way. 
Due to the connection of the $q$-Toda chain, in particular in the two-particle sector,  with certain operators arising as quantisation of mirror curves, 
there has been recently a renewed  interest in describing the spectrum of the chain \cite{HaMa15,KaSe17,KP16,Se05}. 
The works \cite{KP16,Se05} describe the quantisation conditions for the spectrum by solving the finite difference scalar $\op{t}-\op{Q}$ equation in non-explicit terms, in the spirit of the early quantisation conditions
for the quantum mechanical Toda chain \cite{GaPa92,Gu80,Gu81}. The approach of \cite{HaMa15,KP16} takes its roots in the work \cite{NS10} which proposed a heuristic scheme
allowing one to provide quantisation conditions of many quantum integrable models by means of TBA-like equations which involve auxiliary solutions to non-linear integral equations. 
The conjectured quantisation conditions were established, for the quantum mechanical Toda chain, in \cite{KozTes09}. The recent works \cite{HaMa15,KP16}
proposed a conjectural form of quantisation conditions appropriate for describing the spectrum of the $q$-Toda chain. 
These results were checked numerically in the case of a low number of particles.

In the recent work \cite{BaKoPa18}, the authors have obtained a characterisation of the  spectrum of the Ruijsenaars-Schneider $q$-Toda and the Toda$_2$ chain
by means of a scalar $\op{t}-\op{Q}$ equation. In the present paper, we  show that this scalar $\op{t}-\op{Q}$ equation can be solved 
in terms of a one parameter family of Fredholm determinants, or generalisations thereof, acting on $\ell^2(\mathbb{N})$
generalising the construction obtained in \cite{GaPa92,Gu80,Gu81} for the Toda chain. We then show
that this equation may as well be solved within the non-linear integral equation approach developed, for the Toda chain, in \cite{KozTes09}. This leads to 
quantisation conditions for these two models of interest in the form of thermodynamic Bethe Ansatz-like equations.
So far, we were not able to compare our form of the quantisation conditions with the ones obtained in \cite{HaMa15,KP16}. This may be due to the fact that the works  \cite{HaMa15,KP16}
consider the $|q|=1$ case while we focus on the $|q|<1$ regime.

The paper is organised as follows. Section \ref{Section definition du modele} introduces the two models of interest: the $q$-Toda and the Toda$_2$ chain. 
The Hamiltonians of these models are given in Subsection \ref{SousSection presentation directe des modeles}. 
Subsection \ref{SousSection presentation de eqn TQ} recalls the main result of \cite{BaKoPa18}, which are the quantisation conditions for the 
spectrum of these models in terms of a scalar Baxter $\op{t}-\op{Q}$ equation. This is the starting point of the analysis that is carried out in this work. 
 Section \ref{Section Construction des solutions} provides an explicit construction of the solutions to the scalar $\op{t}-\op{Q}$  equations. 
A set of elementary solutions, built in terms of Fredholm determinants or generalisations thereof, to this equation is constructed in 
Subsection \ref{subsection building blocks q pm}. Then, in Subsection \ref{Subsection rewriting through NLIE}, these are recast in terms 
of solutions to non-linear integral equations. The rewriting of quantisation conditions for these models in terms of Bethe equations is carried out in Subsection \ref{Subsection Quantisation conditions by HLBAE}. 
Section \ref{Section equivalence NLIE et TQ} establishes the equivalence of the non-linear integral equation based description of the spectrum with the one
based on $\op{t}-\op{Q}$ equations. Finally, Section \ref{Section Spectrum of transfer matrices} contains results allowing one to reconstruct, starting from 
a solution of a non-linear integral equation parameterising a given generalised Eigenvalue of either model, the model's spectrum. 
The paper contains several appendices where technical results are postponed.
Appendix \ref{Appendix Sous Section determinants infinis} discusses infinite determinants. 
Appendix \ref{Appendix NLIE} gathers various results related to the non-linear integral equation of interest to this work. 
Subappendix \ref{Appendix SousSection solvabilite unique des NLIE pour Y} establishes solvability results for the 
class of non-linear integral equations of interest. Subappendices \ref{Appendice Sous Section Fct Y delta} and \ref{Subappendix Auxiliary fcts nu up down} 
establish  properties of auxiliary function built in terms of solutions to the non-linear integral equation of interest. 
Appendix \ref{Appendix ptes gnles Baxter eqn} establishes various general properties of solutions to the scalar $\op{t}-\op{Q}$  equations of interest to this work. 
Subappendix \ref{Appendix SousSection Wronskiens} computes various associated Wronskians 
while Subappendix \ref{Appendix Subsection forme generale solutions tq} establishes the overall form of solutions. 
Finally,  Appendix \ref{Appendix Special functions} discusses the main properties and definitions of the 
special functions of interest to the present study. Subappendix \ref{Appendix q products} focuses on $q$-products and $\th$-functions, 
Subappendix \ref{Appendix Double Sine} recalls properties of the double sine function and Subappendix \ref{Appendix Section quantum dilog}
reviews properties of interest of the quantum dilogarithm.

\section{The models and their quantisation conditions}
\label{Section definition du modele}

\subsection{The Toda$_2$ and $q$-Toda chains}
\label{SousSection presentation directe des modeles}

Let $\op{X}_n$, $\op{x}_n$ be canonically conjugated operators
$$
[ \op{X}_n,\op{x}_n ] = -\i 
$$
acting on a Hilbert space $V_{x_n} \simeq L^2(\R)$. Thus, their properly scaled exponents form a Weyl pair:
\begin{empheq}{equation}
\ex{-{2\pi\over \omega_2}  \op{x}_n} \ex{- \omega_1 \op{X}_n} = q^2 \ex{- \omega_1 \op{X}_n} \ex{-{2\pi\over \omega_2}  \op{x}_n}, \quad 
\text{with} \quad q=\ex{ \i\pi {\omega_1\over \omega_2} } \;. 
 \label{weyl}
\end{empheq}
 Here, $\om_1, \om_2$ are two auxiliary complex parameters which satisfy the constraint $\Im(\tf{\om_1}{\om_2})>0$
 so that $|q|<1$. Throughout this work, we shall always consider $\om_1,\om_2$ to be generic. 
 
 The $q$-Toda chain refers to the below Hamiltonian on $\mf{h}=\otimes_{n=1}^{N}V_{x_n} \simeq L^2(\R^N)$:
\[
\op{H}_1^{\mathrm{q-Toda}}= \sum\limits_{n=1}^N \left[ 1+ q^{-1} \ex{-{2\pi \over \omega_2} \kappa_1}\ex{-{2\pi \over \omega_2}(\op{x}_n-\op{x}_{n-1})}  \right]\ex{-\omega_1 \op{X}_n} 
\]
while the Toda$_2$ denotes the Hamiltonian:
\[
\op{H}_1^{\mathrm{Toda}_2} =  \sum\limits_{n=1}^N \ex{-\omega_1 \op{X}_n} + \ex{-{2\pi \over \omega_2} \kappa_2}\ex{-{2\pi \over \omega_2}(\op{x}_n-\op{x}_{n-1})} \;. 
\]
$\kappa_1,\kappa_2$ appearing above are some coupling constants.  

By means of the quantum inverse scattering method, these Hamiltonians can be embedded into a larger family of commuting operators 
$\mc{H}=\big\{ \op{H}_0^{\bullet},\op{H}_1^{\bullet}, \dots, \op{H}_N^{\bullet} \big\}$, with $\bullet=\mathrm{q-Toda}$ or $\mathrm{Toda}_2$. 
Furthermore, elements of $\mc{H}$ commute with their respective duals which are obtained by the substitution $(\om_1,\om_2)\leftrightarrow (\om_2,\om_1)$. 
These \textit{a priori} provide one with an independent set of operators so that, in order to have a well-posed spectral problem, one should 
consider the joint diagonalisation of the family $\mc{H}$ and its dual family $\wt{\mc{H}}$. 
Here and in the following, dual quantities will be denoted with a tilde, \textit{viz}. 
$\wt{\op{H}}_k^{\mathrm{q-Toda}}$ and $\wt{\op{H}}_k^{\mathrm{Toda}_2}$. 
In particular, the dual parameter $\tilde{q}$ to $q$ is 
\beq
\tilde{q} \, = \,   \ex{ \i\pi {\omega_2 \over \omega_1} }  \qquad \e{so} \; \e{that} \quad |\tilde{q}|>1 \;. 
\enq

It is convenient to gather the members of the commuting family of operators into a single operator valued hyperbolic polynomial, closely related to the model's
transfer matrix
\beq
\op{t}(\lambda) \, = \,  (-1)^N\ex{ \tfrac{\om_1}{2}\op{P}_{\e{tot}} } \ex{{\pi\over \omega_2} N\lambda }  \sul{j=0}{N} (-1)^j\cdot \ex{-{2\pi\over \omega_2} (N-j)\lambda}  \cdot  \op{H}_j^{\mathrm{q-Toda}}  
\enq
in the case of the $q$-Toda chain and 
\beq
\op{t}(\lambda) \, = \,  \sul{j=0}{N} (-1)^j\cdot \ex{-{2\pi\over \omega_2} (N-j)\lambda}  \cdot  \op{H}_j^{\mathrm{Toda}_2}  
\enq
for the Toda$_2$ chain. One defines analogously the dual quantities.

\subsection{The Baxter equation quantising the spectrum} 
\label{SousSection presentation de eqn TQ}

Irrespectively of the considered model, $\op{t}(\la)$, $\wt{\op{t}}(\la)$ both  commute with the total momentum operator 
\[
\op{P}_{\e{tot}}=\sul{a=1}{N} \op{X}_a \, . 
\]
Thus one may immediately reduce the dimensionality of the spectral problem by projecting onto the subspace $\mf{h}_{p_0}$, 
\beq
\mf{h} \; = \; \oplus \Int{}{} \mf{h}_{p_0} \dd p_0 \;,
\enq
associated with the Fourier mode $p_0$ of the total momentum operator $\op{P}_{\e{tot}}$. 
Upon projection, the operator $\op{t}(\lambda)$ restricts to the operator $\op{t}(\lambda;p_0)$ on the reduced space $\mf{h}_{p_0}$. The reduced operator is expected to have a pure point-wise spectrum
although this property has not yet been established. 
Taken the hyperbolic polynomial form of $\op{t}(\la)$, resp. $\wt{\op{t}}(\la)$, the Eigenvalues of $\op{t}(\lambda;p_0)$, resp. $\wt{\op{t}}(\lambda;p_0)$,  can be parameterised by $N$ complex roots
$\bs{\tau} = (\tau_1,\dots, \tau_N)$, 
resp. $\tilde{\bs{\tau}} = (\tilde{\tau}_1,\dots, \tilde{\tau}_N)$ and take the form: 
\begin{itemize}
\item $q$-Toda
\begin{empheq}{equation}
 t_{\bs{\tau}}(\lambda ) = \prod_{k=1}^N \Big\{ 2\sinh {\pi \over \omega_2}(\lambda -\tau_k)  \Big\} \; ,  
 \qquad \e{and} \qquad 
 \tilde{t}_{ \tilde{\bs{\tau}} }(\lambda ) = \prod_{k=1}^N \Big\{ 2\sinh {\pi \over \omega_1}(\lambda - \tilde{\tau}_k)  \Big\} \; ;
\label{ecriture polynomes t tau q Toda}
\end{empheq}
\item Toda$_2$
\begin{empheq}{equation}
 t_{\bs{\tau}}(\lambda ) =  \prod_{k=1}^{N}\Big\{  \ex{-{2\pi  \over \omega_2} \lambda } - \ex{-{2\pi  \over \omega_2} \tau_k} \Big\} \; , 
\qquad \e{and} \qquad 
\tilde{t}_{ \tilde{\bs{\tau}} }(\lambda ) =  \prod_{k=1}^{N}\Big\{  \ex{-{2\pi  \over \omega_1} \lambda } - \ex{-{2\pi  \over \omega_1} \tilde{\tau}_k} \Big\} \;. 
\label{ecriture polynomes t tau Toda 2}
\end{empheq}
\end{itemize}

Any collection of roots $\bs{\tau}, \tilde{\bs{\tau}} $ parameterising the Eigenvalues of $\op{t}(\lambda;p_0)$, resp. $\wt{\op{t}}(\lambda;p_0)$,
are subject to the constraint 
\beq
\pl{k=1}{N} \ex{ -{\pi \over \omega_2}  \tau_k} = \ex{ - {1\over 2} \omega_1 p_0}  \qquad \e{and} \qquad 
\pl{k=1}{N} \ex{ -{\pi \over \omega_1} \tilde{\tau}_k} = \ex{ - {1\over 2} \omega_2 p_0}  \;, 
\label{ecriture contrainte directe sur les taus}
\enq
for the $q$-Toda chain and 
\beq
\pl{k=1}{N}  \ex{-{2\pi \over \omega_2}  \tau_k} = \ex{- \omega_1 p_0} + \ex{- \f{2\pi N}{\om_2}\kappa_2  }  \qquad \e{and} \qquad  
\pl{k=1}{N} \ex{-{2\pi \over \omega_1}  \tilde{\tau}_k} = \ex{- \omega_2 p_0} + \ex{- \f{2\pi N}{\om_1}\kappa_2  } 
\label{ecriture contrainte duale sur les taus}
\enq
for the Toda$_2$ chain. These constraints translate the fact that the Eigenvalues are associated with Eigenfunctions belonging to $\mf{h}_{p_0}$.

We are now in position to formulate precisely the spectral problem associated with the $q$-Toda and Toda$_2$ chains. 
The latter consists in finding a collection of roots $\bs{\tau}, \tilde{\bs{\tau}} $, satisfying to the constraints 
\eqref{ecriture contrainte directe sur les taus} or \eqref{ecriture contrainte duale sur les taus} 
so that there exists a joint, self-dual,  entire solution $q$ to the set of self-dual Baxter equations
subordinate to the associated hyperbolic polynomials $t_{\bs{\tau}}$ and  $\tilde{t}_{ \tilde{\bs{\tau}} }$
\begin{empheq}{align}
t_{\bs{\tau}}(\lambda ) q(\lambda ) &= g^{N\om_1} \big( \sigma \,  \varkappa^{\om_1 } q(\lambda -\i\omega_1) + \sigma^{-1} q(\lambda + \i\omega_1) \big)  \;, \vspace{3mm} \label{ecriture ensemble autodual eqns Baxter 1}\\
\tilde{t}_{ \tilde{\bs{\tau}} }(\lambda ) q(\lambda ) &= g^{N\om_2} \big( \sigma \,  \varkappa^{\om_2 } q(\lambda -\i\omega_2) + \sigma^{-1} q(\lambda + \i\omega_2) \big)   \;, 
\label{ecriture ensemble autodual eqns Baxter 2}
\end{empheq}
where the polynomials  are defined in \eqref{ecriture polynomes t tau q Toda} or \eqref{ecriture polynomes t tau Toda 2}, depending on the model of interest. 
The parameters appearing in the above equations take the form :

\begin{itemize}
\item $q$-Toda
\begin{empheq}{equation}
\sigma= (-\i)^{N}, \quad  g  = \ex{ -{ \pi \kappa_1 \over \omega_1\omega_2} } , \quad \varkappa= 1 \;; 
\label{ecriture parametres eqn Baxter qToda}
\end{empheq}
\item Toda$_2$
\begin{empheq}{equation}
\sigma =(-1)^N, \quad  g  = \ex{ -{ \pi  \kappa_2 \over \omega_1\omega_2} }  , \quad \varkappa=  \ex{-p_0} \;. 
\label{ecriture parametres eqn Baxter Toda2}
\end{empheq}
\end{itemize}

The main difference between the $q$-Toda and Toda$_2$ chains is that, in the former model, $\varkappa$ depends explicitly on the zero mode $p_0$ and the transfer matrix Eigenvalue polynomial only grows in the direction 
$\Re\big( \tf{\la}{\om_2} \big) \rightarrow  -\infty$. For further applications, it will be important to study the analyticity properties  in $p_0$ of the solution $q$ to the $t-q$ equations governing the spectrum of the 
Toda$_2$ chain. We leave this to a subsequent publication.

\section{Solution of the Baxter equation.}
\label{Section Construction des solutions}

We shall construct solutions $(t_{\bs{\tau}}, \tilde{t}_{ \tilde{\bs{\tau}} }, q)$  to the spectral problem described in Subsection \ref{SousSection presentation de eqn TQ} by following the below procedure:
\begin{itemize}

 \item[i)] we first construct two fundamental, meromorphic and linearly independent, solutions $q_{\pm}$ to the set of self-dual Baxter equations \eqref{ecriture ensemble autodual eqns Baxter 1}-\eqref{ecriture ensemble autodual eqns Baxter 2}, this for any value of the 
roots $\bs{\tau}$ and $\tilde{\bs{\tau}}$ subject to the constraints \eqref{ecriture contrainte directe sur les taus}, \eqref{ecriture contrainte duale sur les taus}.
Their expression involves certain determinants of infinite matrices, in the spirit of Gutzwiller \cite{Gu80,Gu81}. 
 
 \item[ii)] We show that these determinants can be re-expressed in terms of certain solutions to non-linear integral equations and that this induces a re-parametrisation of the 
 solutions $q_{\pm}$ in terms of a new set of variables $\bs{\de}$, $\tilde{\bs{\de}}$ which can be constructed from $\bs{\tau}$ and $\tilde{\bs{\tau}}$ occurring in $\e{i)}$. 
 Here, we follow the strategy devised in \cite{KozTes09} for the Toda chain model. 
 
 \item[iii)] We show that producing a linear combination solving the set of self-dual Baxter equations $q(\la)=\mc{P}_+(\la)q_+(\la)+ \mc{P}_{-}(\la) q_-(\la)$, with $\mc{P}_{\pm}$ being elliptic functions on the lattice $(\i \om_1,\i\om_2)$,
and such that $q$ is entire can only be possible if $\bs{\de} = \tilde{\bs{\de}}$ and $\bs{\de}$ satisfy a set of Bethe equations. 
 
 \item[iv)] We show that the characterisation of the solutions to the spectral problem in terms of solutions to non-linear integral equation is complete, in that with every such solution
 one can associate  two functions $q_{\pm}$ which constitute a set of fundamental solutions to a set of self-dual Baxter equations \eqref{ecriture ensemble autodual eqns Baxter 1}-\eqref{ecriture ensemble autodual eqns Baxter 2}
 associated with some polynomials $t_{\bs{\tau}}, \tilde{t}_{ \tilde{\bs{\tau}} }$. 
 
\end{itemize}

The fact that $|q|<1$, $|\tilde{q}|>1$ play an important role in the analysis developed below. In particular, taking the $|q|, |\tilde{q}| \rightarrow 1$ limits, even on the level of the final formulae 
does not appear evident despite the fact that this limit is formally smooth on the level of the scalar $\op{t}-\op{Q}$ equation. 
The extension of the techniques developed in this work so as to deal with the $|q|=1$ case demands a separate investigation and will not be considered here.

\subsection{The fundamental system of solutions}
\label{subsection building blocks q pm}

From now on, it will appear convenient to introduce the parameter
\beq
\rho= \varkappa g^{2N} 
\enq
and two constants $\zeta , \tilde{\zeta}$ such that 
\beq
\ex{ \om_1 \zeta }  \; = \; \ex{\om_1 p_0 } \pl{a=1}{N} \ex{ - \f{ 2\pi   }{ \om_2}  \tau_a} \qquad \e{and} \qquad 
\ex{ \om_2 \tilde{\zeta} }  \; = \; \ex{\om_2 p_0 } \pl{a=1}{N} \ex{ - \f{ 2\pi   }{ \om_1 } \tilde{\tau}_a  } 
\enq
Note that, for the $q$-Toda chain, owing to the form of the constraints on $\bs{\tau}$, $\tilde{\bs{\tau}}$ \eqref{ecriture contrainte directe sur les taus}, one may take 
 $\zeta=\tilde{\zeta}=0$, and it is this choice that will be made in the following.

Finally, in the case of the Toda$_2$ chain, we shall restrict ourselves to the regime of parameters 
\beq
\e{max}_{a=1,2} \big| \big(\ex{2p_0}  \rho  \big)^{\om_a} \big| \, < \, 1 \;. 
\enq

\subsubsection{The infinite determinants $K_{\pm}$, $\wt{K}_{\pm}$}

By taking suitable infinite size limits of finite size determinants, we construct in Subsection \ref{Appendix Sous Section determinants infinis}
four meromorphic functions $K_{\pm}$ and their dual $\wt{K}_{\pm}$ on $\Cx$. 
The functions $K_{\pm}$ 
\begin{itemize}
\item are $\i \om_2$ periodic;
\item have simple poles at $\{ \tau_k \mp \i \om_1 \mathbb{N} + \i \om_2 \mathbb{Z} \}$; 
\item given $ \la= \i x \om_1+\i y \om_2$, they have the asymptotic expansion 
\begin{eqnarray}
K_+(\la) & = & \left\{  \ba{cc} 1 \, + \,  \e{O}\Big( \ex{ \tfrac{ \pi }{ \om_2 } N \la } \Big) & q-\e{Toda}  \vspace{2mm}  \\  
				1 \, + \,  \e{O}\Big( \ex{ \tfrac{ 2 \pi }{ \om_2 } N \la } \Big)	 & \e{Toda}_2 			      \ea \right.  \qquad     x \rightarrow + \infty    \label{ecriture asymptotiques de K+} \\
K_-(\la) & = & \left\{  \ba{cc} 1 \; \;\;  +  \; \;\;  \e{O}\Big( \ex{ -\tfrac{ \pi }{ \om_2 } N \la } \Big) & q-\e{Toda}   \vspace{2mm}   \\  
				\f{ 1 }{ 1 - \rho^{\om_1} \ex{2\om_1 p_0} } \, + \,  \e{O}\Big( \ex{ -\tfrac{ 2 \pi }{ \om_2 }  \la }	\Big) & \e{Toda}_2 			      \ea \right.      x \rightarrow - \infty   \;;
\label{ecriture asymptotiques de K-}
\end{eqnarray}
\item solve the second order, finite-difference, equations:
\begin{eqnarray}
K_+(\la-\i\om_1) &=& K_+(\la) \, -  \,  \f{ \rho^{\om_1} K_+(\la+\i\om_1) }{ t_{\bs{\tau}}  (\la) t_{\bs{\tau}}  (\la+\i\om_1) }
\label{equation de reccurrence K+}\\
K_-(\la+\i\om_1) &=&  \ex{ \om_1 \zeta} K_-(\la) \,  - \,   \f{ \rho^{\om_1} \ex{2\om_1 \zeta}  K_-(\la-\i\om_1) }{ t_{\bs{\tau}}  (\la) t_{\bs{\tau}}  (\la-\i\om_1) } \;.
\label{equation de reccurrence K-}
\end{eqnarray}
The hyperbolic polynomials  $t_{ \bs{\tau}}$ are given by eq. \eqref{ecriture polynomes t tau q Toda} or eq. \eqref{ecriture polynomes t tau Toda 2}, depending on the model of interest.
We do stress that the parameters $ \bs{\tau}$ defining the polynomials $t_{\bs{\tau}}(\la)$ are taken arbitrary here, it particular they don't necessarily correspond to a parametrisation of an 
Eigenvalue of the transfer matrix $\op{t}(\la)$.

\end{itemize}

Similarly, the dual functions $\wt{K}_{\pm}$ 
\begin{itemize}
\item are  $\i \om_1$ periodic;
\item have simple poles at $\{ \tilde{\tau}_k \mp \i \om_2 \mathbb{N} + \i \om_1 \mathbb{Z} \}$; 
\item given $ \la= \i x \om_1+\i y \om_2$, have the asymptotic expansion 
\begin{eqnarray}
\wt{K}_-(\la) & = & \left\{  \ba{cc} 1 \, + \,  \e{O}\Big( \ex{ \tfrac{ \pi }{ \om_2 } N \la } \Big) & q-\e{Toda}  \vspace{2mm}  \\  
				1 \, + \,  \e{O}\Big(\ex{ \tfrac{ 2 \pi }{ \om_2 } N \la } \Big)	 & \e{Toda}_2 			      \ea \right.    y \rightarrow - \infty  \vspace{3mm} \label{ecriture asymptotiques de tilde K+}\\
\wt{K}_+(\la) & = & \left\{  \ba{cc} 1 \; \;\;  +  \; \;\;  \e{O}\Big( \ex{ -\tfrac{ \pi }{ \om_2 } N \la } \Big) & q-\e{Toda}   \vspace{2mm}  \\  
				\f{ 1 }{ 1 - \rho^{\om_2} \ex{2\om_2 p_0} } \, + \,  \e{O}\Big( \ex{ -\tfrac{ 2 \pi }{ \om_2 }  \la }	\Big) & \e{Toda}_2 			      \ea \right.    y \rightarrow  + \infty \;; 
\label{ecriture asymptotiques de tilde K-}
\end{eqnarray}
\item  $\wt{K}_{\pm}$ solve the second order, finite-difference, equations:
\begin{eqnarray}
\wt{K}_+(\la-\i\om_2) &=&  \ex{\om_2 \tilde{\zeta} } \wt{K}_+(\la)  \, - \,   \f{ \rho^{\om_2}  \ex{2 \om_2 \tilde{\zeta} }  \wt{K}_+(\la+\i\om_2) }{ \tilde{t}_{ \tilde{\bs{\tau}} }  (\la)  \tilde{t}_{ \tilde{\bs{\tau}} } (\la+\i\om_2) } 
\label{equation de reccurrence tilde K+}\\
\wt{K}_-(\la+\i\om_2) &=&  \wt{K}_-(\la)  \, - \,   \f{ \rho^{\om_2}  \wt{K}_-(\la-\i\om_2) }{  \tilde{t}_{ \tilde{\bs{\tau}} }  (\la)  \tilde{t}_{ \tilde{\bs{\tau}} } (\la-\i\om_2)  } \;.
\label{equation de reccurrence tilde K-}
\end{eqnarray}
The hyperbolic polynomials  $ \tilde{t}_{  \tilde{\bs{\tau}} }$ are given by eq. \eqref{ecriture parametres eqn Baxter qToda} or eq. \eqref{ecriture parametres eqn Baxter Toda2} and, 
again, the parameters $\tilde{\bs{\tau}}$ defining these polynomials  are taken arbitrary, so that the connection with 
Eigenvalues of the transfer matrix $\tilde{\op{t}}(\la)$ is irrelevant.

\end{itemize}

It seems also worthy to stress that, in the Toda$_2$ chain case, the finite difference equations satisfied by $\wt{K}_{\pm}$ are \textit{not} the dual ones of those satisfied by $K_{\pm}$.

There is a nice relationship between $K_{\pm}$ and $\wt{K}_{\pm}$ in case when the modular parameters $(\om_1,\om_2)$ and coupling constants $\kappa_1,\kappa_2$ 
satisfy the reality condition $\overline{\om_1}=\om_2$ and $\kappa_a\in \R$. 
In this case, the transfer matrices are related as $\big( \op{t}(\la)\big)^{\dagger}=\wt{\op{t}}( \overline{\la})$, so that $\overline{ t_{\bs{\tau}}(\la) } = \tilde{t}_{ \tilde{\bs{\tau}} } (\overline{\la})$. 
Then, by using that $\overline{\ex{\om_1 \zeta}}=\ex{\om_2 \tilde{\zeta}}$, the explicit determinant representation for $K_{\pm}$ entail that
\beq
\overline{ K_{\pm}(\la) } \, = \, \wt{K}_{\mp}(\overline{\la})\;. 
\label{ecriture condition realite K pm et Tilde K pm}
\enq

\subsubsection{The Hill determinants and its zeroes $\bs{\de}, \tilde{\bs{\de}}$ }

By rearranging the finite difference equations satisfied by $K_{\pm}$ eqns. \eqref{equation de reccurrence K+}-\eqref{equation de reccurrence K-} 
and $\wt{K}_{\pm}$ eqns. \eqref{equation de reccurrence tilde K+}-\eqref{equation de reccurrence tilde K-} one obtains that the quantities 
\begin{eqnarray}
\mc{H}(\la) & = &  K_+(\la) K_-(\la+\i\om_1) \, -  \,  \rho^{\om_1} \, \ex{ \om_1 \zeta} \cdot \f{  K_+(\la+\i\om_1)  K_-(\la) }{ t_{\bs{\tau}}  (\la) t_{\bs{\tau}}  (\la+\i\om_1) } \label{definition det Hill} \\
\wt{\mc{H}}(\la) & = &  \wt{K}_+(\la) \wt{K}_-(\la+\i\om_2) \, -  \,  \rho^{\om_2} \, \ex{ \om_2 \tilde{\zeta} } \cdot \f{   \wt{K}_+(\la+\i\om_2)  \wt{K}_-(\la) }{ \tilde{t}_{ \tilde{\bs{\tau}} }  (\la) 
\tilde{t}_{ \tilde{\bs{\tau}} } (\la+\i\om_2) } \;, 
\label{definition det hill dual}
\end{eqnarray}
solve 
\beq
\left\{ \ba{ccc} \mc{H}(\la+\i\om_2) \, = \, \mc{H}(\la) \qquad & \e{and} & \qquad \mc{H}(\la-\i\om_1) \, =  \,  \ex{ -\om_1 \zeta} \, \mc{H}(\la)   \vspace{3mm} \\ 
\wt{\mc{H}}(\la+\i\om_1) \, = \, \wt{\mc{H}}(\la) \qquad & \e{and} & \qquad \wt{\mc{H}}(\la-\i\om_2) \, =  \,  \ex{ \om_2 \tilde{\zeta} } \,  \wt{\mc{H}}(\la) \ea \right. \;. 
\nonumber 
\enq
It is easy to see that $\mc{H}$, resp. $\wt{\mc{H}}$, have simple poles on the lattice $\big\{ \tau_k \, + \, \i \om_1 \mathbb{Z} \, + \, \i \om_2 \mathbb{Z} \big\}$. Thus, 
there exists constants $\mf{h}$, $\wt{\mf{h}}$ and $N$ roots $\bs{\de} \, = \, (\de_1,\dots,\de_N)$ along with their duals $\tilde{\bs{\de}}  \, = \,  (\tilde{\de}_1, \dots, \tilde{\de}_N )$
satisfying, independently of the model, to the constraints 
\beq
p_0 \, = \, \f{ 2\pi }{ \om_1 \om_2 } \sul{a=1}{N} \de_k \qquad \e{and} \qquad 
p_0 \, = \, \f{ 2\pi }{ \om_1 \om_2 } \sul{a=1}{N} \tilde{\de}_k \;, 
\label{contraintes sur les zeros dek tilde dek des dets de Hill}
\enq
such that one has the zero/pole factorisation 
\beq
\mc{H}(\la) = \mf{h} \cdot \f{ \th_{\bs{\de}}(\la)  }{ \th_{\bs{\tau}}(\la) }  \quad , \quad 
\wt{\mc{H}}(\la) = \wt{\mf{h}} \cdot \f{ \wt{\th}_{ \tilde{\bs{\de}} }(\la)  }{ \wt{\th}_{ \tilde{\bs{\tau}} }(\la) } \;.
\label{ecriture factorisation determinants de Hill et son dual}
\enq
Here, given any set of $N$-parameters $\bs{\la}=(\la_1,\dots, \la_N)$,  we introduced the shorthand notation 
\beq
\th_{\bs{\la}}(\la) \, = \, \pl{k=1}{N}\th(\la-\la_k)\qquad \e{and} \qquad 
\wt{\th}_{\bs{\la}}(\la) \, = \, \pl{k=1}{N} \tilde{\th}(\la- \la_k) \;, 
\enq
where $\th(\la)$ is closely related to the Jacobi $\theta_1$ function and $\wt{\th}$ to its dual. See  Appendix \ref{Appendix q products}, eqn. \eqref{definition des fonction theta et theta duale} in particular,  for more details.

$\bs{\de}$, $\tilde{\bs{\de}}$ are representatives of the lattice of zeroes of the quasi-elliptic functions $\mc{H}$, $\wt{\mc{H}}$. 
In fact, by using the infinite determinant representation for the functions $K_{\pm}$, $\wt{K}_{\pm}$, one may interpret 
$\mc{H}$ and $\wt{\mc{H}}$ as being double-sided infinite determinants of Hill type.  This interpretation is rather direct for
the $q$-Toda chain but demands some work in the Toda$_2$ chain.

Finally, when the modular parameters $(\om_1,\om_2)$ and coupling constants $\kappa_1,\kappa_2$ 
satisfy the reality condition $\overline{\om_1}=\om_2$ and $\kappa_a\in \R$, the two Hill determinants are related as 
\beq
\overline{\mc{H}(\la)} \; = \; \wt{\mc{H}}(\overline{\la} -\i \om_2) \;,
\label{ecriture conjugaison cas realite des dets de Hill}
\enq
what is a direct consequence of the definitions \eqref{definition det Hill}, \eqref{definition det hill dual} and the conjugation properties for $K_{\pm}$ \eqref{ecriture condition realite K pm et Tilde K pm}.

\subsubsection{The solutions $q_{\pm}$}

We are now in position to present the explicit form of the fundamental system of solutions $q_{\pm}$ to the set of dual Baxter 
equations \eqref{ecriture ensemble autodual eqns Baxter 1}-\eqref{ecriture ensemble autodual eqns Baxter 2}. The formulae describing $q_{\pm}$ involve $p$-infinite product 
$(z;p)$ for $|p|<1$ and it is convenient to introduce the compact notations
\beq
(w(\la);p)_{\bs{\de}}=\pl{k=1}{N}(w(\la-\de_k);p) \qquad \e{for} \;\e{any} \; \e{function} \; w \; .
\enq
We refer to Appendix \ref{Appendix q products} for more details on these functions. Finally, recall 
that to \textit{any} collection of variables $\bs{\tau}$, $\wt{\bs{\tau}}$ one can associate, through the Hill determinant construction, 
two constants $\mf{h}$, $\wt{\mf{h}}$ and a collection of variables $\bs{\de}$, $\wt{\bs{\de}}$.

Then, we introduce the functions 
\beq
q_\pm(\la)\, = \, \f{ Q_{\pm}(\la) }{  \th_{ \pm\bs{\de} }(\pm\la) } 
\label{definition fcts q pm solutions de TQ self dual}
\enq
with
\beq 
Q_{+}(\la) \, = \,  \big( \varkappa g^N \big)^{-\i\la}   \, \psi_+(\la) \, \wt{\psi}_{+}(\la) \, f_{p_0}^{(+)}(\la)  \quad \e{and} \quad
Q_-(\la)\, = \,   \big( g^N \big)^{\i\la }\, \psi_-(\la) \, \wt{\psi}_{-}(\la) \, f_{p_0}^{(-)}(\la)   \;. 
\label{definition Q pm grand determinant}
\enq
The building blocks of $Q_{\pm}$ are constructed with the help of the functions $K_{\pm}$ and $\wt{K}_{\pm}$ as:
\beq
\psi_+(\la) \, = \, K_+(\la)  \; \Big( q^2 \ex{\f{2\pi}{\om_2}\la}; q^2 \Big)_{\bs{\tau}} \quad , \quad 
\wt{\psi}_+(\la) \, = \, \wt{\mf{h}}^{-1} \cdot \wt{K}_{+}(\la) \;  \Big( \wt{q}^{\,-2} \ex{-\f{2\pi}{\om_1}\la}; \wt{q}^{\, -2} \Big)_{ \tilde{\bs{\tau}} }
\enq
and 
\beq
\psi_-(\la) \, = \, \mf{h}^{-1} \cdot K_{-}(\la)  \; \Big( q^2 \ex{-\f{2\pi}{\om_2}\la}; q^2 \Big)_{\bs{\tau}} \quad , \quad 
\wt{\psi}_-(\la) \, = \, \wt{K}_{-}(\la) \, \Big( \wt{q}^{\, -2} \ex{\f{2\pi}{\om_1}\la}; \wt{q}^{\, -2} \Big)_{ \tilde{\bs{\tau}} } \;. 
\enq

The expression for the functions $ f_{  p_0  }^{(\pm)}(\la)$ depends on whether one considers the $q$-Toda or the Toda$_2$ chain. In the $q$-Toda case, they take the form 
\beq
f_{ p_0 }^{(\pm)}(\la) \; = \; \exp\bigg\{ \i \f{N \pi \la^2 }{2\om_1\om_2} \mp  \f{N \pi  \Om  }{ 2 \om_1\om_2  }\la  \bigg\} \cdot  \ex{ -\i \tfrac{p_0}{2} \la} \; , 
\quad \e{with} \quad \Om = \om_1 + \om_2 \;, 
\label{definition fde pm pour q Toda}
\enq
while, for the Toda$_2$ chain, they read
\beq
f_{ p_0 }^{(+)}(\la) \; = \;  \ex{- \i p_0 \la}
\qquad \e{and} \qquad 
f_{ p_0 }^{(-)}(\la) \; = \; \ex{ \i \f{N \pi \la^2 }{ \om_1\om_2} + \f{N \pi  }{ \om_1\om_2  }\la \Om   } \;. 
\label{definition fde pm pour Toda2}
\enq

 $Q_{\pm}$ given in \eqref{definition Q pm grand determinant} are entire functions. Also, $f_{ p_0 }^{(\pm)}$ solve the finite difference equations \eqref{eqn diff finie f pm Toda2} or \eqref{eqn diff finie f pm qToda}, depending on the model. 

By using the finite difference equations satisfied by $K_{\pm}$, $\wt{K}_{\pm}$ and elementary transformation properties under $\i \om_1, \i \om_2$
shifts for the $q^2$, $\wt{q}^{\,-2}$ products and the $\th$, $\wt{\th}$ functions, one readily checks that $q_{\pm}$ defined above 
solve eqns. \eqref{ecriture ensemble autodual eqns Baxter 1}-\eqref{ecriture ensemble autodual eqns Baxter 2}.

\subsubsection{Some heuristics leading to the construction of $q_{\pm}$}

We now briefly discuss the role played by the various building blocks of the functions $q_{\pm}$. The two solutions to the set of dual Baxter equations  \eqref{ecriture ensemble autodual eqns Baxter 1}-\eqref{ecriture ensemble autodual eqns Baxter 2} can be factorised as 
\begin{align*}
q_{+}(\la) & = \;\f{ \th_{ \bs{\tau} }(\la) }{ \th_{ \bs{\de} }(\la)  } \cdot q_{+}^{(0)}(\la) \cdot K_{+}(\la)\cdot \wt{K}_{+}(\la) \\
q_{-}(\la) & = \; \f{ \th_{ -\bs{\tau} }(-\la) }{ \th_{ -\bs{\de} }(-\la)  } \cdot q_{-}^{(0)}(\la) \cdot  K_{-}(\la) \cdot \wt{K}_{-}(\la) \;. 
\end{align*}
The prefactors involving $\th$-function correspond to a particular choice of a  quasi-elliptic function. multiplying a given solution. They should be thought of as corresponding
to a particularly convenient normalisation of solutions. The functions 
\beq
q_{+}^{(0)}(\la) \; = \;    
\f{ \Big( \wt{q}^{\,-2} \ex{-\f{2\pi}{\om_1}\la}; \wt{q}^{\, -2} \Big)_{ \tilde{\bs{\tau}} } }{  \big( \varkappa g^{N}\big)^{ \i\la}   \Big(   \ex{-\f{2\pi}{\om_2}\la}; q^2 \Big)_{\bs{\tau}}     } \, f_{p_0}^{(+)}(\la)
\quad \e{and} \quad
q_{-}^{(0)}(\la) \; = \;   
\f{ \Big( \wt{q}^{\,-2} \ex{-\f{2\pi}{\om_1}\la}; \wt{q}^{\, -2} \Big)_{ \tilde{\bs{\tau}} } }{   g^{-\i N \la}  \Big(   \ex{ \f{2\pi}{\om_2}\la}; q^2 \Big)_{\bs{\tau}}     } \, f_{p_0}^{(-)}(\la)
\enq
correspond to the solutions to the "asymptotic" $t-q$ equations, where the second $(+)$ or the first $(-)$ term in the set of self-dual $t-q$ equations \eqref{ecriture ensemble autodual eqns Baxter 1}-\eqref{ecriture ensemble autodual eqns Baxter 2} has been dropped. 
Such equations describe the spectrum of appropriate open Toda chains, and take the form 
\beq
t_{\bs{\tau}}(\lambda ) q_+^{(0)}(\lambda )  = g^{N\om_1} \sigma \,  \varkappa^{\om_1 } q_+^{(0)}(\lambda -\i\omega_1) \qquad \e{and} \qquad 
\tilde{t}_{ \tilde{\bs{\tau}} }(\lambda ) q_+^{(0)}(\lambda ) = g^{N\om_2}  \sigma \,  \varkappa^{\om_2 } q_+^{(0)}(\lambda -\i\omega_2)  \;. 
\enq
as well as 
\beq
t_{\bs{\tau}}(\lambda ) q_-^{(0)}(\lambda )  = g^{N\om_1} \sigma^{-1} q_-^{(0)}(\lambda +\i\omega_1) \qquad \e{and} \qquad 
\tilde{t}_{ \tilde{\bs{\tau}} }(\lambda ) q_-^{(0)}(\lambda ) = g^{N\om_2}  \sigma^{-1} \,  q_-^{(0)}(\lambda +\i\omega_2)  \;. 
\enq
The functions $K_{\pm}$, resp. $\wt{K}_{\pm}$, correct the behaviour of the full solution for the $\i \om_1$, resp. dual $\i \om_2$, regime 
so that one may hope to obtain a solution to the full self-dual $t-q$ equations \eqref{ecriture ensemble autodual eqns Baxter 1}-\eqref{ecriture ensemble autodual eqns Baxter 2}.

\subsection{A rewriting in terms of solutions to non-linear integral equations}
\label{Subsection rewriting through NLIE}

 \subsubsection{The auxiliary functions $Y_{\bs{\de}}$ and $ \wt{Y}_{ \tilde{\bs{\de}} }$ }
 
For further purpose, we single out two strips in the complex plane:
\beq
\mc{B}=  \Big\{ z \in \mathbb{C} :  z = \i x\om_1 + \i y \om_2 \;  \; (x,y) \in \R \times  ]-1/2;1/2[  \Big\}
\label{definition bande B}
\enq
and its dual 
\beq
\wt{\mc{B}} = \Big\{ z \in \mathbb{C} :  z = \i x \om_1 + \i y \om_2 \;  \; (x,y) \in   ]-1/2;1/2[  \times \R  \Big\}\;. 
\label{definition bande B tilde}
\enq
They are depicted in Fig. \ref{domaine B et B dual}. 
\begin{figure}[h]
\begin{center}

\begin{pspicture}(8,4)

\psline{<->}(6.35,2.75)(5.65,1.25)

\psline{->}(6,2)(6.5,0.7)

\psline(5.85,4.05)(6.85,1.45)

\psline(6.15,-0.05)(5.15,2.55)

\rput(6.7,2.8){$\tfrac{\i\om_1}{2}$}
\rput(5.2,1){$-\tfrac{\i\om_1}{2}$}

\rput(6.5,1.5){$\i\om_2$}

\rput(6.7,0.3){$\wt{\mc{B}}_+$}
\rput(5.5,3.5){$\wt{\mc{B}}_-$}

\pscurve[linestyle=dashed](6.35,2.75)(6,2.75)(5.65,1.25)

\psline[linewidth=2pt]{->}(6,2.75)(5.95,2.75)

\rput(5.4,2.7){$\wt{\msc{C}}_{ \tilde{\bs{\de}}; \rho }$}




\psline{->}(1.75,1.3)(2.45,2.8)


\rput(1.1,2.2){$-\tfrac{\i\om_2}{2}$}
\rput(2.3,0.5){$\tfrac{\i\om_2}{2}$}

\rput(1.9,2.2){$ \i \om_1$}


\psline{<->}(1.5,2,65)(2,0.75)

\psline(2.2,3.5)(0.8,0.5)

\psline(2.7,2.3)(1.65,-0.1)

\rput(0.8,0){$\mc{B}_-$}
\rput(2.5,3.3){$\mc{B}_+$}

\pscurve[linestyle=dashed](1.5,2,65)(1.8,0.7)(2,0.75)
\psline[linewidth=2pt]{->}(1.8,0.75)(1.85,0.7)
\rput(1.3,0.8){$\msc{C}_{ \bs{\de} ; \rho }$}

\end{pspicture}

\caption{The strips $\mc{B}$ and $\wt{\mc{B}}$ and the curves $\msc{C}_{ \bs{\de} }$ and $\wt{\msc{C}}_{ \tilde{\bs{\de}} }$.\label{domaine B et B dual}}
\end{center}
\end{figure}

 Further, we observe that the Hill determinants $\mc{H}(\la)$ and $\wt{\mc{H}}(\la)$ can be decomposed as
\beq
\mc{H}(\la) \, = \,  u_+(\la) \, u_-(\la)  \qquad \e{and} \qquad \wt{\mc{H}}(\la) \, = \,  \wt{u}_+(\la)\,  \wt{u}_-(\la) \; ,
\enq
where we have set 
\beq
u_-(\la) = \mf{h} \f{  \Big(  \ex{-\f{2\pi}{\om_2}\la}; q^2 \Big)_{\bs{\de}}   }{   \Big(   \ex{-\f{2\pi}{\om_2}\la}; q^2 \Big)_{\bs{\tau}}    }    
\qquad \e{and} \qquad 
u_+(\la) =  \f{  \Big( q^2 \ex{\f{2\pi}{\om_2}\la}; q^2 \Big)_{\bs{\de}}   }{   \Big( q^2 \ex{\f{2\pi}{\om_2}\la}; q^2 \Big)_{\bs{\tau}}    }  \;.
\label{definition u pm}
\enq
Their duals have a similar representation 
\beq
\wt{u}_+(\la) = \wt{\mf{h}}  \f{  \Big( \wt{q}^{\,-2} \ex{-\f{2\pi}{\om_1}\la}; \wt{q}^{\, -2} \Big)_{ \tilde{\bs{\de}} }   }{  \Big( \wt{q}^{\,-2} \ex{-\f{2\pi}{\om_1}\la}; \wt{q}^{\, -2} \Big)_{ \tilde{\bs{\tau}} } }
\qquad  \e{and}  \qquad 
\wt{u}_-(\la) = \f{  \Big(  \ex{\f{2\pi}{\om_1}\la}; \wt{q}^{\, -2} \Big)_{ \tilde{\bs{\de}} }   }{  \Big(   \ex{ \f{2\pi}{\om_1}\la}; \wt{q}^{\, -2} \Big)_{ \tilde{\bs{\tau}} } } \;.
\label{definition u tilde pm}
\enq

It is easy to see that the functions $\la \,  \mapsto  \, \tf{  K_{\pm}\big(\la\mp\i\tf{\om_1}{2} \big)  }{ u_{\pm}\big(\la-\i\tf{\om_1}{2} \big) }$
have only simple poles and that these form the lattice $\big\{ \de_k \mp \i\tf{\om_1}{2} \mp \i \mathbb{N} \om_1 + \i \mathbb{Z} \om_2 \big\}$. 
Furthermore, owing to the asymptotic expansions of $K_{\pm}$, one has that 
\beq
\f{  K_{+}\big(\la-\i\tfrac{\om_1}{2} \big)  }{ u_{+}\big(\la-\i\tfrac{\om_1}{2} \big) }  \underset{x\rightarrow + \infty }{\longrightarrow} 1 \qquad \e{and} \qquad 
\f{  K_{-}\big(\la+\i\tfrac{\om_1}{2} \big)  }{ u_{-}\big(\la-\i\tfrac{\om_1}{2} \big) }  \underset{x\rightarrow - \infty }{\longrightarrow}  \f{ \mf{h}^{-1} }{ 1 - \rho^{\om_1} \ex{2\om_1 p_0} \bs{1}_{\e{Toda}_2} }
\label{ecriture asymptotiques de ratios K pm dans B pm}
\enq
for $\la=\i x \om_1+\i y \om_2 \in \mc{B}$ and where $\bs{1}_{\e{Toda}_2}=1$ for the Toda$_2$ model and  $\bs{1}_{\e{Toda}_2}=0$ for the $q$-Toda chain.  Thus, there exists some $x_{0}\geq 0$ such that 
these functions have no zeroes in the region corresponding to $\pm x > x_0$. 

Thus, one may introduce a curve $\msc{C}_{\bs{\de}; \rho}$, starting at $\i x_0^{\prime} \om_1-\i \tf{\om_2}{2}$
and ending at $\i x_0^{\prime} \om_1+\i \tf{\om_2}{2}$, for some $x_0^{\prime}$, that divides the strip $\mc{B}$ in two disjoint domains $\mc{B}_{\pm}$
such that $\pm \i t \in \mc{B}_{\pm}$ when $t>0$ and large enough. The curve is oriented in such a way that $\mc{B}_+$ is to its left, see Fig. \ref{domaine B et B dual}. 
The curve $\msc{C}_{\bs{\de}; \rho}$ is chosen in such a way that: 
\begin{itemize}
 
\item the poles and zeroes of $  \f{  K_{+}\big(\la-\i\tfrac{\om_1}{2} \big)  }{ u_{+}\big(\la-\i\tfrac{\om_1}{2} \big) }$ are all in $\mc{B}_{ - }$;   

\item the poles and zeroes of $  \f{  K_{-}\big(\la+\i\tfrac{\om_1}{2} \big)  }{ u_{-}\big(\la-\i\tfrac{\om_1}{2} \big) }$ are all in $\mc{B}_{ + }$. 

\end{itemize}

This property ensures that there exist holomorphic and $\i \om_2$ periodic determinations of the logarithms 
\beq
\mf{l}_+(\la) \, = \, \ln\bigg[  \f{  K_{+}\big(\la-\i\tf{\om_1}{2} \big)  }{ u_{+}\big(\la-\i\tf{\om_1}{2} \big) }\bigg] \quad \e{in} \quad \mc{B}_{+}
\quad \e{and} \quad 
\mf{l}_-(\la) \, = \,  \ln\bigg[ \f{  K_{-}\big(\la+\i\tf{\om_1}{2} \big)  }{ u_{-}\big(\la-\i\tf{\om_1}{2} \big) } \bigg] \quad \e{in} \quad \mc{B}_{-} \;,  
\enq
which, owing to \eqref{ecriture asymptotiques de ratios K pm dans B pm}, are bounded functions in $\mc{B}_{\pm}$. 
In fact one can even choose the determination so that $\mf{l}_+( \i x \om_1 + \i y \om_2) \rightarrow 0  $ for $x \rightarrow +\infty$, and uniformly in $y\in \R$. 

Further, we introduce two functions on $\msc{C}_{\bs{\de}; \rho}$:
\begin{eqnarray}
Y_{\bs{\de}}(\la) & = & \mf{l}_+(\la+\i \om_1) \, + \, \mf{l}_-(\la-\i \om_1) \; ,  \label{definition fct Y delta via matrices K pm}    \\
\mc{V}_{\bs{\de}}(\la) &=& \f{  K_+(\la-\i \tf{\om_1}{2} ) K_-(\la+\i \tf{\om_1}{2} )  }{  \mc{H}(\la-\i \tf{\om_1}{2} )  }   \;. 
\label{definition fct Vdelta via matrices K pm}
\end{eqnarray}
It is easy to see that $\mc{V}_{\bs{\de}}$ admits a continuous determination of its logarithm on 
$\msc{C}_{\bs{\de}; \rho}$ given by 
\beq
\ln \big[\mc{V}_{\bs{\de}}\big](\la)  \, = \,  \mf{l}_+(\la) \, + \, \mf{l}_-(\la) 
\label{ecriture determination du log de V delta}
\enq
and that one has the explicit representation
\beq
\ex{ Y_{\bs{\de}}(\la)  } \; = \; \ex{\om_1 \zeta} \cdot \f{ K_+(\la+\i \tfrac{\om_1}{2} )K_-(\la-\i \tfrac{\om_1}{2} ) }{ \mc{H}(\la-\i \tfrac{\om_1}{2} ) }  \cdot
\f{ t_{ \bs{\de}}(\la-\i \tfrac{\om_1}{2} ) \, t_{ \bs{\de}}(\la+\i \tfrac{\om_1}{2} ) }   
			{ t_{\bs{\tau}}(\la-\i \tfrac{\om_1}{2} ) \, t_{\bs{\tau}}(\la+\i \tfrac{\om_1}{2} ) } \;. 
\label{expression explicite  pour Y delta en terms K pm etc}
\enq
The latter is a consequence of the identity
\beq
u_+\big( \la+\i\tfrac{\om_1}{2} \big)\, u_-\big( \la-3 \i\tfrac{\om_1}{2} \big)  \; = \; \ex{-\zeta \om_1 } 
\f{ t_{\bs{\tau}}(\la-\i \tfrac{\om_1}{2} ) t_{\bs{\tau}}(\la+\i \tfrac{\om_1}{2} ) }
		      { t_{ \bs{\de}}(\la-\i \tfrac{\om_1}{2} ) t_{ \bs{\de}}(\la+\i \tfrac{\om_1}{2} ) }    \mc{H}\big( \la-\i\tfrac{\om_1}{2} \big) \;. 
\enq

Finally, the functional relation between the Hill determinant and $K_{\pm}$ entails that $Y_{\bs{\de}}$ and $\mc{V}_{\bs{\de}}$ are related as 
\beq
\mc{V}_{\bs{\de}}(\la) \; = \; 1 \, + \, \f{ \rho^{\om_1} \ex{Y_{ \bs{\de} }(\la)}  }{ t_{ \bs{\de}}(\la-\i\tfrac{\om_1}{2}) t_{ \bs{\de}}(\la + \i\tfrac{\om_1}{2})    } \;. 
\enq

Analogously, one introduces the dual quantities.  We only insist on the differences occurring in the dual setting. There, one has the 
 asymptotic expansions 
\beq
\f{  \wt{K}_{+}\big(\la-\i\tfrac{\om_2}{2} \big)  }{ \wt{u}_{+}\big(\la-\i\tfrac{\om_2}{2} \big) }  \underset{y\rightarrow + \infty }{\longrightarrow} \f{ \wt{\mf{h}}^{-1} }{ 1 - \rho^{\om_2} \ex{2\om_2 p_0} \bs{1}_{\e{Toda}_2} } \qquad \e{and} \qquad 
\f{  \wt{K}_{-}\big(\la+\i\tfrac{\om_2}{2} \big)  }{ \wt{u}_{-}\big(\la-\i\tfrac{\om_2}{2} \big) }  \underset{y\rightarrow - \infty }{\longrightarrow}  1
\label{ecriture asymptotiques de ratios duaux K pm dans tilde B pm}
\enq
for $\la=\i x \om_1+\i y \om_2 \in \wt{\mc{B}}$. One then defines the curve $\wt{\msc{C}}_{ \tilde{\bs{\de}}; \rho}$ analogously to $\msc{C}_{ \bs{\de}; \rho}$, although one should pay attention to the fact that 
the dual curve has to be oriented so that the domain $ \wt{\mc{B}}_+$ is to its left, see Fig. \ref{domaine B et B dual}. 
One likewise defines dual, $\i \om_1$ periodic, holomorphic determinations of the logarithms 
\beq
\wt{\mf{l}}_+(\la) \, = \, \ln\bigg[  \f{  \wt{K}_{+}\big(\la-\i\tf{\om_2}{2} \big)  }{ \wt{u}_{+}\big(\la-\i\tf{\om_2}{2} \big) }\bigg] \quad \e{in} \quad \wt{\mc{B}}_{+}
\quad \e{and} \quad 
\wt{\mf{l}}_-(\la) \, = \,  \ln\bigg[ \f{  \wt{K}_{-}\big(\la+\i\tf{\om_2}{2} \big)  }{ \wt{u}_{-}\big(\la-\i\tf{\om_2}{2} \big) } \bigg] \quad \e{in} \quad \wt{\mc{B}}_{-} \;,  
\enq
which, owing to \eqref{ecriture asymptotiques de ratios duaux K pm dans tilde B pm}, are bounded functions in $\mc{B}_{\pm}$. 
In fact, one can even choose the determination so that $\wt{\mf{l}}_-( \i x \om_1 + \i y \om_2) \rightarrow 0  $ for $y \rightarrow -\infty$, and uniformly in $x\in \R$.

Define two kernels $K$, $\wt{K}$ as
\beq
K(\la)= \f{1}{2\i\om_2} 
\Big\{  \coth\Big[\f{\pi}{\om_2}(\la -\i\om_1)\Big] - \coth\Big[\f{\pi}{\om_2}(\la +\i\om_1)\Big]  \Big\} 
\qquad \e{and} \qquad 
\wt{K}=K_{\mid \om_1 \leftrightarrow \om_2} \, .
\label{definition noyau K}
\enq

Then, the functions $Y_{\bs{\de}}$ introduced in \eqref{definition fct Y delta via matrices K pm} and its dual counterpart $\wt{Y}_{ \tilde{\bs{\de}} }$ satisfy to the below non-linear integral equations:
\beq
 Y_{ \bs{\de} } (\la)= \Int{ \msc{C}_{\bs{\de};\rho} }{}    K(\la-\tau)
\ln\bigg[  1+  \f{ \rho^{\om_1}  \ex{ Y_{ \bs{\de}} (\tau)}  }{ t_{ \bs{\de}}\big( \tau - \i\tfrac{\om_1}{2}) t_{ \bs{\de}}\big(\tau + \i\tfrac{\om_1}{2}\big)   }    \bigg] \cdot  \dd \tau 
\label{eqn NLIE pour Y}
\enq
and its dual
\beq
 \wt{Y}_{ \tilde{\bs{\de}}} (\la) =  \Int{ \wt{\msc{C}}_{ \tilde{\bs{\de}};\rho} }{}     \wt{K}(\la-\tau)
\ln\bigg[  1+  \f{ \rho^{\om_2} \ex{\wt{Y}_{ \tilde{\bs{\de}} } (\tau)}  }{ \tilde{t}_{ \tilde{\bs{\de}}}\big( \tau -\i\tfrac{\om_2}{2} \big) \tilde{t}_{ \tilde{\bs{\de}}}\big( \tau + \i\tfrac{\om_2}{2} \big)    }    \bigg] \cdot \dd \tau  \;. 
\label{eqn NLIE pour Y dual}
\enq
The logarithms in the integrand are to be understood as being given by the determinations of $\ln \big[\mc{V}_{\bs{\de}}\big](\la)$ and 
$\ln \big[ \wt{\mc{V}}_{ \tilde{\bs{\de}}}\big](\la)$ that have been provided above.

The non-linear integral equation property can be established as follows. Let $ \la \in \msc{C}_{\bs{\de};\rho} $ and  set
\beq
\Psi(\la)=\Int{ \msc{C}_{\bs{\de};\rho} }{} 
 \dd \tau K(\la-\tau) \ln \big[\mc{V}_{\bs{\de}}\big](\la)\;.
\enq
By using the expression \eqref{ecriture determination du log de V delta} and the fact that $\la \mapsto \mf{l}_{\pm}(\la)$ are analytic in $\mc{B}_{\pm}$, $\i\om_2$ periodic and 
bounded when $\la\rightarrow  \infty$, $\la \in \mc{B}_{\pm}$, one may recast $\Psi$ in the form 
\beq
\Psi(\la)=\Int{ \partial\mc{B}_+ }{}  \dd \tau K(\la-\tau) \mf{l}_{+}(\tau) +
\Int{ -\partial\mc{B}_- }{}  \dd \tau K(\la-\tau) \mf{l}_{-}(\tau)  \;. 
\enq
Here $\partial \mc{B}_{\pm}$ stands for the canonically oriented boundary of $\mc{B}_{\pm}$. 
Indeed, the contribution of the lateral boundaries does not contribute to each of the integrals due to the $\i\om_2$ periodicity of the integrand while the
boundaries at infinity produce vanishing contributions since, for $\la \in \msc{C}_{\bs{\de};\rho}$ the kernel $\tau \mapsto K(\la-\tau)$
decays when $\tau \rightarrow \infty$, $\tau \in \mc{B}$. Then, since $\tau \mapsto K(\la-\tau)$ is meromorphic on $\mc{B}$ and, for $\la \in \msc{C}_{\bs{\de};\rho}$,
has simple poles at $\tau=\la \pm \i \om_1 \in \mc{B}_{\pm}$ with residues $\tf{\pm 1}{ (2\i\pi) }$, one concludes that 
\beq
\Psi(\la)= \mf{l}_+(\la+\i \om_1) \, + \, \mf{l}_-(\la-\i \om_1)  \, = \, Y_{\bs{\de}}(\la) \;, 
\enq
hence the claim. The dual case is dealt with in much the same way.

\subsubsection{Integral representations for $K_{\pm}$, $\wt{K}_{\pm}$}

To start with, one introduces two auxiliary functions  $v_{\ua}(\la), v_{\da}(\la-\i\om_1)$ by means of the below integral representations 
\beq
v_{\ua}(\la) \, = \, \exp\Bigg\{ -\Int{ \msc{C}_{\bs{\de};\rho} }{} \Big\{ \coth\Big[ \f{\pi}{\om_2}(\la-\tau +\i\tfrac{\om_1}{2} ) \Big] +1 \Big\}\ln\big[ \mc{V}_{\bs{\de}}\big] (\tau) \cdot  \f{\dd \tau }{ 2\i \om_2 }  \Bigg\}
\label{definition v up}
\enq
and
\beq
v_{\da}(\la-\i\om_1) \, = \, \exp\Bigg\{  \Int{\msc{C}_{\bs{\de};\rho}}{}  \Big\{ \coth\Big[ \f{\pi}{\om_2}(\la-\tau -\i\tfrac{\om_1}{2} ) \Big] +1  \Big\}  \ln\big[ \mc{V}_{\bs{\de}}\big] (\tau) \cdot  \f{\dd \tau }{ 2\i \om_2 }  \Bigg\} \;. 
\label{definition v down}
\enq
Here, $\mc{V}_{\bs{\de}}$ is as defined through \eqref{definition fct Vdelta via matrices K pm} and $\ln\big[ \mc{V}_{\bs{\de}}\big]$ is the appropriate determination for its logarithm on $\msc{C}_{\bs{\de};\rho}$
given in \eqref{ecriture determination du log de V delta}. 

Similarly, one introduces the dual functions on $\wt{\msc{C}}_{ \tilde{\bs{\de}};\rho}$
\beq
\wt{v}_{\ua}(\la) \, = \, \exp\Bigg\{ -\Int{ \wt{\msc{C}}_{ \tilde{\bs{\de}};\rho} }{}   \Big\{ \coth\Big[ \f{\pi}{\om_1}(\la-\tau +\i\tfrac{\om_2}{2} ) \Big] +1  \Big\}  \ln\big[ \wt{\mc{V}}_{ \tilde{\bs{\de}} }\big] (\tau) \cdot  \f{\dd \tau }{ 2\i \om_1 }  \Bigg\}
\label{definition v up tilde}
\enq
and
\beq
\wt{v}_{\da}(\la-\i\om_2) \, = \, \exp\Bigg\{  \Int{ \wt{\msc{C}}_{ \tilde{\bs{\de}};\rho} }{}  \Big\{ \coth\Big[ \f{\pi}{\om_1}(\la-\tau -\i\tfrac{\om_2}{2} ) \Big] + 1  \Big\}  \ln\big[\wt{\mc{V}}_{ \tilde{\bs{\de}} } \big] (\tau) \cdot  \f{\dd \tau }{ 2\i \om_1 }  \Bigg\} \;. 
\label{definition v down tilde}
\enq

These functions are closely related to $K_{\pm}$ and $\wt{K}_{\pm}$. Indeed, it holds for $\la \in \msc{C}_{\bs{\de};\rho}$ that
\beq
K_+(\la) \, = \, u_+(\la) \, v_{\ua}(\la) \; , \quad K_-(\la) \, = \, u_-(\la-\i\om_1) \, v_{\da}(\la-\i\om_1) 
\label{ecriture K pm en fct de nu up down}
\enq
and, similarly for the dual analogues, given $\la \in \wt{\msc{C}}_{ \tilde{\bs{\de}};\rho}$, 
\beq
\wt{K}_+(\la) \, = \, \wt{u}_+(\la) \, \wt{v}_{\ua}(\la) \; , \quad \wt{K}_-(\la) \, = \, \wt{u}_-(\la-\i\om_2)\,  \wt{v}_{\da}(\la-\i\om_2)  \;. 
\label{ecriture tilde K pm en fct de tilde nu up down}
\enq
We remind that the functions $u_{\pm}$, resp. $\wt{u}_{\pm}$, have been defined in \eqref{definition u pm}, resp. \eqref{definition u tilde pm}. 
Note that \eqref{ecriture K pm en fct de nu up down}-\eqref{ecriture tilde K pm en fct de tilde nu up down} hold in fact everywhere by meromorphic continuation.

These factorisations can be established by residue calculations analogous to those used in the proof that $Y_{\bs{\de}}$ and $\wt{Y}_{\tilde{\bs{\de}}}$ solve  the non-linear integral equations
\eqref{eqn NLIE pour Y} and \eqref{eqn NLIE pour Y dual}. 
We only specify that the $+1$ term in the integrands are chosen in such a way that 
\beq
\coth\Big[ \f{\pi}{\om_2}(\la-\tau \pm \i\tfrac{\om_1}{2} ) \Big] +1  \underset{ \tau \rightarrow \infty, \tau \in \mc{B}_- }{\longrightarrow} 0 
\qquad \e{and} \qquad \coth\Big[ \f{\pi}{\om_1}(\la-\tau \pm \i\tfrac{\om_2}{2} ) \Big] + 1   \underset{ \tau \rightarrow \infty, \tau \in \wt{\mc{B}}_+ }{\longrightarrow} 0 \;. 
\enq
This allows one to ensure that when transforming $\msc{C}_{\bs{\de};\rho}$, resp. $\wt{\msc{C}}_{ \tilde{\bs{\de}};\rho}$, into 
$ \pm \partial \mc{B}_{\pm}$, resp. $ \pm \partial \wt{\mc{B}}_{\pm}$, the boundary at infinity of $ - \partial \mc{B}_{-}$, resp. $ \partial \wt{\mc{B}}_{+}$
does not produce additional contributions stemming from the fact that $\mf{l}_{-}(\tau)$, resp. $\wt{\mf{l}}_+(\tau)$, does not approach zero 
when $\tau \rightarrow \infty$ with $\tau \in \mc{B}_{-}$, resp. with $\tau \in \wt{\mc{B}}_{+}$.

Various properties of the functions  $v_{\ua/\da}$ which follow from their definition \eqref{definition v up}-\eqref{definition v down} and the fact that $Y_{\bs{\de}}$ solve \eqref{eqn NLIE pour Y} are obtained in 
Subsection \ref{Subappendix Auxiliary fcts nu up down}. In particular, it is shown there that  $\la \mapsto v_{\ua}(\la)$ has simple poles at 
$\de_k - \i\mathbb{N}^{*}\om_1 + \i\mathbb{Z} \om_2$ with non-vanishing residues. Likewise, $\la \mapsto v_{\da}(\la-\i\om_1)$
has simple poles at $\de_k + \i\mathbb{N}^{*}\om_1 + \i\mathbb{Z} \om_2$ with non-vanishing residues.
Although the pole property is rather clear on the level of the representations \eqref{ecriture K pm en fct de nu up down}, the fact that these are genuine poles, 
\textit{viz}. that the residues are non-vanishing, is, however, not directly apparent.

It is interesting to observe the form of the finite-difference equations satisfied by the $v_{\ua/\da}$ and their duals. By using their relation to $K_{\pm}$, $\wt{K}_{\pm}$
and the finite difference equations satisfied by the latter one gets that 
\begin{eqnarray}
t_{\bs{\tau}}(\la) v_{\ua}(\la) & = & t_{\bs{\de}}(\la) v_{\ua}(\la-\i \om_1)  \, + \; \rho^{\om_1} \f{  v_{\ua}(\la+\i\om_1 )  }{ t_{\bs{\de}}(\la+\i\om_1)  } \\
t_{\bs{\tau}}(\la) v_{\da}(\la-\i \om_1) & = & t_{\bs{\de}}(\la) v_{\da}(\la )  \, + \; \rho^{\om_1} \f{  v_{\da}(\la-2\i\om_1 )  }{ t_{\bs{\de}}(\la-\i\om_1)  } \;. 
\end{eqnarray}
The dual equations follow by the duality transformation. One should note that, in these equations, the parameters  $\bs{\tau}$ and $\bs{\de}$
are not independent but related through \eqref{ecriture factorisation determinants de Hill et son dual}, \textit{viz}. the $\de_{a}$'s are the representatives of the lattice of zeroes
of the Hill determinant that is built up from the parameters $\tau_a$. Observe also that these equation allow one to express 
the polynomial $t_{\bs{\tau}}(\la)$ solely in terms of $v_{\ua/\da}$ and of the polynomials $t_{\bs{\de}}$. By using the Wronskian relation satisfied by $v_{\ua/\da}$ \eqref{ecriture relation Wronskien}, 
one gets that 
\beq
 t_{\bs{\tau}}(\la)\;  =  \;   t_{ \bs{\de}}(\la) v_{\ua}(\la-\i\om_1) v_{\da}(\la) 
\, - \, \rho^{2\om_1} \f{ v_{\ua}(\la+\i\om_1) v_{\da}(\la - \i\om_1)  }{ \pl{\eps=0, \pm 1}{} t_{ \bs{\de}}(\la+\i\eps \om_1 )   } \;. 
\label{ecriture reconstruction polynome t tau via les nus}
\enq
An analogous equation holds for the dual case.

\subsubsection{New representation for the functions $\psi_{\pm}$, $\wt{\psi}_{\pm}$}

The results of the previous analysis allow one to recast the building blocks $\psi_{\pm}$, $\wt{\psi}_{\pm}$ in terms of quantities 
solely depending on the parameters $\bs{\de}$, $\tilde{\bs{\de}}$ and hence drop the connection with the parameters $\bs{\tau}, \tilde{\bs{\tau}}$. 
These functions are expressed as 
\beq
\psi_+(\la) \, = \, v_{\ua}(\la)  \; \Big( q^2 \ex{\f{2\pi}{\om_2}\la}; q^2 \Big)_{\bs{\de}} \quad , \quad 
\wt{\psi}_+(\la) \, = \, \wt{v}_{\ua}(\la) \;  \Big( \wt{q}^{\,-2} \ex{-\f{2\pi}{\om_1}\la}; \wt{q}^{\, -2} \Big)_{ \tilde{\bs{\de}} }
\enq
and 
\beq
\psi_-(\la) \, = \, v_{\da}(\la-\i\om_1)  \, \Big( q^2 \ex{-\f{2\pi}{\om_2}\la}; q^2 \Big)_{\bs{\de}} \quad , \quad 
\wt{\psi}_-(\la) \, = \, \wt{v}_{\da}(\la-\i\om_2) \, \Big( \wt{q}^{\, -2} \ex{\f{2\pi}{\om_1}\la}; \wt{q}^{\, -2} \Big)_{ \tilde{\bs{\de}} } \;. 
\enq

From the preceding discussions, it follows that:
\begin{itemize}

 \item $\psi_{\pm}$ is entire 
 
 \item $\psi_{\pm}$ does not 
vanish on the lattice  $\de_k+\i\mathbb{Z} \om_1 +\i\mathbb{Z} \om_2$ as follows from the properties of $v_{\ua/\da}$ discussed earlier on; 
\item  Dual properties hold for  $\wt{\psi}_{\pm}$; 
 
\end{itemize}

The parametrisation of $q_{\pm}$'s building blocks in terms of the functions $v_{\ua/\da}$, $\wt{v}_{\ua/\da}$ allows one to prove that $q_{\pm}$ are 
 two linearly independent solutions of the set of dual Baxter $t-q$ equations. 
The linear independence is a consequence of $q_{\pm}$ not sharing a common set of zeroes.
Indeed, let $\mc{Z}$, resp. $\wt{\mc{Z}}$, be the set of zeroes of $\mc{V}_{\bs{\de}}$, resp. $\wt{\mc{V}}_{ \tilde{\bs{\de}} }$, and let 
$\mc{Z}_{\pm}=\mc{Z}\cap \mc{B}_{\pm}$ and $\wt{\mc{Z}}_{\pm}= \wt{\mc{Z}}\cap \wt{\mc{B}}_{\pm}$. 
The results obtained in Subsection \ref{Subappendix Auxiliary fcts nu up down} of the appendix ensure that $q_{\pm}$
has zeroes at $\big\{ \mc{Z}_{\pm} \pm \i\tf{\om_1}{2} + \i \mathbb{Z}\om_2 \big\} \cup \big\{ \wt{\mc{Z}}_{ \pm } \pm \i\tf{\om_2}{2} + \i \mathbb{Z}\om_1 \big\} $. 
These sets are obviously disconnected. It remains to  show that  $\mc{Z}_{-}$ is non-empty. This is a consequence of  the fact that $\ex{ Y_{\bs{\de}}(\la) }$ has poles at $\de_{k} \pm \i \tf{\om_1}{2} \pm \i \mathbb{N}^* \om_1 + \i\mathbb{Z} \om_2$.

\subsection{Quantisation conditions in the form of Bethe equations for the $\de_k$'s}
\label{Subsection Quantisation conditions by HLBAE}

For the purpose of stating the result, it is convenient to introduce two functions 
\begin{empheq}{align*}
\mc{I}_{\bs{\de}} (\la) &=  -\Int{ \msc{C}_{ \bs{\de};\rho} }{} 
\Big\{ \coth\Big[ \f{\pi}{\om_2}(\la-\tau + \i \f{\om_1}{2} ) \Big]  +  \coth \Big[\f{\pi}{\om_2}(\la-\tau-\i \f{\om_1}{2} ) \Big]+ 2 \Big\}
\ln[\mc{V}_{\bs{\de}}](\tau) \cdot \f{\dd \tau}{2\i\om_2}   \; , \nonumber \\
\wt{\mc{I}}_{\tilde{\bs{\de}}}(\la) &=  - \Int{ \wt{\msc{C}}_{\tilde{\bs{\de}};\rho} }{} 
\Big\{ \coth\Big[\f{\pi}{\om_1}(\la-\tau + \i \f{\om_2}{2})\Big] +  \coth \Big[ \f{\pi}{\om_1}(\la-\tau-\i\f{\om_2}{2}) \Big] + 2 \Big\}
\ln [ \, \wt{\mc{V}}_{ \tilde{\bs{\de}} }] (\tau) \cdot \f{ \dd \tau }{ 2 \i\om_1 }   \; .
\nonumber
\end{empheq}

Let $q$ be an entire solution of the set of self-dual Baxter equations \eqref{ecriture ensemble autodual eqns Baxter 1}-\eqref{ecriture ensemble autodual eqns Baxter 2} 
associated with the polynomials $t_{\bs{\tau}}$ and $\wt{  t}_{ \tilde{\bs{\tau}} }$. Let $\bs{\de}$ and $\tilde{\bs{\de}}$
be the representatives of the zeroes of the associated Hill determinant \eqref{definition det Hill} and its dual \eqref{definition det hill dual},
such that the coordinates $\de_k$ are pairwise distinct modulo the lattice $\i\om_1 \mathbb{Z}+\i\om_2\mathbb{Z}$

Then it holds that $\bs{\de}=\tilde{\bs{\de}}$ and, up to an overall inessential constant, $q$ can be expressed in terms of the elementary solutions $q_{\pm}$ \eqref{definition fcts q pm solutions de TQ self dual}
as 
\beq
q(\la)=q_+(\la)-\xi q_-(\la) \;. 
\enq
The set of $N+1$ parameters $(\bs{\de},\xi)$ satisfies to the constraint
\beq
\sul{\ell=1}{N} \de_{\ell} \, = \, \f{ \om_1 \om_2 }{ 2\pi } \,  p_0 
\label{ecriture contrainte finale sur les deltas}
\enq
as well as to the Bethe equations which take the form 

\begin{itemize}
 
 \item $q$--Toda
\beq
\ex{ \tfrac{2\i\pi \kappa_1 }{ \om_1 \om_2 } \de_k \, + \,  \mc{I}_{\bs{\de}} (\de_k) \, + \,  \wt{\mc{I}}_{\bs{\de}} (\de_k) } \cdot 
\pl{\ell=1}{N} \f{ \varpi\big( \de_{\ell}-\de_k+\i\tf{\Om}{2} \big)  } { \varpi\big(\de_k-\de_{\ell}+\i\tf{\Om}{2} \big)  }
\, = \, \xi \ex{ \tfrac{\Om}{2} p_0 }
\quad \e{for} \quad  k=1,\dots,N \;. 
\label{ecriture HLBAE qToda}
\enq

  \item Toda$_2$
\beq
\ex{ \tfrac{2\i\pi \kappa_2 }{ \om_1 \om_2 } \de_k \, - \i \f{\pi N }{ \om_1 \om_2 } \de_k^2 \, + \,  \mc{I}_{\bs{\de}} (\de_k) \, + \,  \wt{\mc{I}}_{\bs{\de}} (\de_k) } \cdot
\pl{\ell=1}{N} \f{ \varpi\big( \de_{\ell}-\de_k+\i\tf{\Om}{2} \big)  } { \varpi\big(\de_k-\de_{\ell}+\i\tf{\Om}{2} \big)  }
\, = \, \xi \ex{ \tfrac{\Om}{2} p_0 }
\quad \e{for} \quad  k=1,\dots,N \;. 
\label{ecriture HLBAE Toda2}
\enq
\end{itemize}
These equations are expressed in terms of the quantum dilogarithm defined in \eqref{definition dilogarithme}\;.

Prior to giving a proof of this statement, some remarks are in order.

 \begin{itemize}
  \item[i)] The Bethe equations \eqref{ecriture HLBAE qToda}, \eqref{ecriture HLBAE Toda2} 
are manifestly modular invariant. Furthermore, if the reality condition $\om_1=\overline{\om_2}$ and $\kappa_a \in \R$ holds,
then the Bethe equations are compatible with real roots $\delta_l$ and $\xi \ex{ {\Omega\over 2} p_0}$ being  of modulus one.

Indeed then,   due to \eqref{equation conj complx dilogarithme}, it holds
\begin{empheq}{equation}
\overline{ \f{ \varpi\big( z+\i\tf{\Om}{2} \big)  } { \varpi\big(-z+\i\tf{\Om}{2} \big)  }  } =
  \f{ \varpi\big(-\overline{z}+\i\tf{\Om}{2} \big) }{ \varpi\big( \overline{z}+\i\tf{\Om}{2} \big)  }    \;. 
\end{empheq}
Also in this situation, one can readily check that $ \overline{ \msc{C}_{\bs{\de};\rho} } \, = \, \wt{ \msc{C} }_{ \tilde{\bs{\de}};\rho}$, 
where the complex conjugation also takes care of the change of orientation. 
Further, assuming that $\bs{\de}\in \R^N$, one obtains that $\exp\Big\{ \overline{ Y_{\bs{\de}}(\la) } \Big\} = \exp\Big\{ \wt{Y}_{\bs{\de}}(\overline{\la}) \Big\}$
as follows from the explicit expression \eqref{expression explicite  pour Y delta en terms K pm etc} for these quantities, 
the conjugation properties of the zeroes of the transfer matrices Eigenvalues $\overline{\bs{\tau}}=\tilde{\bs{\tau}}$ and those of the 
Hill determinants \eqref{ecriture conjugaison cas realite des dets de Hill} and of $K_{\pm}$ \eqref{ecriture condition realite K pm et Tilde K pm}. 
From there, one deduces that one may choose branches appropriately so that $\overline{ \ln [\mc{V}_{\bs{\de}}\big](\la) }=\ln [ \wt{\mc{V}}_{ \bs{\de}}\big]( \overline{\la}) $,
and thus, by taking complex conjugation, one gets that  $ \overline{ \mc{I}_{\bs{\de}}(\la)} \, =\, -   \wt{\mc{I}}_{\bs{\de}} (\overline{\la})$.  
This property entails that, in such a case, one has that $\mc{I}_{\bs{\de}} (\la) \, + \,  \wt{\mc{I}}_{\bs{\de}} (\la) $ is purely imaginary for real $\la$. 
\item[ii)] The expressions for $q_{\pm}$ simplify in the case when $\bs{\de}=\tilde{\bs{\de}}$ and read
\beq
q_{+}(\la) \, = \,  \big( \varkappa g^N \big)^{-\i\la}   \, \f{ v_{\ua}(\la) \tilde{v}_{\ua}(\la)  }{ \prod_{\ell=1}^{N} \mc{S}(\la-\de_{\ell}) }\, f_{p_0}^{(+)}(\la)  
\enq
and
\beq
q_-(\la)\, = \,   g^{\i N \la }\,  \f{  v_{\da}(\la-\i\om_1) \tilde{v}_{\da}(\la-\i\om_2)  }{ \prod_{\ell=1}^{N} \mc{S}(\de_{\ell}-\la) }  \, f_{p_0}^{(-)}(\la)    \;, 
\enq
where $\mc{S}$ is the double sine function. It follows immediately from this representation that, when $\bs{\de}=\tilde{\bs{\de}}$, $q_{\pm}$ are manifestly modular invariant. 
\item[iii)] The statement of the quantisation conditions when the coordinates of $\bs{\de}$ are not pairwise distinct is more complicated 
and we chose to omit it here since we believe this situation to be very non-generic. Still, we would like to stress that in the non-pairwise distinct case, one cannot \textit{a priori}
discard the possibility that $\bs{\de} \not= \tilde{\bs{\de}}$. 

\end{itemize}

 \vspace{5mm}
 
 The proof of the above characterisation of the spectrum goes as follows. Since $q_{\pm}$
 is a fundamental system of solutions to the set of self-dual Baxter equations \eqref{ecriture ensemble autodual eqns Baxter 1}-\eqref{ecriture ensemble autodual eqns Baxter 2},  by virtue of the structure of solutions to the 
set of self-dual $t-q$ equations which is obtained in Subsection \ref{Appendix Subsection forme generale solutions tq} of the appendix, 
for any solution to these equations, the entire solution $q$ in particular,  the exists elliptic functions $\mc{P}_{\pm}(\la)$  such that
\beq
q(\la)=  \mc{P}_+(\la) q_+(\la) + \mc{P}_-(\la) q_-(\la) \;.
\enq
As $q$ is entire, the self-dual Wronskian of $q$, \textit{c.f.} \eqref{definition Wronskien self dual}, is also an entire function. 
The functional equations \eqref{ecriture eqns diff finies pour Wronskien Self dual} it satisfies then determines it completely to be $\mc{W}[q]=C_q \varkappa^{-\i\la}$ for some normalisation constant $C_q$. 
On the other hand, the self-dual Wronskian can be computed explicitly by means of \eqref{calcul explicite du Wronskien} leading to 
\beq
\mc{W}[q](\la) \, = \,   \varkappa^{-i\la} g^{-N \Om}  C_{ \tilde{\bs{\de}} }  \f{  \th_{ \tilde{\bs{\de}} }(\la) }{ \th_{\bs{\de}}(\la)  } \mc{P}_+(\la)\, \mc{P}_-(\la) \;,
\enq
where $C_{ \tilde{\bs{\de}} }$ is a constant defined through \eqref{definition cste C tilde delta qToda}-\eqref{definition cste C tilde delta Toda2}.

Both information entail that  
\beq
\mc{P}_+(\la) \, \mc{P}_-(\la)   =    - \xi \cdot  \f{ \th_{\bs{\de}}(\la)  }{  \th_{ \tilde{\bs{\de}} }(\la) }  \quad \e{with} \quad \xi \, = \, g^{ N \Om}   \f{ C_q }{ C_{ \tilde{\bs{\de}} }  }\;.
\enq
Suppose that $\bs{\de}\not=\tilde{\bs{\de}}$ and let $\de_{k}\in \bs{\de}\setminus \tilde{\bs{\de}}$. 
Then, $ \mc{P}_+(\la)\mc{P}_-(\la) $ has simple zeroes at $  \de_k + \i\om_1\mathbb{Z} + \i\om_2\mathbb{Z}$. Each zero being simple,
only one of the two elliptic functions may vanish on the lattice $  \de_k + \i\om_1\mathbb{Z} + \i\om_2\mathbb{Z}$, say $\mc{P}_{+}$. 
Therefore, one has $\mc{P}_{-}(\de_k) \not=0$. Recall  that both $q_{\pm}$ have 
simple poles at $\{ \de_a + \i\om_1\mathbb{Z} + \i\om_2\mathbb{Z}, a \in [\![1;N ]\!] \}$. Thence, should such a situation occur, one would have 
\beq
\e{Res}\big( q(\la) \dd \la, \la=\de_k \big) \not= 0 ,
\enq
hence contradicting that $q$ is entire. Accordingly, it holds that $\bs{\de}=\tilde{\bs{\de}}$ so that:
\beq
\mc{P}_+(\la) \, \mc{P}_-(\la)   =   -  \xi  \;. 
\enq
If the elliptic functions are non-constant, then they necessarily have zeroes and poles. Suppose that $z $ is a zero of $\mc{P}_{+}$. Then $\mc{P}_{+}$
vanishes on the lattice $z+\i\mathbb{Z}\om_1+\i\mathbb{Z}\om_2$. Thus, $\mc{P}_{-}$ has poles on this lattice. However, it follows from the characterisation 
of $v_{\da}(\la-\i\om_1)$, resp. $\tilde{v}_{\da}(\la-\i \om_2)$, given in Subsection \ref{Subappendix Auxiliary fcts nu up down}
that, for a bounded $|\la|$, $\psi_-(\la-\i n\om_1-\i m \om_2) \wt{\psi}_{-}(\la-\i n\om_1-\i m \om_2)$  
does not vanish provided that $n$ and $m$ are large enough. Thence, it holds that the function $q_-\mc{P}_-$ has a pole at $z-\i n\om_1-\i m \om_2$
for some $n,m$ large enough. Since $q_+\mc{P}_+$ is regular\footnote{If $z\in \bs{\de}+\i\mathbb{Z}\om_1+\i\mathbb{Z}\om_2$, then the pole at $z$ of $q_+$ is compensated by the zero of $\mc{P}_+$.} 
at  $z-\i n\om_1-\i m \om_2$, this contradicts that $q$ is entire. 
Thus, $\mc{P}_+$ and $\mc{P}_-$ are both constant, and up to an overall inessential constant one may take $\mc{P}_+=1$ and $\mc{P}_-=-\xi$.

Thus $q$ is expressed by the below linear combination
\beq
q(\lambda)   = q_+(\lambda) - \xi q_-(\lambda) = {1\over \theta_{\bs{\delta}}(\lambda) } \left[ Q_+(\lambda) - \xi (-1)^N \pl{\ell=1}{N} \Big\{ \ex{- \f{2\pi}{\om_2}(\la-\de_{\ell}) } \Big\}   Q_-(\lambda) \right] \;. 
\label{ecriture explicite q}
\enq
where we used that
\beq
{ \theta_{\bs{\delta}} (\lambda) \over \theta_{-\bs{\delta}}(-\lambda) } \, = \, (-1)^N \pl{\ell=1}{N} \Big\{ \ex{- \f{2\pi}{\om_2}(\la-\de_{\ell}) } \Big\} \;. 
\enq
The denominator in \eqref{ecriture explicite q} has simple zeroes on the lattice $\delta_k + i\omega_1 \mathbb{Z} + i\omega_2 \mathbb{Z}$, $k \in  [\![1;N]\!]$. 
Since $\bs{\de}=\tilde{\bs{\de}}$, $Q_{\pm}$ do not vanish on this lattice. This ensures that the parameters $\bs{\de}$ 
characterising an entire solution $q$ have to satisfy to the constraints
\begin{empheq}{equation}
{Q_+(\delta_k + \i n \omega_1 + \i m \omega_2) \over Q_-(\delta_k + \i n \omega_1 + \i m \omega_2)}  = \xi (-1)^N q^{-2Nn} \pl{\ell=1}{N} \Big\{ \ex{- \f{2\pi}{\om_2}(\de_k - \de_{\ell}) } \Big\}
\label{bethenm}
\end{empheq}
for $k=1,\dots,N$ and any $(n,m)\in \mathbb{Z}^2$. By using eqns. \eqref{ecriture check W om1}-\eqref{ecriture check W om2}
which ensure that 
\beq
\check{W}_{\om_1}[Q_+,Q_-](\delta_k + \i n\omega_1 +  \i m\omega_2) \, = \, \check{W}_{\om_2}[Q_+,Q_-](\delta_k + \i n\omega_1 +  \i m\omega_2) \, = \, 0 \;, 
\enq
 one gets that 
\begin{empheq}{equation}
{Q_+(\delta_k + \i n\omega_1 +\i m \omega_2) \over Q_-(\delta_k + \i n\omega_1 + \i m\omega_2)}  =  q^{-2Nn}  { Q_+(\delta_k  ) \over Q_-(\delta_k  )}   \;. 
\end{empheq}
In other words, all constraints \eqref{bethenm} will be satisfied, provided that they holds for $n=m=0$ and $k=1,\dots, N$. 
Thus, $q$ defined as $q=q_+-\xi q_-$ will be entire, provided that it holds 
\beq
\lim_{\la \rightarrow \de_k} \f{ q_+(\la) }{ q_-(\la) } \, = \, \xi \quad \e{for} \quad k=1,\dots, N \;. 
\enq
A straightforward calculation shows that 
\bem
\f{ q_+(\la) }{ q_-(\la) } \, = \, \rho^{-\i\la} \ex{ \mc{I}_{\bs{\de}}(\la) \, + \,  \wt{\mc{I}}_{\bs{\de}}(\la)  } \cdot 
\bigg\{ \ex{-\tfrac{\i\pi N}{ \om_1 \om_2} \de_k^2 - \tfrac{2 \i \pi \de_{k} }{ \om_1 \om_2}  \sum_{a=1}^{N}\de_{a} } \bigg\}^{ \bs{1}_{ \e{Toda}_{2} } } \\
  \times    \pl{\ell=1}{N} \bigg\{ \ex{ -\tfrac{\pi \Om }{ \om_1 \om_2 } \de_{a}  } \f{ \varpi\big( \de_{\ell}-\la + \i\tf{\Om}{2} \big)  } { \varpi\big(\la-\de_{\ell}+\i\tf{\Om}{2} \big)  }  \bigg\} \;. 
\end{multline}
Here, $ \bs{1}_{ \e{Toda}_{2} }=1$ in the case of the Toda$_2$ model and $ \bs{1}_{ \e{Toda}_{2} }=0$ for the  $q$-Toda chain. 

To conclude, we remind that the additional constraint \eqref{ecriture contrainte finale sur les deltas} on the roots $\de_a$ is already present from the very start of the 
analysis, \textit{c.f.} \eqref{contraintes sur les zeros dek tilde dek des dets de Hill}.


 \section{Equivalence of $TQ$ equations and non-linear integral equations}
\label{Section equivalence NLIE et TQ}
 
 In this section we show that the description of the spectrum through non-linear integral equations is fully equivalent to the original, dual $t-q$ equation based description \ref{ecriture ensemble autodual eqns Baxter}.
 Namely, we show that starting from a given set of solutions $Y_{ \bs{\de} }$, $\wt{Y}_{ \tilde{\bs{\de}} }$ to the appropriately defined non-linear integral equations \eqref{eqn NLIE pour Y}, \eqref{eqn NLIE pour Y dual},
one can introduce two functions $q_{\pm}$. These functions enjoy certain property that allow one to produce two combinations of these functions giving rise to two hyperbolic polynomials $t_{\bs{\tau}}$ and $\tilde{t}_{\tilde{\bs{\tau}}}$
whose roots depend on the sets of original roots $\bs{\de}, \tilde{\bs{\de}}$ arising in the non-linear integral equations for $Y_{ \bs{\de} }$, $\wt{Y}_{ \tilde{\bs{\de}} }$.
These polynomials are such that the $q_{\pm}$ solve the set of dual $t-q$ equations subordinate to the polynomials $t_{\bs{\tau}}$ and $\tilde{t}_{\tilde{\bs{\tau}}}$.

\subsection{The building blocks $q_+$ and $q_-$}
\label{subsection building blocks q pm NLIE}

\subsubsection{The fundamental building blocks }

Recall the two strips $\mc{B}$ and $\wt{\mc{B}}$ introduced respectively in \eqref{definition bande B}, \eqref{definition bande B tilde}. 
Starting from now, we consider two collections of N parameters   $\bs{\de}=(\de_1,\dots,\de_N) \in \mc{B}^N$ and $\tilde{\bs{\de}} = ( \tilde{\de}_1, \dots, \tilde{\de}_N)\in \tilde{\mc{B}}^N$,
such that each collection satisfies to  the constraints
%
%
%
%
%
%
%
%
%
%
%
%
\beq
\sul{k=1}{N}\de_k  \; = \; \sul{k=1}{N} \tilde{\de}_k  \; = \; \f{\om_1 \om_2}{2\pi} p_0\;. 
\label{ecriture liste contraintes sur les deltas}
\enq
In the discussions to come, we will only consider the case of generic, in particular pairwise distinct sets of roots 
$\bs{\de}$, $\tilde{\bs{\de}}$. However, all properties continue on to the non-generic case, upon evident modifications due to multiplicities, 
by taking limits.

Next, we assume that one is given two solutions $Y_{\bs{\de}}$ and $\wt{Y}_{ \tilde{\bs{\de}} }$ to the non-linear integral equations \eqref{eqn NLIE pour Y}, \eqref{eqn NLIE pour Y dual}. 
Since, these non-linear integral equations are the starting point of the analysis, some care is needed in discussing the contours 
  $\msc{C}_{\bs{\de};\rho}, \wt{\msc{C}}_{ \tilde{\bs{\de}};\rho}$. These are defined as follows. 
 $\msc{C}_{\bs{\de};\rho}$ is a contour in $\mc{B}$  starting at $\i x \om_1-\i \tf{\om_2}{2}$
and ending at $\i x \om_1+\i \tf{\om_2}{2}$ for some well chosen $x \in \R$. This entails that $\msc{C}_{\bs{\de};\rho}$ cuts $\mc{B}$, 
in two disjoint domains $\mc{B}_{\pm}$ such that $\pm \i t \om_1 \in \mc{B}_{\pm}$ for $t>0$ and large.
The contour $\msc{C}_{\bs{\de};\rho}$ is oriented so that the domain $\mc{B}_{+}$ is to its \textit{left}. 

The dual definition holds for the contour  $\wt{\msc{C}}_{ \tilde{\bs{\de}};\rho}$. One should however pay attention that the orientation 
requirement imposes that $\wt{\msc{C}}_{ \tilde{\bs{\de}};\rho}$ starts at $\i \tilde{x} \om_2+\i \tf{\om_1}{2}$
and ends at $\i \tilde{x} \om_2-\i \tf{\om_1}{2}$ for some well chosen $ \tilde{x} \in \R$.

The contour $\msc{C}_{\bs{\de};\rho}$, resp. $\wt{\msc{C}}_{ \tilde{\bs{\de}};\rho}$, has to satisfy the three properties:
\begin{itemize}
\item[i)] it is such that the points $\de_k \, \pm \, \i\f{\om_1}{2}\pm \i\mathbb{N} \om_1 $, resp.  $\tilde{\de}_k \, \pm \, \i\f{\om_2}{2}\pm \i\mathbb{N} \om_2 $,  
are all in $\mc{B}_{\pm}$, resp. $\wt{\mc{B}}_{\pm}$;
\item[ii)] the functions  
\beq
\mc{V}_{\bs{\de}}(\la) \; = \; 1 \, + \, \f{ \rho^{\om_1} \ex{Y_{ \bs{\de} }(\la) }   }{ t_{ \bs{\de}}\big(\la-\i\tfrac{\om_1}{2}\big) t_{ \bs{\de}}\big(\la + \i\tfrac{\om_1}{2} \big)    } \quad \e{and} \quad
\wt{\mc{V}}_{ \tilde{\bs{\de}} }(\la) \; = \; 1 \, + \, \f{ \rho^{\om_1} \ex{\wt{Y}_{ \tilde{\bs{\de}} }(\la)}  }{ \tilde{t}_{ \tilde{\bs{\de}} }\big( \la-\i\tfrac{\om_1}{2} \big) \tilde{t}_{ \tilde{\bs{\de}} }\big(\la + \i\tfrac{\om_1}{2} \big)    } 
\label{definition V delta}
\enq
do not vanish on $\msc{C}_{\bs{\de};\rho}$, resp. on  $\wt{\msc{C}}_{ \tilde{\bs{\de}};\rho}$, and have continuous logarithms $\ln \big[\mc{V}_{\bs{\de}} \big](\tau)$, resp. $\ln \big[\wt{\mc{V}}_{ \tilde{\bs{\de}} } \big](\tau)$,
on these curves; 

\item[iii)] when shrinking $\rho$ to $0$, no zero of the function $\mc{V}_{\bs{\de}}$, resp. $\wt{\mc{V}}_{ \tilde{\bs{\de}} }$, crosses the contour 
$\msc{C}_{\bs{\de};\rho}$, resp.  $\wt{\msc{C}}_{ \tilde{\bs{\de}};\rho}$.

\end{itemize}

Thus, the non-linear integral equations are \textit{joint} equations for the unknown functions $Y_{ \bs{\de} }$ and  $\wt{Y}_{ \tilde{\bs{\de}}}$
and the contours $\msc{C}_{\bs{\de};\rho}$ and  $\wt{\msc{C}}_{ \tilde{\bs{\de}};\rho}$.

Several comments are in order

\begin{itemize}
 
 \item[i)] It is shown in Appendix \ref{Appendix SousSection solvabilite unique des NLIE pour Y} that, provided $|\rho|$ is small enough, there exist a unique solution $Y_{\bs{\de}}$, resp. $Y_{\tilde{\bs{\de}}}$, 
to \eqref{eqn NLIE pour Y}, resp. \eqref{eqn NLIE pour Y dual}, and that, for $|\rho|$ small enough, one can take the contours $\msc{C}_{\bs{\de};\rho}$, resp.  $\wt{\msc{C}}_{ \tilde{\bs{\de}};\rho}$, to be $\rho$ independent. 


\item[iii)] Appendix \ref{Appendice Sous Section Fct Y delta} discusses the analytic properties of $Y_{\bs{\de}}$ and $\wt{Y}_{\tilde{\bs{\de}}}$, solving \eqref{eqn NLIE pour Y} and \eqref{eqn NLIE pour Y dual}, 
as meromorphic functions on $\Cx$

\end{itemize}

\vspace{3mm}

The two functions $ Y_{ \bs{\de} }$ and  $\wt{Y}_{ \tilde{\bs{\de}}}$ being given, recall the definition of the functions 
$\mc{V}_{\bs{\de}}(\la) $ and $\wt{\mc{V}}_{ \tilde{\bs{\de}} }(\la)$  given in \eqref{definition V delta} and then 
 define the functions $v_{\ua}(\la), v_{\da}(\la-\i\om_1)$ by means of the formulae 
\eqref{definition v up} and \eqref{definition v down}. 
Similarly, one introduces the dual functions on $\wt{\msc{C}}_{ \tilde{\bs{\de}};\rho}$
through eqns. \eqref{definition v up tilde}-\eqref{definition v down tilde}. 

According to Subsection \ref{Subappendix Auxiliary fcts nu up down}, $\la \mapsto v_{\ua}(\la)$ extends to a meromorphic function on $\Cx$. It is non-vanishing in $\mc{B}_+$
and has simple poles at $\de_k - \i\mathbb{N}^{*}\om_1 + \i\mathbb{Z} \om_2$ with non-vanishing residues. Likewise, $\la \mapsto v_{\da}(\la-\i\om_1)$ extends to a meromorphic function on $\Cx$. It is non-vanishing in $\mc{B}_-$
and has simple poles at $\de_k + \i\mathbb{N}^{*}\om_1 + \i\mathbb{Z} \om_2$ with non-vanishing residues.

\subsubsection{The functions of main interest}

Finally, let 
\beq
q_\pm(\la)\, = \, \f{ Q_{\pm}(\la) }{  \th_{ \pm\bs{\de} }(\pm\la) } 
\label{definition fcts q pm base des solutions de TQ}
\enq
with
\beq 
Q_{+}(\la) \, = \,  \big( \varkappa g^N \big)^{-\i\la}   \, \psi_+(\la) \, \wt{\psi}_{+}(\la) \, f_{\bs{\de}}^{(+)}(\la)  \quad \e{and} \quad
Q_-(\la)\, = \,   \big( g^N \big)^{\i\la }\, \psi_-(\la) \, \wt{\psi}_{-}(\la) \, f_{\bs{\de}}^{(-)}(\la)   \;. 
\enq
The building blocks of $Q_{\pm}$ are constructed with the help of the functions $v_{\ua/\da}$ and $\tilde{v}_{\da/\ua}$:
\beq
\psi_+(\la) \, = \, v_{\ua}(\la)  \; \Big( q^2 \ex{\f{2\pi}{\om_2}\la}; q^2 \Big)_{\bs{\de}} \quad , \quad 
\wt{\psi}_+(\la) \, = \, \wt{v}_{\ua}(\la) \;  \Big( \wt{q}^{\,-2} \ex{-\f{2\pi}{\om_1}\la}; \wt{q}^{\, -2} \Big)_{ \tilde{\bs{\de}} }
\enq
and 
\beq
\psi_-(\la) \, = \, v_{\da}(\la-\i\om_1)  \, \Big( q^2 \ex{-\f{2\pi}{\om_2}\la}; q^2 \Big)_{\bs{\de}} \quad , \quad 
\wt{\psi}_-(\la) \, = \, \wt{v}_{\da}(\la-\i\om_2) \, \Big( \wt{q}^{\, -2} \ex{\f{2\pi}{\om_1}\la}; \wt{q}^{\, -2} \Big)_{ \tilde{\bs{\de}} } \;. 
\enq

Some explanations are in order. The functions $ f_{\bs{\de}}^{(\pm)}(\la)$ are as defined through \eqref{definition fde pm pour q Toda} for the $q$-Toda chain and \eqref{definition fde pm pour Toda2} for the Toda$_2$ chain.

It will be of use for the following, to establish a few overall properties of the functions $Q_{\pm}$ and its building blocks which follow from properties of solutions to the non-linear
integral equation for $Y_{\bs{\de}}$ and $\wt{Y}_{\tilde{\bs{\de}}}$:
\begin{itemize}

 \item $\psi_{\pm}$ is entire and does not 
vanish on the lattice  $\de_k+\i\mathbb{Z} \om_1 +\i\mathbb{Z} \om_2$ as follows from the properties of $v_{\ua/\da}$ discussed earlier on; 
\item  dual properties hold for  $\wt{\psi}_{\pm}$; 
\item  this ensures that $Q_{\pm}$ are both entire functions. 
 
\end{itemize}

\subsection{$t_{\tau}$ and $\tilde{t}_{\tilde{\tau}}$ are trigonometric polynomials.}
\label{Subsection t tau tilde t tau tilde are trig plys}

Let $q_{\pm}$ be as given by eq. \eqref{definition fcts q pm base des solutions de TQ} with $\bs{\de}$, $\tilde{\bs{\de}}$ satisfying to \eqref{ecriture liste contraintes sur les deltas}. 
In addition, in the Toda$_2$ case, assume that 
\beq
\big| \rho^{\om_1} \ex{2\om_1 p_0}\big| \,  < \, 1 \qquad \e{and} \qquad \big| \rho^{\om_2} \ex{2\om_2 p_0}\big| \,  < \, 1 \;. 
\label{ecriture hypothese supplementaire Toda2}
\enq

Then, there exists $N$ roots $ \bs{\tau}=\{\tau_a\}_1^N$ and $N$ dual roots $\tilde{\bs{\tau}}=\{ \tilde{\tau}_a\}_1^N$ such that 
\beq
t_{\bs{\tau}}(\la) \; = \; \sg  g^{N\om_1}\varkappa^{\om_1} \f{ q_+(\la-\i\om_1)  q_-(\la+\i\om_1)   - q_-(\la-\i\om_1)  q_+(\la+\i\om_1)  }{  q_+(\la)  q_-(\la+\i\om_1)   - q_-(\la)  q_+(\la+\i\om_1)  } \; , 
\label{definition alternative ply t lambda}
\enq
and  
\beq
\tilde{t}_{ \tilde{\bs{\tau}} }(\la) \; = \; \sg  g^{N\om_2}\kappa^{\om_2} \f{ q_+(\la-\i\om_2)  q_-(\la+\i\om_2)   - q_-(\la-\i\om_2)  q_+(\la+\i\om_2)  }{  q_+(\la)  q_-(\la+\i\om_2)   - q_-(\la)  q_+(\la + \i\om_2)  } \;. 
\label{definition alternative ply t lambda dual}
\enq
The roots $\{\tau_a\}$ and $\{ \tilde{\tau}_a\}$ satisfy to 
\beq
\ba{cccccc} 
\pl{a=1}{N} \ex{-\f{2\pi}{\om_2}\tau_a} \, = &  \ex{-\om_1 p_0} + \ex{ - \f{2\pi}{\om_2}N \kappa_2   } &,& 
\quad \pl{a=1}{N} \ex{-\f{2\pi}{\om_1}\tilde{\tau}_a} \, = &  \ex{-\om_2 p_0} + \ex{ - \f{2\pi}{\om_1}N \kappa_2   }  & \quad \e{Toda}_2   \vspace{3mm} \\
\pl{a=1}{N} \ex{-\f{\pi}{\om_2}\tau_a} \, = &  \ex{- \frac{\om_1}{2} p_0}  &, &
\quad \pl{a=1}{N} \ex{-\f{\pi}{\om_1}\tilde{\tau}_a} \, = &  \ex{- \frac{\om_2}{2} p_0}   & \quad q-\e{Toda}   			\ea    \; . 
\label{ecriture condition sur les racines tau et tilde tau}
\enq

\vspace{5mm}

In the Toda$_2$ case, we are only able to establish the hyperbolic polynomiality of the quantities appearing in the \textit{rhs} of \eqref{definition alternative ply t lambda}-\eqref{definition alternative ply t lambda dual}
under the hypothesis \eqref{ecriture hypothese supplementaire Toda2}. We have no idea whether the upper bound in this hypothesis represents some genuine limitation of the method 
or whether it may be omitted without altering the conclusion of the above statement by a suitable modification of the form of the non-linear integral equation.

\vspace{5mm}
 
  In order to establish the statement, we denote by $\mc{T}_{\bs{\de}}(\la)$  the \textit{rhs} of \eqref{definition alternative ply t lambda}. Then a direct calculation shows that 
\beq
\mc{T}_{\bs{\de}}(\la) \, = \, \sg g^{N \om_1} \varkappa^{\om_1}  \pl{k=1}{N} \Big\{ -\ex{-\f{2\pi}{\om_2}(\la-\de_k) }\Big\}  \, 
\f{ \check{U}_{\om_1}[Q_+^{(\om_1)},Q_-^{(\om_1)}](\la) }{  \check{W}_{\om_1}[Q_+^{(\om_1)},Q_-^{(\om_1)}](\la) }
\label{definition mathcal T delta}
\enq
where 
\beq
\check{U}_{\om_1}[Q_+,Q_-](\la)\, = \, Q_+(\la-\i\om_1)Q_-(\la+\i\om_1)\, - \, q^{4N} Q_+(\la+\i\om_1)Q_-(\la-\i\om_1)\;,  
\enq
and 
\beq
\check{W}_{\om_1}[Q_+,Q_-](\la)\, = \, Q_+(\la)Q_-(\la+\i\om_1)\, - \, q^{2N} Q_+(\la+\i\om_1)Q_-(\la)\; . 
\enq
 $\check{W}_{\om_1}[Q_+,Q_-]$ can be computed in a closed form; the explicit expression can be found in \eqref{ecriture check W om1}. 
Also, we have introduced 
\beq
Q_{+}^{(\om_1)}(\la) \, = \,  \big( \varkappa g^N \big)^{-\i\la}   \, \psi_+(\la) \,f_{p_0}^{(+)}(\la)  \quad \e{and} \quad
Q_-^{(\om_1)}(\la)\, = \,   \big( g^N \big)^{\i\la }\, \psi_-(\la) \, f_{p_0}^{(-)}(\la)   
\enq
Note that the functions $\wt{\psi}_{+}$ and $\wt{\psi}_{-}$ could have been simplified between the numerator and denominator in \eqref{definition mathcal T delta} by using their $\i\om_1$
periodicity. 

The functions $Q_{\pm}^{(\om_1)}$ are entire functions what entails that $\mc{T}_{\bs{\de}}$ is a meromorphic function whose poles correspond to the zeroes of  
\beq
\check{W}_{\om_1}[Q_+^{(\om_1)},Q_-^{(\om_1)}](\la)\, = \,\varkappa^{-\i\la} g^{-N\om_1 }  f_{p_0}^{(+)}(\la)f_{p_0}^{(-)}(\la+\i\om_1) \th_{\bs{\de}}(\la) \, . 
\enq
This expression in a consequence of \eqref{ecriture check W om1} and the $\i\om_1$ periodicity of $\wt{\psi}_{+}$ and $\wt{\psi}_{-}$. Thus, 
the  potential poles of  $\mc{T}_{\bs{\de}}(\la)$  can only be located at the zeroes of $\th_{\bs{\de}}(\la)$, \textit{viz}. belong to the lattice
$\de_k+\i\om_1\mathbb{Z}+\i\om_2 \mathbb{Z}$. It follows from the properties justified at the end of Subsection \ref{subsection building blocks q pm NLIE}
that the functions $Q_{\pm}^{(\om_1)}$ do not vanish on this lattice. Writing out explicitly that $\check{W}_{\om_1}[Q_+^{(\om_1)},Q_-^{(\om_1)}](\de_k+\i n \om_1+\i m \om_2)=0$, one infers that 
\bem
Q_+^{(\om_1)}(\de_k+\i n \om_1+\i m \om_2)\, Q_-^{(\om_1)}(\de_k+\i (n+1) \om_1+\i m \om_2) \\
= \, q^{2N} Q_+^{(\om_1)}(\de_k+\i (n+1) \om_1+\i m \om_2) \, Q_-^{(\om_1)}(\de_k+\i n \om_1+\i m \om_2)
\end{multline}
what ensures that 
\beq
\f{ Q_+^{(\om_1)}(\de_k+\i n \om_1+\i m \om_2) }{Q_-^{(\om_1)}(\de_k+\i n \om_1+\i m \om_2)  } \, = \,  q^{2 \ell N} \f{ Q_+^{(\om_1)}(\de_k+\i (n+\ell) \om_1+\i m \om_2) }{  Q_-^{(\om_1)}(\de_k+\i (n+\ell) \om_1+\i m \om_2)} \;. 
\enq
In particular, with $\ell=2$ this entails that $ \check{U}_{\om_1}[Q_+,Q_-](\de_k+\i (n-1) \om_1+\i m \om_2)=0$. Since $\de_k+\i (n-1) \om_1+\i m \om_2$ is a simple 
zero of the denominator in \eqref{definition mathcal T delta}, this entails that $\mc{T}_{\bs{\de}}$ is entire.

One can further simplify the expression for $\mc{T}_{\bs{\de}}$ and recast it as
\beq
\mc{T}_{\bs{\de}}(\la) \, = \,  t_{ \bs{\de}}(\la) v_{\ua}(\la-\i\om_1) v_{\da}(\la) 
\, - \, \rho^{2\om_1} \f{ v_{\ua}(\la+\i\om_1) v_{\da}(\la - \i\om_1)  }{ \pl{\eps=0, \pm 1}{} t_{ \bs{\de}}(\la+\i\eps \om_1 )   } \;. 
\enq
From there it becomes clear that $\mc{T}_{\bs{\de}}$ is $\i\om_2$ periodic and, owing to the asymptotic behaviour of $v_{\ua/\da}$ at $\infty$
which is established in Sub-section \ref{Subappendix Auxiliary fcts nu up down} of the appendix, we infer that 
\beq
\big| \mc{T}_{\bs{\de}}(\la)  \big| \; \leq \; C_{\eps} \cdot  \left\{ \ba{cc}  1+ \big|  \ex{- \f{2\pi}{\om_2}N \la } \big|  &  \e{Toda}_2 \vspace{3mm} \\ 
									 \big|  \ex{ \f{\pi}{\om_2}N \la } \big|  + \big|  \ex{- \f{\pi}{\om_2}N \la } \big|  & q-\e{Toda}      \ea \right. 
\quad \e{for} \quad \la \in \mc{B}\setminus \bigcup\limits_{ \substack{ a\in [\![1; n ]\!] \\ k,\ell\in \mathbb{Z}} }^{}  \mc{D}_{\de_{a}+\i\om_1 k + \i \om_2 \ell , \eps} \;. 
\label{ecriture borne cal T delta en dehors des poles}
\enq
Here, $\eps>0$ is arbitrary but small enough, $C_{\eps}$ is some $\eps$-dependent constant and 
\beq
\mc{D}_{z,\eps}=\Big\{ s \in \Cx \, : \,  |s-z|\leq \eps \Big\} \;. 
\label{definition disque centre z taille eps}
\enq
Since $ \mc{T}_{\bs{\de}}(\la)$ is entire, it admits the integral representation 
\beq
\mc{T}_{ \bs{\de} }(\la) \, = \, \Int{ \partial \mc{D}_{a}  }{} \f{ \mc{T}_{\bs{\de}}(z) }{z-\la} \cdot \f{ \dd z }{2\i\pi}
\qquad \e{for} \qquad  \la \in \mc{D}_{a } \equiv \mc{D}_{\de_{a}+\i\om_1 k + \i \om_2 \ell , 2\eps  } \;. 
\enq
Thus, immediate bounds based on this representation and \eqref{ecriture borne cal T delta en dehors des poles} leads to 
\beq
\big| \mc{T}_{\bs{\de}}(\la)  \big| \; \leq \; \f{C_{\eps}}{\eps} \cdot  \left\{ \ba{cc}  1+ \big|  \ex{- \f{2\pi}{\om_2}N \la } \big|  \vspace{3mm} \\ 
									 \big|  \ex{ \f{\pi}{\om_2}N \la } \big|  + \big|  \ex{- \f{\pi}{\om_2}N \la } \big|        \ea \right. 
\quad \e{for} \quad \la \in  \overline{\mc{D}}_{\de_{a}+\i\om_1 k + \i \om_2 \ell , \eps} \;. 
\enq
Thus, upon replacing $C_{\eps} \hookrightarrow \tf{ C_{\eps} }{ \eps}$, the bounds \eqref{ecriture borne cal T delta en dehors des poles} hold, in fact, on $\Cx$. 
By the hyperbolic variant of Liouville's theorem, we do get that there exist N roots $\tau_{k}$ such that $\mc{T}_{\bs{\de}} = t_{\bs{\tau}}$.

Furthermore, upon looking at the leading $\ex{ \f{\pi}{\om_2}  \la} \rightarrow +\infty$
asymptotics in the $q$-Toda case one gets that 
\beq
\mc{T}_{\bs{\de}}(\la) =  \ex{ \f{\pi}{\om_2} N  \la} \underbrace{\pl{a=1}{N} \ex{ -\f{\pi}{\om_2}  \de_a} }_{= \ex{-\frac{\om_1}{2} p_0} } \; \cdot (1+ \e{o}(1)) \qquad \e{while} \qquad 
t_{\bs{\tau}}(\la) =  \ex{ \f{\pi}{\om_2} N  \la}  \pl{a=1}{N} \ex{ -\f{\pi}{\om_2}  \tau_a}  \; \cdot (1+ \e{o}(1)) \; .
\enq
This obviously yields \eqref{ecriture condition sur les racines tau et tilde tau}. 

In the Toda$_2$ case, one rather computes the $\ex{-\f{2\pi}{\om_2}  \la} \rightarrow 0$  limit. By using the form of the asymptotics of $v_{\ua/\da}$
given in Subsection \ref{Subappendix Auxiliary fcts nu up down}, one gets that 
\begin{multline}
\mc{T}_{\bs{\de}}(\la) =  (-1)^N \f{ \prod_{a=1}^{N} \ex{-\f{2\pi}{\om_2}  \de_a}   }{   1- \rho^{\om_1}  \prod_{a=1}^{N} \ex{\f{4\pi}{\om_2}  \de_a}  } 
\; - \; (-1)^N \f{ \rho^{2\om_1} \prod_{a=1}^{N} \ex{\f{6\pi}{\om_2}  \de_a}   }{   1- \rho^{\om_1}  \prod_{a=1}^{N} \ex{\f{4\pi}{\om_2}  \de_a}  }  \; + \; \e{o}(1) \\
\; = \;  (-1)^N\prod_{a=1}^{N} \ex{-\f{2\pi}{\om_2}  \de_a}   \Big\{ 1 + \rho^{\om_1}  \prod_{a=1}^{N} \ex{\f{4\pi}{\om_2}  \de_a}  \Big\}  \; + \; \e{o}(1) 
\; = \; (-1)^N \Big\{ \ex{-\om_1 p_0} + \rho^{\om_1}  \ex{ \om_1 p_0}  \Big\}  \; + \; \e{o}(1)  \;. 
\end{multline}
One may then conclude by using that $t_{\bs{\tau}}(\la) =  (-1)^N  \pl{a=1}{N} \ex{ -\f{2\pi}{\om_2}  \tau_a}  \; \cdot (1+ \e{o}(1)) $
and recalling the explicit expression for $\rho$ in the Toda$_2$ model.

The reasonings in the dual case are quite similar, with the sole difference that, if $\wt{\mc{T}}_{\bs{\de}}(\la)$ denotes the \textit{rhs} of \eqref{definition alternative ply t lambda dual} one gets 
\beq
\tilde{\mc{T}}_{\bs{\de}}(\la) \, = \, \sg g^{N \om_2} \varkappa^{\om_2}    \, 
\f{ \check{U}_{\om_2}[Q_+^{(\om_2)},Q_-^{(\om_2)}](\la) }{  \check{W}_{\om_2}[Q_+^{(\om_2)},Q_-^{(\om_2)}](\la) }
\label{definition mathcal T delta dual}
\enq
where $Q_{+}^{(\om_2)}(\la) \, = \,  \big( \varkappa g^N \big)^{-\i\la}   \, \wt{\psi}_+(\la) \,f_{\bs{\de}}^{(+)}(\la)$, 
$Q_-^{(\om_2)}(\la)\, = \,   \big( g^N \big)^{\i\la }\, \wt{\psi}_-(\la) \, f_{\bs{\de}}^{(-)}(\la)$, 
\beq
\check{U}_{\om_2}[Q_+,Q_-](\la)\, = \, Q_+(\la-\i\om_2)Q_-(\la+\i\om_2)\, - \,  Q_+(\la+\i\om_2)Q_-(\la-\i\om_2)\;,  
\enq
and 
\beq
\check{W}_{\om_2}[Q_+,Q_-](\la)\, = \, Q_+(\la)Q_-(\la+\i\om_2)\, - \,  Q_+(\la+\i\om_2)Q_-(\la)\; . 
\enq
The $\om_2$ reduced Wronskian appearing in \eqref{definition mathcal T delta dual}, can be expressed as 
\beq
\check{W}_{\om_2}[Q_+^{(\om_2)},Q_-^{(\om_2)}](\la)\, = \,\varkappa^{-\i\la} g^{-N\om_2 }  f_{\bs{\de}}^{(+)}(\la)f_{\bs{\de}}^{(-)}(\la+\i\om_2) \th_{\tilde{\bs{\de}}}(\la) \;. 
\enq
The rest of the reasonings in the dual case are formally the same.

\subsection{The Baxter equation associated with $q_{\pm}$}

 Let $q_{\pm}$ be as given by \eqref{definition fcts q pm base des solutions de TQ} and let the hyperbolic polynomials 
 $t_{\bs{\tau}}$ and $\tilde{t}_{ \tilde{\bs{\tau}} }$ be as given by \eqref{definition alternative ply t lambda}-\eqref{definition alternative ply t lambda dual}. Then
$q_{\pm}$ provide one with the two linearly independent solutions of the self-dual $t-q$ equations \eqref{ecriture ensemble autodual eqns Baxter 1}-\eqref{ecriture ensemble autodual eqns Baxter 2} associated with these polynomials.

The proof of the statement goes as follows. The fact that  $q_{\pm}$ are linearly independent is a consequence of $q_{\pm}$ not sharing a common set of zeroes.
Indeed, let $\mc{Z}$, resp. $\wt{\mc{Z}}$, be the set of zeroes of $\mc{V}_{\bs{\de}}$, resp. $\wt{\mc{V}}_{ \tilde{\bs{\de}} }$, and let 
$\mc{Z}_{\pm}=\mc{Z}\cap \mc{B}_{\pm}$ and $\wt{\mc{Z}}_{\pm}= \wt{\mc{Z}}\cap \wt{\mc{B}}_{\pm}$. 
The results obtained in Subsection \ref{Subappendix Auxiliary fcts nu up down} of the appendix ensure that $q_{\pm}$
has zeroes at $\big\{ \mc{Z}_{\pm} \pm \i\tf{\om_1}{2} + \i \mathbb{Z}\om_2 \big\} \cup \big\{ \wt{\mc{Z}}_{ \pm } \pm \i\tf{\om_2}{2} + \i \mathbb{Z}\om_1 \big\} $. 
These sets are obviously disconnected. It remains to  show that  $\mc{Z}_{-}$ is non-empty. This is a consequence of  $\ex{Y_{\bs{\de}}(\la)}=1+\e{o}(1)$ when $\la \rightarrow \infty$, $\la \in \mc{B}_{+}$, and 
 of the fact that $\ex{ Y_{\bs{\de}}(\la) }$ has poles at $\de_{k} \pm \i \tf{\om_1}{2} \pm \i \mathbb{N}^* \om_1 + \i\mathbb{Z} \om_2$. 
 
Next, introduce the $\i \om_1$-Wronskian 
\beq
W_{\omega_1}[q_+, q_-](\lambda) \, = \,  q_+(\lambda) q_-(\lambda+\i \omega_1) - q_-(\lambda) q_+(\lambda + \i\omega_1) \,. 
\enq
It holds,  
\bem
t_{\bs{\tau}}(\la) q_{+}(\la)   W_{\om_1}[q_+,q_-](\la)  \\ 
\,= \, \sg \cdot  \big(g^{N} \varkappa\big)^{\om_1}\Big\{  q_+(\la-\i \om_1) q_-(\la+\i \om_1) q_{+}(\la) \, - \, q_+(\la+\i \om_1) q_-(\la-\i \om_1) q_{+}(\la) \Big\}  \vspace{4mm} \\
\, =\, q_+(\la-\i \om_1) \Big\{  W_{\om_1}[q_+,q_-](\la) +  q_+(\la+\i \om_1) q_-(\la)  \Big\} \cdot  \sg \,   \big(g^{N} \varkappa\big)^{\om_1}   \vspace{4mm} \\
\, - \, q_+(\la+\i \om_1) \Big\{  -W_{\om_1}[q_+,q_-](\la-\i\om_1) +  q_+(\la-\i \om_1) q_-(\la)  \Big\} \cdot  \sg \,  \big(g^{N} \varkappa\big)^{\om_1}    \;. 
\end{multline}
The last two terms in each bracket cancel out. 
It is discussed in Subsection \ref{Appendix SousSection Wronskiens} of the appendix that $W_{\om_1}[q_+,q_-](\la)$ satisfies to the finite difference equation 
$W_{\om_1}[q_+,q_-](\la-\i\om_1) \, = \,  \sg^{-2} \varkappa^{-\om_1} W_{\om_1}[q_+,q_-](\la) $. Thus, the handlings can be simplified further, leading to 
\beq
t_{\bs{\tau}}(\la) q_{+}(\la)   W_{\om_1}[q_+,q_-](\la) \,= \,g^{N \om_1 } W_{\om_1}[q_+,q_-](\la) 
\Big\{ \sg  \varkappa^{\om_1} q_+(\la-\i \om_1)   \, + \,  \sg^{-1}   q_+(\la+\i \om_1)  \Big\} \;. 
\enq
The closed formula \eqref{ecriture formule close Wronskiens q pm}-\eqref{ecriture check W om1}  for $ W_{\om_1}[q_+,q_-]$ ensures that this Wronskian is 
a meromorphic function on $\Cx$ that, furthermore, is not identically zero. It can thus vanish at most on a locally finite set. Thus, it holds that $q_+$
solves the $\i\om_1$ $t-q$ equation everywhere, with the exception of an at most locally finite set and away from its pole. By continuity, it satisfies 
this equation on $\Cx \setminus \{ \de_k + \i \mathbb{Z} \om_1 + \i \mathbb{Z} \om_2 \}$. 
The reasoning is much the same regarding to the dual,  $\i\om_2$ $t-q$ equation, as well as  for $q_-$. We leave the details to the reader.

\section{Spectrum of the transfer matrices}
\label{Section Spectrum of transfer matrices}

In this last section, we establish how to reconstruct the spectrum of the transfer matrix and its dual from the knowledge of the parameters 
$\bs{\de}$ describing a given joint Eigenvector of these matrices. One could, of course, use the relation \eqref{ecriture reconstruction polynome t tau via les nus}. 
However, the latter seems rather impractical from the point of view of concrete applications. The reconstruction we propose below is rather direct and 
relies on objects that are natural from the point of view of the non-linear integral equation approach.

Let $\bs{\de}$ be a collection of pairwise distinct parameters solving the Bethe equations \eqref{ecriture HLBAE qToda} or \eqref{ecriture HLBAE Toda2}. 
Assume that the contours $\msc{C}_{\bs{\de};\rho}$ and  $\wt{\msc{C}}_{\bs{\de};\rho}$ are such that 
the roots $\bs{\tau}$ and $\tilde{\bs{\tau}}$ associated with the Eigenvalues $t_{\bs{\tau}}$ of $\op{t}(\la)$ and $\tilde{t}_{ \tilde{\bs{\tau}} }$ of $ \wt{\op{t}}(\la)$
are such that 
\beq
\tau_k \pm \i\f{\om_1}{2} \in \mc{B}_{\pm} \qquad \e{and} \qquad \tilde{\tau}_k \pm \i\f{\om_2}{2} \in \wt{\mc{B}}_{\pm} \;. 
\label{ecriture contrainte sur les tau pour reconstruire energie}
\enq
Define the auxiliary functions 
\beq
\left\{ \ba{ccc} \a_k(\la)  & = &   \ex{- \f{2 \pi}{\om_2} k \la }   \vspace{2mm}  \\ 
\wt{\a}_k(\la) & = & \ex{- \f{2 \pi}{\om_1} k \la }  \ea   \right.  \;. 
\nonumber \enq

Then, given $\mc{V}_{\bs{\de}}$, resp. $\wt{\mc{V}}_{ \tilde{\bs{\de}}}$, as defined in \eqref{definition V delta}, it holds 
\begin{eqnarray}
\sul{a=1}{N}\a_k(\tau_a) & = & \sul{a=1}{N}\a_k(\de_a)  \; - \; 
\Int{ \msc{C}_{\bs{\de};\rho} }{} \Big\{ \a_k^{\prime}\big(\mu-\i \tfrac{\om_1}{2}\big) \, - \,  \a_k^{\prime}\big(\mu +\i \tfrac{\om_1}{2}\big)  \Big\} \ln \big[ \mc{V}_{\bs{\de}}\big](\mu) \cdot \f{ \dd \mu }{ 2\i \pi }\; 
\label{ecriture reconstruction tau k}\\
\sul{a=1}{N} \wt{\a}_k( \tilde{\tau}_a) & = & \sul{a=1}{N}\wt{\a}_k(\de_a)  \; - \; 
\Int{ \wt{\msc{C}}_{\bs{\de};\rho} }{} \Big\{ \wt{\a}_k^{\prime}\big(\mu-\i \tfrac{\om_1}{2}\big) \, - \,  \wt{\a}_k^{\prime}\big(\mu +\i \tfrac{\om_1}{2}\big)  \Big\} \ln \big[ \wt{\mc{V}}_{ \bs{\de} }\big](\mu) \cdot \f{ \dd \mu }{ 2\i \pi } \;,  
\label{ecriture reconstruction tau tilde k}
\end{eqnarray}
where $k=1,\dots, N-1$. 
%
%
%
%
%
%
%
%
%
%

Prior to discussing the proof of this representation, some remarks are in order. 

\begin{itemize}

 \item[i)] The integral representation reconstructs the Newton polynomials 
\beq
P_k(X_1,\dots, X_N) \, = \, \sul{a=1}{N} X_a^k
\enq
in the variables $\ex{- \f{2 \pi}{\om_2} k  \tau_{1} }, \dots , \ex{- \f{2\pi}{\om_2} k \tau_{N} } $ for $k=1,\dots, N-1 $. 
Since the newton polynomials are an algebraic basis in the space of symmetric polynomials,
it means that one can reconstruct the elementary symmetric functions $\sg_k$ in the variables $\ex{- \f{2 \pi}{\om_2} \tau_{1} }, \dots , \ex{- \f{2\pi}{\om_2} k \tau_{N} } $ for $k=1,\dots, N-1 $
from the answer given above. The value of the last elementary symmetric polynomial $\sg_N$ is simply deduced from the constraints on the parameters \eqref{ecriture condition sur les racines tau et tilde tau}. 
Thus, the obtained integral representation allows one to fully reconstruct the two hyperbolic polynomials  $t_{\bs{\tau}}$ and $\tilde{t}_{ \tilde{\bs{\tau}} }$.

\item[ii)] The way of reconstructing  $\bs{\tau}$ and $ \tilde{\bs{\tau}}$ solely involves the parameters $\bs{\de}$ and the solutions to the non-linear integral equations
\eqref{definition NLIE}, \eqref{definition NLIE dual}, which occur in these expressions through $\mc{V}_{\bs{\de}}$ and $\wt{\mc{V}}_{ \tilde{\bs{\de}}}$. 
Thus, the present result allows one to fully describe the spectrum of the models solely on the level of non-linear integral equations. 

\item[iii)]  The hypotheses \eqref{ecriture contrainte sur les tau pour reconstruire energie} may appear as \textit{a posteriori} data which are hard to verify \textit{a priori}. 
However, one should note that, taken the definition of  $\mc{H}$, for $|\rho|$ small enough, the sets $\bs{\de}$ will be close to the poles of $\mc{H}$, \textit{viz}. the $\bs{\tau}$. 
Hence, the very definition of the contour $\msc{C}_{\bs{\de};\rho}$ ensures that  hypotheses \eqref{ecriture contrainte sur les tau pour reconstruire energie} does hold. 
For greater values of $|\rho|$, the validity of  \eqref{ecriture contrainte sur les tau pour reconstruire energie} is less clear, although, in practice, one may always reach 
the large values of $|\rho|$ by first starting from $|\rho|$ small enough, and then deforming the parameter. That would then allow one to find, for the value of $\rho$ of interest,
 $p_k, \wt{p}_k\in \mathbb{Z}$ such that 
\beq
\tau_k \pm \i\f{\om_1}{2} + \i p_k \om_1\in \mc{B}_{\pm} \qquad \e{and} \qquad \tilde{\tau}_k \pm \i\f{\om_2}{2} + \i \wt{p}_k \om_2 \in \wt{\mc{B}}_{\pm} \;. 
\enq
The reconstruction \eqref{ecriture reconstruction tau k}-\eqref{ecriture reconstruction tau tilde k} would then still hold, provided one makes the substitution 
$$\tau_k\hookrightarrow \tau_k  + \i p_k \om_1 \; , \quad \e{resp}. \quad \tilde{\tau}_k \hookrightarrow \tilde{\tau}_k  + \i \wt{p}_k \om_2 \; , $$
in the corresponding formulae.  
 
\end{itemize}

We now establish the result. 
We only focus on the Toda$_2$ model, and in that case solely establish eqn. \eqref{ecriture reconstruction tau k}. All the other cases are dealt with in a similar fashion, up to small, evident, modifications.

Let 
\beq
\mc{I}_k\; = \;  - \; 
\Int{ \msc{C}_{\bs{\de};\rho} }{} \Big\{ \a_k^{\prime}\big(\mu-\i \tfrac{\om_1}{2}\big) \, - \,  \a_k^{\prime}\big(\mu +\i \tfrac{\om_1}{2}\big)  \Big\} \ln \big[ \mc{V}_{\bs{\de}}\big](\mu) \cdot \f{ \dd \mu }{ 2\i \pi }\;. 
\enq
Upon using that $\ln \big[ \mc{V}_{\bs{\de}}\big]$ is smooth on the integration contour and that it is $\i\om_2$ periodic there, one may integrate by parts 
and, upon using that the $\a_k$'s are also $\i\om_2$ periodic one gets that the boundary terms cancel out, so that:
\beq
\mc{I}_k\; = \;   
\Int{ \msc{C}_{\bs{\de};\rho} }{} \Big\{ \a_k\big(\mu-\i \tfrac{\om_1}{2}\big) \, - \,  \a_k\big(\mu +\i \tfrac{\om_1}{2}\big)  \Big\} \f{ \mc{V}^{\prime}_{\bs{\de}}(\mu) }{ \mc{V}_{\bs{\de}}(\mu)  } \cdot \f{ \dd \mu }{ 2\i \pi }\;. 
\enq
The functions $v_{\ua/\da}$, resp. $\wt{v}_{\ua/\da}$, built from the given solution to the non-linear integral equation define polynomials 
$t_{\bs{\tau}}$ and $\tilde{t}_{ \tilde{\bs{\tau}} }$. One builds then, just as in Section \ref{Section Construction des solutions}, two elementary solutions 
$q_{\pm}$. From these one builds $\mc{V}_{\bs{\de}}$ as in \eqref{definition fct Vdelta via matrices K pm}. Thus, one has the explicit representation 
\beq
\f{ \mc{V}^{\prime}_{\bs{\de}}(\tau) }{ \mc{V}_{\bs{\de}}(\tau)  } \;  = \; 
\f{K^{\prime}_+\big(\tau-\i\tfrac{\om_1}{2} \big)}{K_+\big(\tau-\i\tfrac{\om_1}{2} \big)} \, + \,  \f{K^{\prime}_-\big(\tau+\i\tfrac{\om_1}{2} \big)}{K_-\big(\tau+\i\tfrac{\om_1}{2} \big)}   \, - \,  
\f{\mc{H}^{\prime}\big(\tau-\i\tfrac{\om_1}{2} \big)}{ \mc{H} \big(\tau-\i\tfrac{\om_1}{2} \big) } \;. 
\enq
Observe that one has the below bounds in the $\la \rightarrow  \infty$ regime
\beq
\Big| \a_k\big(\la-\i \tfrac{\om_1}{2}\big) \, - \,  \a_k\big(\la +\i \tfrac{\om_1}{2}\big)  \Big| \; \leq \; C_k 
\left\{ \ba{cc}  \big| \ex{-\f{2\pi}{\om_2} k \la  } \big| &    , \;\; \la \in \mc{B}_+ \vspace{3mm} \\  
 1 &    , \;\; \la \in \mc{B}_-  \ea \right. 
\enq

for some constant $C_k>0$, as well as
\beq
\Big| \f{K^{\prime}_+\big(\la-\i\tfrac{\om_1}{2} \big)}{K_+\big(\la-\i\tfrac{\om_1}{2} \big)}  \bigg| \; \leq \; C \big| \ex{ \f{2\pi}{\om_2} N \la  } \big|  \;, \;\; \la \in \mc{B}_+ \; \quad \e{and} \quad 
\Big|  \f{K^{\prime}_-\big(\la+\i\tfrac{\om_1}{2} \big)}{K_-\big(\la+\i\tfrac{\om_1}{2} \big)} \bigg| \; \leq \; C \big| \ex{ - \f{2\pi}{\om_2}  \la  } \big|   \;, \;\; \la \in \mc{B}_- 
\enq
for some constant $C$.

Thus, the decay at infinity and the $\i\om_2$ periodicity of the integral allow one to split the integral into three pieces
\bem
\mc{I}_k\; = \;   \Int{ \partial \mc{B}_+  }{} \Big\{ \a_k\big(\mu-\i \tfrac{\om_1}{2}\big) \, - \,  \a_k\big(\mu +\i \tfrac{\om_1}{2}\big)  \Big\} 
\f{K^{\prime}_+\big(\mu-\i\tfrac{\om_1}{2} \big)}{K_+\big(\mu-\i\tfrac{\om_1}{2} \big)}  \cdot \f{ \dd \mu }{ 2\i \pi } \\
-  \Int{ \partial  \mc{B}_-  }{} \Big\{ \a_k\big(\mu-\i \tfrac{\om_1}{2}\big) \, - \,  \a_k\big(\mu +\i \tfrac{\om_1}{2}\big)  \Big\}
\f{K^{\prime}_-\big(\mu+\i\tfrac{\om_1}{2} \big)}{K_-\big(\mu+\i\tfrac{\om_1}{2} \big)}  \cdot \f{ \dd \mu }{ 2\i \pi }  \\
  -  \hspace{-2mm}  \Int{ \substack{ \msc{C}_{\bs{\de};\rho} \cup\\  \{-\msc{C}_{\bs{\de};\rho} +\i \om_1  \} }   }{}  \hspace{-4mm} \a_k\big(\mu-\i \tfrac{\om_1}{2}\big)  
 \f{\mc{H}^{\prime}\big(\mu-\i\tfrac{\om_1}{2} \big)}{ \mc{H} \big(\mu-\i\tfrac{\om_1}{2} \big) }  \cdot \f{ \dd \mu }{ 2\i \pi } \;. 
\end{multline}
The first two parts can be computed, at this stage, by residues and each yields $0$ since the integrand is holomorphic.
Relatively to the third integral, one may close the contours by virtue of $\i\om_2$-periodicity. 
$\ln^{\prime}\mc{H} \big(\mu-\i\tfrac{\om_1}{2} \big)$ has simple poles at $\mu\in \{ \i\tfrac{\om_1}{2}+\de_k + \i \mathbb{Z} \om_1 + \i \mathbb{Z} \om_2  \}$
with $2\i\pi$ residue $+1$ and simple poles at $\mu\in \{ \i\tfrac{\om_1}{2}+\tau_k + \i \mathbb{Z} \om_1 + \i \mathbb{Z} \om_2  \}$
with $2\i\pi$ residue $-1$. 

By construction of the curve $\msc{C}_{\bs{\de};\rho}$, one has that $\de_k \pm \i\tf{\om_1}{2} \in \mc{B}_{\pm}$. Since $\de_k + 3\i\tf{\om_1}{2}$ is already at distance 
greater than $1$ along the $\i \om_1$ axis and in units of $\i \om_1$, solely  $\de_k + \i\tf{\om_1}{2}$ lies inside the integration contour of the 
third integral. Likewise, by virtue of hypothesis \eqref{ecriture contrainte sur les tau pour reconstruire energie}, solely the poles at  $\tau_k + \i\tf{\om_1}{2}$ 
are located inside of the integration contour. Hence, 

\beq
\mc{I}_k\; = \;   \sul{a=1}{N} \Big\{ \a_k(\tau_a) \, - \, \a_k(\de_a)\Big\} \;, 
\enq
what entails the claim.

\section*{Conclusion}

In this work, we constructed the explicit solution to the scalar $\op{t}-\op{Q}$ equations describing the spectrum of the $q$-Toda and Toda$_2$ chains. 
The solution $q$ was expressed through Fredholm determinants and we showed that it is equivalently expressed in terms of solutions to non-linear integral 
equations. Ultimately, we obtained a set of quantisation conditions, in the form of Thermodynamic Bethe Ansatz like equations.

\section*{Acknowledgments}

K.K.K. acknowledges support from  CNRS and ENS de Lyon. The authors are indebted to R. Kashaev, M. Mari\~{n}o, G. Niccoli and J. Teschner  for stimulating discussions
at various stages of the project.

\begin{appendix}

 \section{Infinite determinants}
\label{Appendix Sous Section determinants infinis}

 Let $K_{\pm}^{(n)}(\la)$ be defined as the below determinants of $n\times n$ tridiagonal matrices
\begin{eqnarray}
K_+^{(n)}(\la) &=& \det \left[ \ba{cccccc}  1  & b_1^{(+)} & 0 & \cdots  &0  \\
            c_2^{(+)} & 1 &    b_2^{(+)} & \ddots &  \vdots\\   
            0 & \ddots & \ddots & \ddots  &     0 \\
	    0 & \ddots & \ddots & \ddots  &     b_{n-1}^{(+)} \\
            0 & \ddots & 0   &  c_n^{(+)}  & 1
                                    \ea \right] 
 \quad \e{with}  \quad \left\{ \ba{cc} b_k^{(+)} & = \, \f{1}{t_{\bs{\tau}} (\la +  \i k \om_1) }  \vspace{2mm} \\ 
 c_k^{(+)} & = \, \f{\rho^{\om_1}}{ t_{\bs{\tau}}(\la+\i k \om_1) } \ea \right. \\
K_-^{(n)}(\la) &=& \det \left[ \ba{cccccc}  \ex{\om_1 \zeta}  & b_1^{(-)} & 0 & \cdots & 0    \\
            c_2^{(-)} & \ex{\om_1 \zeta}  &    b_2^{(-)} & \ddots &  \vdots  \\   
            0 & \ddots & \ddots & \ddots  &     0 \\
	    0 & \ddots & \ddots & \ddots  &     b_{n-1}^{(-)} \\
            0 & \ddots & 0   &  c_n^{(-)}  & \ex{\om_1 \zeta}
                                    \ea \right] 
 \quad \e{with}  \quad \left\{ \ba{cc} b_k^{(-)} & = \, \f{1}{t_{\bs{\tau}} (\la -  \i k \om_1) }  \vspace{2mm} \\ 
 c_k^{(-)} & = \, \f{\rho^{\om_1} \ex{ 2 \om_1 \zeta} }{ t_{\bs{\tau}}(\la  - \i k \om_1) } \ea \right. 
\end{eqnarray}
Here, we remind that $\ex{\om_1 \zeta}= \ex{\om_1 p_0} \pl{a=1}{N} \ex{-\tfrac{2\pi}{\om_2} \tau_a}$. Analogously, define the dual determinants as 

\begin{eqnarray}
\wt{K}_+^{(n)}(\la) &=& \det \left[ \ba{cccccc}  \ex{\om_2 \tilde{\zeta} }  & \tilde{b}_1^{(+)} & 0 & \cdots & 0    \\
            \tilde{c}_2^{(+)} & \ex{\om_1 \tilde{\zeta} }  &    \tilde{b}_2^{(+)} & \ddots &  \vdots  \\   
            0 & \ddots & \ddots & \ddots  &     0 \\
	    0 & \ddots & \ddots & \ddots  &     \tilde{b}_{n-1}^{(+)} \\
	    0 & \ddots & 0   &   \tilde{c}_n^{(+)}  & \ex{\om_2 \tilde{\zeta} }
                                    \ea \right] 
 \quad \e{with}  \quad \left\{ \ba{cc} \tilde{b}_k^{(+)} & = \, \f{1}{ \tilde{t}_{ \tilde{\bs{\tau}} } (\la +  \i k \om_2) }  \vspace{2mm} \\ 
  \tilde{c}_k^{(+)} & = \, \f{ \rho^{ \om_2 } \ex{ 2 \om_2 \tilde{\zeta} }  }{  \tilde{t}_{ \tilde{\bs{\tau}} } (\la+\i k \om_2) } \ea \right.  \\
\wt{K}_-^{(n)}(\la) &=& \det \left[ \ba{cccccc}  1  &  \tilde{b}_1^{(-)} & 0 & \cdots  &0  \\
            \tilde{c}_2^{(-)} & 1 &    \tilde{b}_2^{(-)} & \ddots &  \vdots\\   
            0 & \ddots & \ddots & \ddots  &     0 \\
	    0 & \ddots & \ddots & \ddots  &     \tilde{b}_{n-1}^{(-)} \\
	    0 & \ddots & 0   &  \tilde{c}_n^{(-)}  & 1
							\ea \right] 
 \quad \e{with}  \quad \left\{ \ba{cc} \tilde{b}_k^{(-)} & = \, \f{1}{ \tilde{t}_{ \tilde{\bs{\tau}} } (\la - \i k \om_2) }  \vspace{2mm} \\ 
	      \tilde{c}_k^{(-)} & = \, \f{  \rho^{\om_2}   }{  \tilde{t}_{ \tilde{\bs{\tau}}} (\la - \i k \om_2)   } \ea \right.
\end{eqnarray}
Further, in the dual case, one has $\ex{\om_2\tilde{\zeta} } = \ex{\om_2 p_0} \pl{a=1}{N} \ex{-\tfrac{2\pi}{\om_1} \tilde{\tau}_a}$.

Define 
\beq
\mc{D}^{(\pm)}_{\bs{\tau}} \, = \, \Cx \setminus \Big\{ \tau_k  \mp \i \om_1 \mathbb{N}^* +\i \om_2 \mathbb{Z} \Big\}_{k=1}^{N}
\qquad \e{and} \qquad 
\wt{\mc{D}}^{(\pm)}_{ \tilde{\bs{\tau}} } \, = \, \Cx \setminus \Big\{ \tilde{\tau}_k  \mp \i \om_2 \mathbb{N}^* + \i \om_1 \mathbb{Z} \Big\}_{k=1}^{N} \;. 
\enq

Then, under the hypothesis 
\beq
 \underset{a\in \{ 1,2 \} }{\e{max}}  \Big| \rho^{\om_a}  \ex{2 \om_a p_0 } \Big| \; < \; 1 
\enq
for the Toda$_2$ chain and under no additional condition for the $q$-Toda chain, $K_{\pm}^{(n)}$, resp. $\wt{K}_{\pm}^{(n)}$, converge 
on compact subsets of $\mc{D}^{(\pm)}_{\bs{\tau}} $, resp. $\wt{\mc{D}}^{(\pm)}_{ \tilde{\bs{\tau}} }$, to holomorphic $\i \om_2$ periodic functions $K_{\pm}$, resp. $\i \om_1$ periodic functions $\wt{K}_{\pm}$, 
on $\mc{D}^{(\pm)}_{\bs{\tau}}$, resp. $\wt{\mc{D}}^{(\pm)}_{ \tilde{\bs{\tau}} }$. These functions extend to meromorphic functions on $\Cx$.  
The functions $K_{\pm}$: 
\begin{itemize}
\item  have simple poles at $\big\{ \tau_k  \mp \i \om_1 \mathbb{N}^* +\i \om_2 \mathbb{Z} \big\}_{k=1}^{N}$;
\item have the asymptotics \eqref{ecriture asymptotiques de K+}, \eqref{ecriture asymptotiques de K-}; 
\item solve the second order finite difference equations \eqref{equation de reccurrence K+}, \eqref{equation de reccurrence K-}. 

\end{itemize}

Analogously, the dual functions $\wt{K}_{\pm}$:
\begin{itemize}
\item   $\wt{K}_{\pm}$ have simple poles at $\big\{ \tilde{\tau}_k  \mp \i \om_2 \mathbb{N}^* + \i \om_1 \mathbb{Z} \big\}_{k=1}^{N}$;
\item have the asymptotics \eqref{ecriture asymptotiques de tilde K+}, \eqref{ecriture asymptotiques de tilde K-}; 
\item solve the second order finite difference equations \eqref{equation de reccurrence tilde K+}, \eqref{equation de reccurrence tilde K-}. 
 
\end{itemize}

\vspace{5mm}

The proof of these properties goes as follows. 
We first assume that convergence holds and establish periodicity and the second order finite difference equation. We treat the case of $K_{+}$
as all others are dealt similarly.

We first prove the periodicity. One obviously has $K_+^{(n)}(\la + \i\om_2) = K_+^{(n)}(\la)$
%
%
%
%
%
%
%
%
Hence, taking the limit, one gets $K_+(\la + \i\om_2) = K_+(\la)$.
The recurrence equations \eqref{equation de reccurrence K+} follows by developing in respect to the first column:
\beq
K_+^{(n)}(\la-\i\om_1) = K_+^{(n-1)}(\la) - \f{ \rho^{\om_1} K_+^{(n-2)}(\la+\i\om_1)  }{ t_{\bs{\tau}}(\la+\i\om_1) t_{\bs{\tau}}(\la) } \;,
\enq
and then taking the limit on the level of this representation. 
Finally, the presence/absence of poles at some point $z \in \Cx$ 
follows upon taking the $n \rightarrow +\infty$ limit on the level of $\e{Res}(K_+^{(n)}(\la) \dd \la, \la=z) $.

Hence, it remains to establish convergence and asymptotics. Here, we treat the case of $K_+$ and $K_-$ as both determinants demands 
slightly different techniques to deal with.

To start with, we observe that given an $n\times n$ matrix with entries $M_{ab}^{(n)}$ and the identity matrix $I_n$, it holds 
\beq
\det\big[ I_n \, + \, M^{(n)} \big] \; = \; \sul{k=0}{n}\f{1}{k!} \sul{ \substack{ a_1,\dots, a_k \\ a_i \in [\![ 1 ; n ]\!]} }{} \det_k\big[ M_{a_{\ell}a_{p}}^{(n)}  \big]
\enq

Further, one has the bound ensured by Haddamard's inequality and the fact that it holds $\sul{\ell=1}{n} |w_{\ell}|^2 \leq \Big( \sul{\ell=1}{n} |w_{\ell}| \Big)^2 $:
\beq
\Big| \sul{ \substack{ a_1,\dots, a_k \\ a_i \in [\![ 1 ; n ]\!]} }{} \det_k\big[ M_{a_{\ell}a_{p}}^{(n)}  \big] \Big| \; \leq \; 
\sul{ \substack{ a_1,\dots, a_k \\ a_i \in [\![ 1 ; n ]\!]} }{}  \pl{s=1}{k} \sqrt{ \sul{p=1}{n} |M_{a_{s}a_{p}}^{(n)} |^2  }
\; \leq \: \Big( \op{C}[ M  ] \Big)^k
\enq
where 
\beq
\op{C}[ M  ]  \; = \; \limsup_{n\rightarrow + \infty} \sul{a,b \in \mathbb{N}^*}{} |M_{a b}^{(n)} | \;. 
\label{definition operateur C}
\enq
Thus, the sequence of determinants converges whenever $\op{C}[ M  ]<+\infty$. Furthermore, in such a case, one has the uniform in $n$ bound
\beq
\Big| \det\big[  I_n \, + \, M^{(n)} \big]  - 1 \Big| \; \leq \; \op{C}[ M  ]  \ex{  \op{C}[ M  ]   }  \;.
\label{ecriture borne determinant id plus M}
\enq

One can represent  $K_+^{(n)}(\la)=\det\big[ I_n \, +\,   \de \mc{K}_{+}^{(n)} \big]$ where 
\beq
  \Big(\de \mc{K}_{+}^{(n)}\Big)_{ab} \; = \; b_a^{(+)} \de_{a,b-1} \, + \, c_a^{(+)} \de_{a,b+1}
\enq
and $\de_{ab}$ is the Kronecker symbol. Further, take $\eps>0$ and small enough and define
\beq
\mc{D}^{(\pm)}_{\bs{\tau};\eps} \; = \; \Cx \setminus \bigcup\limits_{k=1}^{N} \bigcup_{\ell \in \mathbb{N}^* }\bigcup_{ p \in \mathbb{Z}} \mc{D}_{\tau_k\mp\i\om_1\ell+\i p \om_2,\eps}
\enq
where $\mc{D}_{z,\eps}$ is the open disk of radius $\eps$ centred at $z$. Then, uniformly in $\la \in \mc{D}^{(+)}_{\bs{\tau};\eps}$ of the form $\la=\i x \om_1+\i y \om_2$, with $x \geq x_0$ for some $x_0 \in \R$, 
it holds, for some $x_0$ and $\eps$-dependent constant $C$ , that  
\beq
   \Big| \f{1}{ t_{\bs{\tau}}(\la+\i k \om_1) } \Big| \; \leq \; C \Big| \ex{\f{  \ga  \pi}{\om_2} N \la   }   q^{k \ga }\Big| 
   \quad \e{with} \quad \left\{ \ba{cc} \ga=1 & q-\e{Toda}  \\ \ga=2 & \e{Toda}_2 \ea  \right. \; . 
\enq
Thence, for some constant $C^{\prime}$, one can bound $\op{C}[ \de \mc{K}_{+}^{(n)}  ] $,  with $\op{C}$ defined in \eqref{definition operateur C}, as
\beq
\op{C}[ \de \mc{K}_{+}^{(n)}  ]  \, \leq \; \sul{a \in \mathbb{N}^* }{}  \big| b_a^{(+)} \big| \,  + \, \sul{a \in \mathbb{N}^* }{}  \big| c_a^{(+)} \big|  \, \leq \, 
C^{\prime} \Big| \ex{\f{  \ga  \pi}{\om_2} N \la   }   \Big| \;. 
\enq
This bounds ensures that $K_n^{(+)}(\la)$ converges on compact subsets of $\mc{D}^{(\pm)}_{\bs{\tau}}$. Since the limit issues from a uniform convergence on compacts of a sequence 
on holomorphic functions, one gets that $K_+$ is holomorphic on  $\mc{D}^{(\pm)}_{\bs{\tau}}$. Finally, the asymptotics follow from the estimates on $\op{C}[ \de \mc{K}_{+}^{(n)}  ] $ given above
and the bound \eqref{ecriture borne determinant id plus M}.

In the case of the $q$-Toda chain, the case of $K_-^{(n)}$ is dealt with by very analogous handlings. However, the Toda$_2$ case demands more work. 
In that case, one can represent $K_-^{(n)}(\la) \, = \, \det\big[ \mc{T} \, + \, \de \mc{K}^{(n)}_-   \big] $, where $T$
is a tri-diagonal symmetric $n\times n$ Toeplitz matrix
\beq
\mc{T}\, =\,  \left[ \ba{cccccc}  \ex{\om_1 \zeta} & b  & 0 &0   \\
            c & \ex{\om_1 \zeta} &  \ddots &   0   \\   
            0 & \ddots &     \ddots &    b \\
	    0 & 0 &    c   &    \ex{\om_1 \zeta} 
                                    \ea \right] 
\quad \e{with} \quad  b \, = \, c \, = \,  \rho^{ \f{\om_1}{2} }\ex{\om_1 p_0}  \;, 
\enq
while $ \big(\de \mc{K}^{(n)}_-  \big)_{ab} \; = \; \mf{b}_a \de_{a,b-1} + \mf{c}_a \de_{a,b+1}  $ and the coefficients take the form 
\beq
\mf{b}_k \, = \,\mf{c}_k \, = \, \rho^{\f{\om_1}{2}  } \ex{\om_1 p_0} \bigg\{ \f{ (-1)^N  \ex{-\f{2\pi}{\om_2} \sul{a=1}{N}\tau_a}   }{t_{\bs{\tau}} (\la -  \i k \om_1) }-1 \bigg\}  
\quad \e{so} \; \e{that} \quad 
\big| \mf{b}_k  \big| \, \leq \, C \, \big| \ex{-\f{2\pi}{\om_2} \la } \big| \, |q^{2k} |\;, 
\label{ecriture bornes sur les b goth}
\enq
for some constant $C$ and uniformly in $\la\in \mc{D}^{(-)}_{\bs{\tau};\eps}$, with $\la=\i x \om_1+\i y \om_2$ with $x \leq x_0$ and $x_0$ fixed. 
The inverse of $\mc{T}$ can be computed explicitly, see \textit{e.g.} \cite{FoPe01} where the inversion of general tri-diagonal matrices is performed.
Set $\a$ such that $\ex{-\a}=b$. Observe that the hypothesis $|b|<1$ implies that $\Re(\a)>0$. Then, one has 
\beq
\mc{T}^{-1}_{k\ell} \; = \; (-1)^{k+\ell} \ex{\a} \left\{ \ba{cc} \f{ \sinh(k \a) \sinh[(n+1-\ell)\a] }{ \sinh(\a) \sinh[(n+1)\a]  } & k \leq \ell  \\
						     \f{ \sinh(\ell \a) \sinh[(n+1-k)\a] }{ \sinh(\a) \sinh[(n+1)\a]  } & k > \ell    \ea \right. \;. 
\enq
Then, it is straightforward to obtain, that, uniformly in $k,\ell \in [\![1; n ]\!]$, one has
\beq
\Big| \mc{T}^{-1}_{k\ell} \Big| \; \leq \; 2 \Big| \f{ \ex{\a} }{ \sinh(\a) }\Big| \cdot  \big| \ex{ \a}\big|^{-|k-\ell|}  \;. 
\label{ecriture bornes T inverse}
\enq
Thus, since one has 
\beq
 \Big(\mc{T}^{-1} \cdot \de \mc{K}^{(n)}_-  \Big)_{k\ell} \; = \; \mf{b}_{\ell-1} \mc{T}^{-1}_{k\ell-1} + \mf{c}_{\ell+1} \mc{T}^{-1}_{k\ell+1}
\enq
it holds
\beq
\Big| \Big(\mc{T}^{-1} \cdot \de \mc{K}^{(n)}_-  \Big)_{k\ell} \Big| \; \leq \; C      \big| \ex{ \a}\big|^{-|k-\ell|} \big|  \big| \ex{-\f{2\pi}{\om_2} \la } \big| \, |q^{2 \ell} | \;, 
\enq
 as ensured by eqns. \eqref{ecriture bornes sur les b goth}, \eqref{ecriture bornes T inverse} and for some constant $C$. Thus, it holds that 
\beq
 \op{C}\Big[\mc{T}^{-1} \cdot \de \mc{K}^{(n)}_-  \Big] \, \leq \, C^{\prime}  \big| \ex{-\f{2\pi}{\om_2} \la } \big|  \;. 
\enq
Since 
$$
\det[ \mc{T}] =   \f{1-\ex{-2 \a (n+1) } }{  1-\ex{-2\a} }
\quad \e{one} \; \e{has} \quad 
K_-(\la) \;  =\; \f{1-\ex{-2 \a (n+1) } }{  1-\ex{-2\a} } \cdot \det\big[ I_n \, + \, \mc{T}^{-1}\de \mc{K}_-^{(n)} \big]
$$
what allows one to conclude regarding to convergence to $K_-$ as well as to its asymptotics along $\mc{B}_-$.

It is useful to remark that, in the $q$-Toda chain case, all quantities $K_{\pm}$, $\wt{K}_{\pm}$ correspond to 
one parameter $\la$ family of  Fredholm determinants of $\e{id}$ + trace class operators  on $\ell^{2}(\mathbb{N})$. They can thus be immediately defined
as the corresponding infinite determinants, see \textit{e.g.} \cite{GoGoKr00}. However, this is no longer the case for the Toda$_2$ chain 
as the approximants of $K_{-}$ and $\wt{K}_{+}$ do not have a standard form. Of course, upon multiplication by $\mc{T}^{-1}$ one still recovers 
an expression for the form $\e{id}$ + rank $n$ approximation of a trace class operator, but the structure of the obtained matrix makes the 
second order finite difference equation structure much less apparent.

 \section{Non-linear integral equations: solvability and properties}
\label{Appendix NLIE} 
 
 Recall the definitions of the strips $\mc{B}$ given in  \eqref{definition bande B} and $\wt{\mc{B}}$ given in \eqref{definition bande B tilde}. 
 
Furthermore, given a collection of points $\de_k  \in \mc{B} $ and $\tilde{\de}_k\in \wt{\mc{B}}$ one introduces a 
non-intersecting contour
\begin{itemize}
 
 \item $\Ga_{\bs{\de} }$ in $\mc{B}$ starting at $\i x \om_1-\i \tf{\om_2}{2}$ and ending at $\i x \om_1+\i \tf{\om_2}{2}$ for some $x  \in \R$.
  The contour  $\Ga_{\bs{\de}}$ cuts $\mc{B}$ in two disjoint domains $\mc{B}_{\pm}$ such that $\pm \i t \om_1 \in \mc{B}_{\pm}$ for $t>0$ and large. 
 The contour is oriented in such a way that $\mc{B}_+$ is on its $+$ side, where the $+$ side is to the left of the contour along its orientation. 
 Finally, the contour $\Ga_{\bs{\de}}$ is chosen such that the points $\de_k \, \pm \, \i\f{\om_1}{2}\pm \i\mathbb{N} \om_1 $ are all in $\mc{B}_{\pm}$. 
 
 \item $\wt{\Ga}_{ \tilde{\bs{\de}} }$ in $\wt{\mc{B}}$ starting at $\i \tilde{x} \om_2 + \i \tf{\om_1}{2}$ and ending at $\i \tilde{x} \om_2+\i \tf{\om_1}{2}$ for some $\tilde{x}  \in \R$.
  The contour  $\wt{\Ga}_{ \tilde{\bs{\de}} }$ cuts $ \wt{\mc{B}}$ in two disjoint domains $\wt{\mc{B}}_{\pm}$ such that $\pm \i t \om_2 \in \wt{\mc{B}}_{\pm}$ for $t>0$ and large. 
 The contour is oriented in such a way that $\wt{\mc{B}}_+$ is on its $+$ side, where the $+$ side is to the left of the contour along its orientation. 
 Finally, the contour $\wt{\Ga}_{ \tilde{\bs{\de}} }$ is chosen such that the points  $\tilde{\de}_k \, \pm \, \i\f{\om_2}{2}\pm \i\mathbb{N} \om_2 $ are all in $\wt{\mc{B}}_{\pm}$.

\end{itemize}

\subsection{Unique solvability at small $\rho$ of the NLIEs of interest}
\label{Appendix SousSection solvabilite unique des NLIE pour Y}

Given   $f \in \mc{C}^{(0)}\big( \Ga_{\bs{\de}} \big) $, resp. $f \in \mc{C}^{(0)}\big( \wt{\Ga}_{ \wt{\bs{\de}}} \big)  $, denote
\beq
 \norm{f}_{\infty} \; = \; \sup\Big\{ |f(s)|\, : \, s \in \Ga_{\bs{\de}} \Big\}\; \; ,  \qquad \e{resp}. \qquad 
 \norm{f}_{ \wt{\infty} } \; = \; \sup\Big\{ |f(s)|\, : \, s \in\wt{\Ga}_{ \wt{\bs{\de}}} \Big\} \; . 
\enq
 Fix $R>0$ and introduce two Banach spaces 
\[
\mc{S}_{R}=\Big\{ f \in \mc{C}^{(0)}\big( \Ga_{\bs{\de}} \big)  \; : \; \norm{f}_{\infty}< R   \Big\}  \qquad \e{and} \qquad 
\wt{\mc{S}}_{R}=\Big\{ f \in \mc{C}^{(0)}\big( \wt{\Ga}_{ \wt{\bs{\de}}} \big)     \; : \; \norm{f}_{ \wt{\infty} }< R   \Big\}  \;.  
\]

Then, there exists $\rho_0>0$ and small enough such that, for any $|\rho|<\rho_0$,

\begin{itemize}

 \item[i)] for any  $f \in \mc{S}_{R}$ and $\wt{f} \in \wt{\mc{S}}_{R}$, the functions 
\beq
\tau \mapsto  1+  \f{ \rho^{\om_1} \ex{f(\tau)}  }{  \pl{ \eps= \pm 1 }{}t_{ \bs{\de}}\big(\tau+ \eps \i\tfrac{\om_1}{2}\big)     }   
\quad \e{and} \quad
\tau \mapsto  1 +  \f{ \rho^{\om_2} \ex{ \wt{f}(\tau)}  }{  \pl{ \eps= \pm 1 }{} \tilde{t}_{ \tilde{\bs{\de}} }\big(\tau+ \eps \i\tfrac{\om_2}{2}\big)     }   
\enq
have no zeroes on $\Ga_{\bs{\de}}$, resp. $\wt{\Ga}_{ \wt{\bs{\de}}}$, and have continuous principal value logarithms on these curves; 

\item[ii)] the non-linear integral equation
\beq
g(\la)= \Int{ \Ga_{\bs{\de}} }{} K(\la-\tau)
\ln\bigg[  1+  \f{ \rho^{\om_1} \ex{g(\tau)}  }{ t_{ \bs{\de}}\big(\tau-\i\tfrac{\om_1}{2}\big)  t_{ \bs{\de}}\big(\tau+\i\tfrac{\om_1}{2}\big)   }    \bigg] \cdot  \dd \tau 
\label{definition NLIE}
\enq
has a unique solution in $\mc{S}_R$.  Similarly, the dual non-linear integral equation
\beq
\wt{g}(\la)= - \Int{  \wt{\Ga}_{ \wt{\bs{\de}}} }{}  
\wt{K}(\la-\tau)
\ln\bigg[  1+  \f{ \rho^{\om_2} \ex{\wt{g}(\tau)}  }{ \tilde{t}_{ \tilde{\bs{\de}}}\big(\tau -\i\tfrac{\om_2}{2}\big) \tilde{t}_{ \tilde{\bs{\de}}}\big(\tau +\i\tfrac{\om_2}{2}\big)    }    \bigg] \cdot  \dd \tau 
\label{definition NLIE dual}
\enq
has a unique solution in $\wt{\mc{S}}_R$. We remind that $K$ has been introduced in eqn. \eqref{definition noyau K}, while its dual is defined as $\wt{K}=K_{\mid \om_1 \leftrightarrow \om_2}$.

\end{itemize}

Here, we give the proof in the case of equation \eqref{definition NLIE}. The dual case can be dealt with in much the same way. 
The first statement is clear provided that $|\rho|$ is small enough. 
Thence,  the operator
\beq
\op{O}[f](\la) = \Int{ \Ga_{\bs{\de}} }{}  \dd \mu \;  K(\la-\mu)
\ln\bigg\{ 1+ \f{  \rho^{\om_1} \ex{f(\mu)}  }{   t_{ \bs{\de}}\big(\tau-\i\tfrac{\om_1}{2}\big)  t_{ \bs{\de}}\big(\tau+\i\tfrac{\om_1}{2}\big)    } \bigg\} 
\enq
 is well-defined on $\mc{S}_{R}$. Also it stabilises this set provided that $|\rho|$ is small enough (its maximal magnitude depends on $R$). Indeed,
by using that 
\beq
 \Big|\ln (1+z) \Big| \,\leq \, 2 |z| \qquad \e{for} \quad |z|<\tfrac{1}{2}
\enq
and upon taking $|\rho|$ small enough, direct bounds lead to 
\beq
\big| \op{O}[f](\la) \big|  \, \leq \,  \Int{ \Ga_{\bs{\de}} }{}   |\dd \mu|  \; 
\f{ 2   | K(\la-\mu) | \,  |\rho^{\om_1}| \, \ex{ |f(\mu)| }  }{ \big|  t_{ \bs{\de}}\big(\mu-\i\tfrac{\om_1}{2}\big)  t_{ \bs{\de}}\big(\mu+\i\tfrac{\om_1}{2}\big)    \big|  } \; \leq \; 
2 I_K |\rho^{\om_1}| \ex{ \norm{f}_{\infty} } J^{-1} \; ,
\enq
where
\beq
I_K =  \max_{\la \in \Ga_{\bs{\de}}  }\Int{ \Ga_{\bs{\de}} }{} |\dd \tau| \, |K(\la-\tau)|  \; , 
\enq
and $J=\inf_{\tau \in  \Ga_{\bs{\de}} }  \big|  t_{ \bs{\de}}\big(\tau-\i\tfrac{\om_1}{2}\big)  t_{ \bs{\de}}\big(\tau+\i\tfrac{\om_1}{2}\big) \big|  >0$ since, 
by construction of $ \Ga_{\bs{\de}}$, 
\[
\e{d}\Big( \big\{\de_{k} \pm \i\tfrac{\om_1}{2} +\i\mathbb{Z} \om_1\big\} ,  \Ga_{\bs{\de}}  \Big)>0 \;.
\]
We shall now prove that $\op{O}$ is a contractive mapping on $\mc{S}_R$. This settles the question of existence and uniqueness of solutions to 
\eqref{definition NLIE} since any solution to \eqref{definition NLIE}  is a fixed point of $\op{O}$ in $\mc{S}_{R}$, by virtue of the Banach fixed point theorem.

Let $f, g \in \mc{S}_{R} $, then for $|\rho|$ small enough, 
\bem
\big|\op{O}[f] - \op{O}[g] \big| (\la)  \leq   \Int{0}{1} \!\!\dd t \Int{ \Ga_{\bs{\de}} }{}   \!\! |\dd \tau|  
\f{ |K(\la-\tau)|  \, | \rho^{\om_1}|  \,  \big| \ex{g(\tau)+t(f-g)(\tau) } \big|   \, \cdot \, | (f-g)(\tau) |  } 
{ \big|  t_{ \bs{\de}}\big(\tau-\i\tfrac{\om_1}{2}\big)  t_{ \bs{\de}}\big(\tau+\i\tfrac{\om_1}{2}\big)  \big|  -  |\rho^{\om_1}| \big| \ex{g(\tau)+t(f-g)(\tau) } \big|      }   
  \\
\leq I_K \f{ |\rho^{\om_1}| \ex{R }  } 
{  J -  |\rho^{\om_1}| \ex{ R }  }  \norm{f-g}_{\infty} <  \f{1}{2} \norm{f-g}_{\infty}\;.
\end{multline}

\subsection{The solutions $Y_{\bs{\de}}$ }
\label{Appendice Sous Section Fct Y delta}

In the following, given a collection of variables $\bs{\de}$ along with their duals $\wt{\bs{\de}}$, we assume that we are \textit{given} solutions $Y_{ \bs{\de}} $ and $\wt{Y}_{ \tilde{\bs{\de}} }$ to the two 
non-linear integral equations appearing in the body of the paper, eqns. \eqref{eqn NLIE pour Y} and \eqref{eqn NLIE pour Y dual}, along with their associated contours $\msc{C}_{\bs{\de};\rho}$ and $\wt{\msc{C}}_{ \tilde{\bs{\de}};\rho}$.
Also, below,  we will use the shorthand notations $\mc{V}_{\bs{\de}}$ and $\wt{\mc{V}}_{ \tilde{\bs{\de}} }$ introduced in \eqref{definition V delta}. 
Finally, we assume $\bs{\de}, \tilde{\bs{\de}}$ to be generic and, in particular, pairwise distinct \textit{modulo} the lattice $\i \om_1\mathbb{Z}+\i \om_2 \mathbb{Z}$.

 The function $\ex{ Y_{\bs{\de}}}$ and $ \ex{\wt{Y}_{\tilde{\bs{\de}}}}$, extend to meromorphic functions on $\Cx$. Let $\mc{Z}$, resp. $\wt{\mc{Z}}$, 
denote the set of zeroes of $\mc{V}_{\bs{\de}}$, resp. $\wt{\mc{V}}_{ \tilde{\bs{\de}} }$, 
in $\mc{B}$, resp. $\wt{\mc{B}}$, and let $\mc{Z}_{\pm}=\mc{B}_{\pm}\cap \mc{Z}$, resp. $\wt{\mc{Z}}_{\pm}= \wt{\mc{B}}_{\pm}\cap \wt{\mc{Z}}$. Then it holds that 
\begin{itemize}
 \item $\ex{ Y_{\bs{\de}} }$ has simple poles at $\de_k \, \pm \, 3\i\f{\om_1}{2}\pm \i\mathbb{N} \om_1 + \i \mathbb{Z} \om_2$; 
\item $\ex{ Y_{\bs{\de}} }$ has zeroes at $\mc{Z}_{\pm} \, \pm \, \i\om_1  + \i \mathbb{Z} \om_2$; 
\end{itemize}
and theses are its sole zeroes/poles in $\Cx$. Likewise 
\begin{itemize}
 
 \item $\ex{ \wt{Y}_{ \tilde{\bs{\de}}} }$ has simple poles at $ \tilde{\de}_k \, \pm \, 3\i\f{\om_2}{2}\pm \i\mathbb{N} \om_2 + \i \mathbb{Z} \om_1$; 
\item $\ex{ \wt{Y}_{ \tilde{\bs{\de}} } }$ has zeroes at $\wt{\mc{Z}}_{\pm} \, \pm \, \i\om_2  + \i \mathbb{Z} \om_1$;

\end{itemize}
and theses are its sole zeroes/poles in $\Cx$.

We stress that each of these simple poles is genuine meaning that it has \textit{non-vanishing} residues.

The functions  $\mc{V}_{\bs{\de}}$, resp. $\wt{\mc{V}}_{\tilde{\bs{\de}}}$ are meromorphic on $\Cx$:
\begin{itemize}
 \item $\mc{V}_{\bs{\de}}$ has simple poles at $\de_k \, \pm \, \i\f{\om_1}{2}\pm \i\mathbb{N} \om_1 + \i \mathbb{Z} \om_2$
 \item $\wt{\mc{V}}_{\tilde{\bs{\de}}}$ has simple poles at $\tilde{\de}_k \, \pm \, \i\f{ \om_2 }{2} \pm \i\mathbb{N} \om_2 + \i \mathbb{Z} \om_1$
\end{itemize}
and theses are their sole poles in $\Cx$.

Furthermore, given any $\eps>0$ small enough,  $Y_{\bs{\de}}$ has the asymptotic behaviour 

$\bullet$ $q$-Toda
\beq
\ex{ Y_{\bs{\de}}(\la) } \;  = \;  1+ \e{O}\Big(  \ex{ \pm 2N \f{\pi \la }{\om_2 }} \Big) \quad \e{for} \;\;  \la \rightarrow \infty \quad \la \in \mc{B}_{\pm}\setminus 
\bigcup\limits_{ \substack{ a \in [\![ 1;N ]\!]  \\ k,\ell\in \mathbb{Z} } }^{}  \mc{D}_{\de_{a}+\i\om_1 k + \i \om_2 \ell , \eps} \;, 
\label{ecriture DA de Y delta q toda}
\enq
where $\mc{D}_{z,\eps}$ is the disk of radius $\eps$ and centred at $z$. 

$\bullet$ Toda$_2$
\beq
\ex{ Y_{\bs{\de}}(\la) } \;  = \;  1+ \e{O}\Big(  \ex{  4 N \f{\pi \la }{\om_2 }} \Big) \quad \e{for} \;\; \la \rightarrow \infty \quad \la \in \mc{B}_{+}\setminus 
\bigcup\limits_{ \substack{ a \in [\![ 1;N ]\!]  \\ k,\ell\in \mathbb{Z} } }^{}  \mc{D}_{\de_{a}+\i\om_1 k + \i \om_2 \ell , \eps} \;, 
\label{ecriture DA de Y delta q toda dans B+}
\enq
and 
\beq
\ex{ Y_{\bs{\de}}(\la) } \;  = \;  \f{1}{  1- \rho^{\om_1} \pl{a=1}{N} \ex{\tfrac{4\pi}{\om_2}\de_a} }  \, + \,  \e{o}\big( 1\big)  \quad \e{for}  \;\;  \la \rightarrow \infty \quad \la \in \mc{B}_{-}\setminus 
\bigcup\limits_{ \substack{ a \in [\![ 1;N ]\!]  \\ k,\ell\in \mathbb{Z} } }^{}  \mc{D}_{\de_{a}+\i\om_1 k + \i \om_2 \ell , \eps} \;, 
\label{ecriture DA de Y delta q toda dans B-}
\enq
provided that $ \Big| \rho^{\om_1} \pl{a=1}{N} \ex{\tfrac{4\pi}{\om_2}\de_a} \Big|<1$. 

Analogous asymptotics holds for the dual case. 

\vspace{5mm}

The proof of the meromorphic continuation goes by induction. 

First of all, equation \eqref{eqn NLIE pour Y} defines $ \ex{ Y_{\bs{\de}} }$ as an analytic function on the strip $\mc{B}^{(1)}$, where 
\beq
\mc{B}^{(n)} \; = \; \Big\{ z = x \i \om_1 + y \i \om_2 \in \mc{B} \; : \; \e{for} \; \; w = u\i\om_1+ \i y \om_2 \, \in \,   \msc{C}_{\bs{\de};\rho}  \; \; \e{it} \;  \e{holds} \; \; |u-x|<n \  \Big\} \;. 
\enq
Therefore, $\mc{V}_{\bs{\de}}$ is meromorphic on the same strip, with simple poles at $\de_{a}\pm \i \tf{ \om_1 }{ 2 }$ and with residues
\beq
\e{Res} \Big( \mc{V}_{\bs{\de}}(\la) \dd \la , \la= \de_{k} \pm \i \tfrac{ \om_1 }{ 2 }  \Big) \; = \; 
\f{  \rho^{\om_1} \ex{ Y_{\bs{\de}} \big(\de_{k} \pm \i \tfrac{ \om_1 }{ 2 } \big)  }   }
	{  t_{\bs{\de}}( \de_{k} \pm \i \om_1 ) \,  t_{\bs{\de}}^{\prime}( \de_{k}  )    } \; \not=\; 0 \;. 
\label{ecriture condition non nullite premier residu Y delta}
\enq

Let $\msc{C}_{\bs{\de};\rho}^{(\om_2)} \, = \, \cup_{n \in \mathbb{Z}} \big\{ \msc{C}_{\bs{\de};\rho}+\i n \om_2  \big\}$.
Given $ \la \in \Cx \setminus \msc{C}_{\bs{\de};\rho}^{(\om_2)}$, define 
\beq
\ex{J_{\bs{\de}}(\la) } \; = \;\exp\bigg\{  \Int{ \msc{C}_{\bs{\de};\rho} }{}
\Big(  \coth\Big[ \f{\pi}{\om_2}(\la-\tau-\i\om_1)\Big] - \coth\Big[\f{\pi}{\om_2}(\la-\tau+\i\om_1)\Big]  \Big)
\ln \big[ \, \mc{V}_{\bs{\de}}\, \big](\tau) \cdot \f{\dd \tau}{2\i\om_2}  \bigg\} \;. 
\enq
Then, $\ex{ J_{\bs{\de}}(\la) } $ is holomorphic in $\Cx \setminus\msc{C}_{\bs{\de};\rho}^{(\om_2)}$ and satisfies to the jump conditions
\beq
\ex{J_{\bs{\de};+}(\la+\i\om_1) } \ex{-J_{\bs{\de};-}(\la+\i\om_1) }  \, = \, \mc{V}_{\bs{\de}}^{-1}(\la) \quad \e{and} \qquad
\ex{J_{\bs{\de};+}(\la - \i\om_1) } \ex{-J_{\bs{\de};-}(\la - \i\om_1) }  \, = \, \mc{V}_{\bs{\de}}(\la) \; \;  
\enq
for $ \la \in\msc{C}_{\bs{\de};\rho}^{(\om_2)}$. 
Here, $\ex{J_{\bs{\de};\pm}(\la + \eps \i\om_1) }$ stand for the $\pm$ boundary values of the function on the curve  $  \msc{C}_{\bs{\de};\rho}^{(\om_2)}$.
We remind that the $+$ side of  a curve is found to the left when following its orientation.

Thus, from the already gathered data, one gets that $ \ex{Y_{\bs{\de}}(\la)}$ can be meromorphically continued to $\mc{B}^{(2)}$ as
\beq
\ex{Y_{\bs{\de}}(\la) } \, = \, \ex{ J_{\bs{\de}}(\la) } \left\{  \ba{cc} \mc{V}_{\bs{\de}}(\la-\i\om_1)  & \la \in \mc{B}^{(2)}_{+}\setminus \mc{B}^{(1)}     \vspace{2mm} \\ 
									    1  & \la \in  \mc{B}^{(1)}   \vspace{2mm} \\
									      \mc{V}_{\bs{\de}}(\la+\i\om_1)  & \la \in \mc{B}^{(2)}_{-}\setminus \mc{B}^{(1)}   \ea  \right.   \;. 
\enq
Let  $\mc{Z}^{(1)}$ denote the zeroes of the meromorphic function $\mc{V}_{\bs{\de}}$ in $\mc{B}^{(1)}$. 
Thus, $\ex{Y_{\bs{\de}}(\la)}$, as a meromorphic function on $ \mc{B}^{(2)}$,  has zeroes in $\mc{Z}_{\pm}^{(1)}\pm\i \om_1$, with $\mc{Z}_{\pm}^{(1)}= \mc{Z}^{(1)}\cap \mc{B}_{\pm}$ and simple poles in $\de_{k} \pm \i 3 \tf{\om_1}{2}$.
Since the residues of $\mc{V}_{ \bs{\de} }$ do not vanish at $\de_{k} \pm \i  \tf{\om_1}{2}$, \textit{c.f.} eqn. \eqref{ecriture condition non nullite premier residu Y delta},
the same property holds for the residues of  $\ex{Y_{\bs{\de}}(\la)}$
at  $\de_{k} \pm \i 3 \tf{\om_1}{2}$ since $\ex{J_{ \bs{\de} } }$ does not vanish on $\mc{B}^{(2)} \setminus \mc{B}^{(1)}$. 
The above also entails that $\mc{V}_{\bs{\de}}$ can be meromorphically continued to $\mc{B}^{(2)}$. Its only poles are simple and are located at  
$\de_{k} \pm \i \tf{\om_1(1+p)}{2}$ for $p=0,1$ and all have non-vanishing residues. Furthermore, denote by  $\mc{Z}^{(2)}$ the zeroes of $\mc{V}_{ \bs{\de} }$  in $\mc{B}^{(2)}$.  
In its turn this information allows to meromorphically continue $Y_{\bs{\de}}$ to $\mc{B}^{(3)}$. 
By straightforward induction, the claim follows. The situation is similar for the dual case. 
It thus follows that 
\beq
\ex{Y_{\bs{\de}}(\la)} \, = \, \ex{ J_{\bs{\de}}(\la) } \left\{  \ba{cc} \mc{V}_{\bs{\de}}(\la-\i\om_1)  & \la \in \mc{B}_{+}\setminus \mc{B}^{(1)}       \vspace{2mm}   \\ 
									    1  & \la \in  \mc{B}^{(1)}    \vspace{2mm}  \\
									      \mc{V}_{\bs{\de}}(\la+\i\om_1)  & \la \in \mc{B}_{-}\setminus \mc{B}^{(1)}   \ea  \right.   \;. 
\enq

It remains to establish the statement relative to the asymptotics. The $q$-Toda and Toda$_2$ chains demand a separate treatment. 

$\bullet$ $q$-Toda chain

Let 
\[
\mf{b}^{(n)} \; = \; \Bigg\{ z \in \Cx \; : \; z=x \i \om_1 + y \i \om_2 \; , \; \left\{ \ba{c}  y \in [-\tf{1}{2};\tf{1}{2}]  \\ x \in [n;n+1] \ea \right.  \Bigg\} 
\setminus  \bigcup\limits_{ \substack{ a \in [\![ 1;N ]\!]  \\ k,\ell\in \mathbb{Z} } }^{}  \mc{D}_{\de_{a}+\i\om_1 k + \i \om_2 \ell , \eps} \;.  
\]
We only focus on the case $\la \rightarrow \infty$ with $\la \in \mc{B}_{+;\eps}$ where 
\beq
\mc{B}_{\pm;\eps} \, = \,  \mc{B}_{\pm}\setminus  \bigcup\limits_{ \substack{ a \in [\![ 1;N ]\!]  \\ k,\ell\in \mathbb{Z} } }^{}  \mc{D}_{\de_{a}+\i\om_1 k + \i \om_2 \ell , \eps}  \;,
\enq
the other case being tractable in a similar way.  

First of all, for $n \in \mathbb{N}$, one has that 
\beq
\ex{ J_{ \bs{\de} }( \ga + \i n \om_1 )  } \, = \, 1  \, + \,  \e{O}\big( q^{ 2 n } \big)
\enq
and
\beq
\f{1}{  t_{ \bs{\de} } \big( \ga + \i n \om_1 -3\i\tfrac{\om_1}{2}\big)  t_{ \bs{\de} } \big( \ga + \i n \om_1 - \i\tfrac{\om_1}{2} \big)  } \; = \; \e{O}\big( q^{2Nn}\big)
\enq
uniformly in $\ga \in \mf{g}$, with 
\beq
\mf{g} \; = \; \Bigg\{ \ga = \i x \om_1 + \i y \om_2 \; ,  \; \left\{ \ba{c} x \in [0;1] \\ \; y\in [-\tf{1}{2};\tf{1}{2}]  \ea \right. \;
\e{so}\; \e{that} \quad  d\big( \ga, \big\{ \de_k + \i \mathbb{Z} \om_1+ \i \mathbb{Z} \om_2 \big\} \big)>\eps  \Bigg\} \;. 
\label{definition ensemble borne g pour les gammas}
\enq
Then, upon setting 
\beq
\mf{m}_{n}\; = \; \e{sup}\Big\{  \big| \ex{Y_{\de}(\la) } \big| \; : \; \la \in \mf{b}_n \Big\}\;, 
\enq
one readily infers the bound 
\beq
\mf{m}_n \, \leq \, \Big(1+C_1 |q|^{2n} \Big)\Big( 1 + |\rho^{\om_1}| \mf{m}_{n-1} C_2 |q|^{ 2 n N } \Big) \;, 
\enq
for some constants $C_1, C_2>0$ and provided that $n>n_0$ for some $n_0$ large enough. 
Upon setting 
\beq
\mf{m}_n^{\prime}  \, = \, \f{ \mf{m}_n  }{ 1+C_1 |q|^{2n} }   \quad \e{and} \; \e{observing} \; \e{that} \quad 
\f{  |\rho^{\om_1}|  C_2 |q|^{ 2 n N }   }{ 1+C_1 |q|^{2n} }  < 1
\enq
for any $n>n_0^{\prime}$, with $n_0^{\prime}$ large enough, one gets that $\mf{m}_n^{\prime}-\mf{m}_{n-1}^{\prime}<1$. Thus, 
$\mf{m}_n^{\prime} < n-n_0^{\prime}+1 + \mf{m}_{n_0^{\prime}-1}^{\prime}$. Thus, for $n$ large enough 
\beq
\mf{m}_n \, \leq \, \Big(1+C_1 |q|^{2n} \Big)\Big( 1 + |\rho^{\om_1}| 2 (n-1) C_2 |q|^{ 2 n N } \Big) \;, 
\enq
what ensures that $\ex{ Y_{\bs{\de}} }$ is bounded on $\cup_{n\geq n_0^{\prime \prime }  } \mf{b}^{(n)}$ for some $n_0^{\prime \prime }$ large enough. 
The expression for $\ex{ Y_{\bs{\de}} }$ in $\mc{B}_{+}\setminus \mc{B}^{(1)}$ allows one to conclude regarding to the asymptotics.

$\bullet$ Toda$_{2}$ chain

The regime where $\la \rightarrow \infty$ with $\la \in \mc{B}_{+;\eps}$ is treated in much the same way as for the $q$-Toda chain since,
for $\ga \in \mf{g}$, it holds:
\beq
\f{1}{ t_{ \bs{\de}}\big( \ga + \i n \om_1  \big) } \, = \, q^{2n N} \big( 1+ \e{O}(q^{2n}) \big) \ex{ \tfrac{2\pi}{\om_2} N \ga }
\enq
provided that $d\big( \ga, \big\{ \de_k + \i \mathbb{Z} \om_1+ \i \mathbb{Z} \om_2 \big\} \big)>\eps$.

However, for $\ga \in \mf{g}$ as in \eqref{definition ensemble borne g pour les gammas}, one has 
\beq
\f{1}{ t_{ \bs{\de}}\big( \ga - \i n \om_1  + 3 \i\tfrac{\om_1}{2} \big)  t_{ \bs{\de}}\big( \ga - \i n \om_1 + \i\tfrac{\om_1}{2} \big) } \, = \,   \pl{a=1}{N} \ex{ \tfrac{ 4\pi }{ \om_2 } \de_a } \; \cdot \; 
\big( 1+ q^{2n} \mc{T}_n(\ga)  \big) 
\enq
with $\mc{T}_n(\ga)$ uniformly bounded in $n\in \mathbb{N}$ and $\ga \in \mf{g}$. 
Likewise,
\beq
\ex{ J_{ \bs{\de} }( \ga -  \i n \om_1 )  } \, = \, 1  \, + \,  q^{ 2 n } j_n(\ga) 
\enq
$j_n(\ga)$ uniformly bounded in $n\in \mathbb{N}$ and $\ga \in \mf{g}$. We set 
\[
\mc{Y}_n(\ga) = \ex{ Y_{\bs{\de}}(\ga-\i n \om_1) } \quad \e{and} \quad  \chi = \rho^{\om_1} \pl{a=1}{N} \ex{ \tfrac{ 4\pi }{ \om_2 } \de_a } \; .
\]
One has the induction 
\beq
\mc{Y}_n(\ga)  \, = \, \Big( 1  \, + \,  q^{ 2 n } j_n(\ga)  \Big)\cdot \Big( 1 + \chi \mc{Y}_{n-1}(\ga) \big( 1+ q^{2n} \mc{T}_n(\ga)  \big)  \Big) \;. 
\enq
Upon making the change of unknown function
\beq
\mc{Y}_n(\ga)  \, = \, \f{1}{1-\chi} \, + \, \chi^n \vp_n(\ga) 
\enq
and setting
\beq
 w_n^{(1)}(\ga) \, = \, j_n(\ga) \, + \, \mc{T}_n(\ga) \, + \,  q^{2n} j_n(\ga) \mc{T}_n(\ga)   \; \; , \qquad
w_n^{(2)}(\ga) \, = \, j_n(\ga) + \f{\chi w_n^{(1)}(\ga) }{ 1-\chi }    \; , 
\enq
one gets 
\beq
\vp_{n}(\ga) \; = \; \big( q^{2} \chi^{-1} \big)^n w_n^{(2)}(\ga) \; + \; \big( 1+q^{2n} w_n^{(1)}(\ga) \big) \vp_{n-1}(\ga)\;. 
\enq
Let $\pi_n(\ga) \, = \, \pl{p=n_0}{n}\big( 1+q^{2p} w_p^{(1)}(\ga) \big)$. 
$\pi_n(\ga)$ is bounded in $n$ and $\ga \in \mf{g}$, and is uniformly away from $0$. Upon 
changing the unknown function $\vp_n(\ga) \, = \, \pi_n(\ga) \psi_n(\ga)$ one infers that 
\beq
\psi_n(\ga) \, = \; \psi_{n_0-1}(\ga) \; + \; \sul{k=n_0}{n}  \big( q^{2} \chi^{-1} \big)^k\,  \f{ w_n^{(2)}(\ga) }{ \pi_n(\ga) } \;. 
\enq
If $ \big| q^{2} \chi^{-1} \big| >1$, then 
\beq
\psi_n(\ga) \, \sim \;  \f{ \big( q^{2} \chi^{-1} \big)^n }{ 1-\chi q^{-2} } \,  \f{ w_{\infty}^{(2)}(\ga) }{ \pi_{\infty}(\ga) } 
\enq
what ensures that $\mc{Y}_{n}(\ga)=\f{1}{1-\chi} \, + \, \e{O}\big( q^{2n} \big) $. If $\big|  q^{2} \chi^{-1} \big|<1$, then $\psi_n(\ga) $
is bounded in $n$, uniformly in $\ga \in \mf{g}$. Then  $\mc{Y}_{n}(\ga)=\f{1}{1-\chi} \, + \, \e{O}\big( \chi^{n} \big) $.
This entails the claim for the range of parameters stated in the claim.




\subsection{The auxiliary functions $v_{\ua/\da}$}
 \label{Subappendix Auxiliary fcts nu up down}
 
 Let $Y_{\bs{\de}}$, $\msc{C}_{\bs{\de};\rho}$ and their duals be as introduced in Subsection \ref{Appendice Sous Section Fct Y delta} above and let 
 $\bs{\de}$, $\tilde{\bs{\de}}$ satisfy to the constraint \eqref{ecriture liste contraintes sur les deltas}. 
 Recall the definition \eqref{definition V delta} of $\mc{V}_{\bs{\de}}$ and $\wt{\mc{V}}_{ \tilde{\bs{\de}}}$. 
Let $v_{\ua}(\la), v_{\da}(\la-\i\om_1)$, resp. $\wt{v}_{\ua}(\la), \wt{v}_{\da}(\la-\i\om_2)$, be defined on $\msc{C}_{\bs{\de};\rho}$, 
resp. $\wt{\msc{C}}_{ \tilde{\bs{\de}};\rho}$, by means of the integral representations given in the core of the paper,  
eqns. \eqref{definition v up}, \eqref{definition v down} and eqns. \eqref{definition v up tilde}, \eqref{definition v down tilde}
for their duals. 

Then $ \la \mapsto v_{\ua}(\la)$ has no zeroes and no poles in $\mc{B}_+$. It  extends into a meromorphic function on $\Cx$ with
\begin{itemize}
 
 \item simple poles at $\de_k - \i \mathbb{N}^*\om_1  + \i \mathbb{Z} \om_2$;
 
 \item  zeroes at $\mc{Z}_--\i\tfrac{\om_1}{2}$.

 \end{itemize}
 Similarly, $\la \mapsto  v_{\da}(\la-\i\om_1)$ has no zeroes and poles in $\mc{B}_-$ and extends into a meromorphic function on $\Cx$ with
\begin{itemize}
 
 \item simple poles at $\de_k + \i \mathbb{N}^*\om_1 +\i\mathbb{Z} \om_2$; 
 
 \item zeroes at $\mc{Z}_+ + \i\tfrac{\om_1}{2}$ . 
 
 \end{itemize}

 Furthermore, $v_{\ua/\da}$ are related to $Y_{\bs{\de}}$ as
\beq
v_{\ua}\Big( \la + \i \f{\om_1}{2} \Big) \cdot  v_{\da}\Big( \la -3 \i \f{\om_1}{2} \Big) \, = \, \ex{ Y_{\bs{\de}}(\la) }
\enq
and satisfy to the Wronskian relation 
\beq
v_{\ua}\big( \la\big) \cdot  v_{\da}\big( \la \big) \, = \, 1 \, + \,  \rho^{\om_1} \f{ v_{\ua}\big( \la+\i\om_1\big) \,   v_{\da}\big( \la -\i\om_1 \big)  }{ t_{ \bs{\de}}(\la) \, t_{ \bs{\de}}(\la + \i\om_1)      }  \;. 
\label{ecriture relation Wronskien}
\enq

It is convenient to introduce the compact notation 
\beq
 \mc{D}_{ \bs{\de} , \eps} \; = \;  \bigcup\limits_{ \substack{ a \in [\!  [ 1;N ]\!]  \\ k,\ell\in \mathbb{Z} } }^{}  \mc{D}_{\de_{a}+\i\om_1 k + \i \om_2 \ell , \eps}  
\enq
Finally, for any $\eps>0$ small enough, $v_{\da}$ has the $ \la \rightarrow \infty$ asymptotic behaviour
\beq
v_{\da}(\la) \, = \, \left\{ \ba{cc}  1+\e{O}\Big( \ex{ \f{2\pi}{\om_2} \la}  \Big) \; ,   &    \la \in \mc{B}_{+}\setminus  \mc{D}_{ \bs{\de} , \eps}  \vspace{3mm}  \\ 
\exp\bigg\{\Int{ \msc{C}_{\bs{\de};\rho} }{}  \ln[ \mc{V}_{\bs{\de}}] (\tau)   \f{\dd \tau }{  \i \om_2 }   \bigg\} +\e{O}\Big( \ex{ - \f{2\pi}{\om_2} \la}  \Big)  \; , &  \la \in \mc{B}_{-}\setminus  \mc{D}_{ \bs{\de} , \eps}
  \ea \right.  \;. 
\enq
The asymptotic $ \la \rightarrow \infty$   behaviour of  $v_{\ua}$ depends, however, on the model and takes the form 
\beq
v_{\ua}(\la) \, = \, \left\{  \ba{cc c }  1   + \e{O}\Big( \ex{ \f{2\pi}{\om_2} \la}  \Big)  \; , &  \la \in \mc{B}_{+} \setminus  \mc{D}_{ \bs{\de} , \eps}     \vspace{3mm}   \\ 
 \f{  \exp\bigg\{ - \Int{ \msc{C}_{\bs{\de};\rho} }{}  \ln[ \mc{V}_{\bs{\de}}] (\tau)   \f{\dd \tau }{  \i \om_2 }   \bigg\}   }{  1 - \rho^{\om_1} \ex{ 2 \om_1 p_0}   \bs{1}_{\e{Toda}_2}}     
 + \e{O}\Big( \ex{ -\f{2\pi}{\om_2} \la}  \Big)  \; , &  \la \in \mc{B}_{-} \setminus  \mc{D}_{ \bs{\de} , \eps} 
  \ea \right.\;. 
\enq
Here $ \bs{1}_{\e{Toda}_2}=1$ in the case of the Toda$_2$ model and  $ \bs{1}_{\e{Toda}_2}=0$ in the case of the $q$-Toda chain. 

Dual properties hold for the dual functions, namely 
for any $\eps>0$ small enough, $\wt{v}_{\ua}$ has the $ \la \rightarrow \infty$ asymptotic behaviour
\beq
\wt{v}_{\ua}(\la) \, = \, \left\{ \ba{cc}  \exp\bigg\{-\Int{ \wt{\msc{C}}_{ \tilde{\bs{\de}};\rho} }{}  \ln[ \wt{\mc{V}}_{  \tilde{\bs{\de}} }] (\tau)   \f{\dd \tau }{  \i \om_1 }   \bigg\}  +\e{O}\Big( \ex{ -\f{2\pi}{\om_1} \la}  \Big)  \; ,  &    \la \in \wt{\mc{B}}_{+}\setminus  \mc{D}_{ \tilde{\bs{\de}} , \eps}  \vspace{3mm}  \\ 
1 +\e{O}\Big( \ex{ \f{2\pi}{\om_1} \la}  \Big)  \; , &  \la \in \wt{\mc{B}}_{-}\setminus  \mc{D}_{ \tilde{\bs{\de}} , \eps}
  \ea \right.  \;. 
\enq
The asymptotic $ \la \rightarrow \infty$   behaviour of  $v_{\ua}$ depends, however, on the model and takes the form 
\beq
\wt{v}_{\da}(\la) \, = \, \left\{  \ba{cc c }  \f{  \exp\bigg\{   \Int{ \wt{\msc{C}}_{ \tilde{\bs{\de}};\rho} }{}  \ln[ \wt{\mc{V}}_{ \tilde{\bs{\de}} }] (\tau)   \f{\dd \tau }{  \i \om_1 }   \bigg\}   }
{  1 - \rho^{\om_2} \ex{ 2 \om_2 p_0}   \bs{1}_{\e{Toda}_2}}        + \e{O}\Big(- \ex{ \f{2\pi}{\om_1} \la}  \Big) &  \la \in \mc{B}_{+} \setminus  \mc{D}_{ \bs{\de} , \eps}     \vspace{3mm}   \\ 
 1
 + \e{O}\Big( \ex{  \f{2\pi}{\om_1} \la}  \Big) &  \la \in \mc{B}_{-} \setminus  \mc{D}_{ \bs{\de} , \eps} 
  \ea \right.\;. 
\enq

\vspace{5mm}

The proof of this statement follows from the results obtained in Subsection \ref{Appendice Sous Section Fct Y delta} and the analytic continuation formulae
\bem
v_{\ua}(\la) \, = \, \exp\Bigg\{ -\Int{ \msc{C}_{\bs{\de};\rho} }{} \bigg\{ \coth\Big[ \f{\pi}{\om_2}(\la-\tau +\i\tfrac{\om_1}{2} ) \Big] \, + \, 1 \bigg\}  \ln\big[ \mc{V}_{\bs{\de}}\big] (\tau) \cdot  \f{\dd \tau }{ 2\i \om_2 }  \Bigg\}\\
\times 			\left\{ \ba{cc} 1 & \la \in \mc{B}_+-\i\tfrac{\om_1}{2}  \vspace{2mm} \\  
					\mc{V}_{\bs{\de}}\Big(    \la + \i\tfrac{\om_1}{2} \Big)  & \la \in  \mc{B}_--\i\tfrac{\om_1}{2}			    \ea \right.  \;, 
\end{multline}
and that 
\bem
v_{\da}(\la-\i\om_1) \, = \, \exp\Bigg\{  \Int{ \msc{C}_{\bs{\de};\rho} }{} \bigg\{ \coth\Big[ \f{\pi}{\om_2}(\la-\tau -\i\tfrac{\om_1}{2} ) \Big]  \, + \, 1 \bigg\} \ln\big[ \mc{V}_{\bs{\de}}\big] (\tau) \cdot  \f{\dd \tau }{ 2\i \om_2 }  \Bigg\}\\
\times 			\left\{ \ba{cc}\mc{V}_{\bs{\de}}\Big(    \la - \i\tfrac{\om_1}{2}\Big)   & \la \in \mc{B}_+ + \i\tfrac{\om_1}{2}  \vspace{2mm}  \\  
					1 & \la \in  	\mc{B}_- + \i\tfrac{\om_1}{2}		    \ea \right. \;. 
\end{multline}
From there, the results of Subsection \ref{Appendice Sous Section Fct Y delta} allow one to read out all the claimed properties.  
One reasons similarly in the dual case.

\section{General properties of solutions to the Baxter equation}
\label{Appendix ptes gnles Baxter eqn}

In this section of the appendix, we discuss various overall properties satisfied by solutions $q$ to the system of dual Baxter equations \eqref{ecriture ensemble autodual eqns Baxter 1}-\eqref{ecriture ensemble autodual eqns Baxter 2}. 
We shall assume that $\om_1$ and $\om_2$ are generic and that they satisfy to the condition $\Im( \om_1/\om_2) >0$ so that $|q|<1$.

\subsection{Wronskians}
\label{Appendix SousSection Wronskiens}

Given any two solutions $q_1$ and $q_2$ of the system of dual Baxter equations \eqref{ecriture ensemble autodual eqns Baxter 1}-\eqref{ecriture ensemble autodual eqns Baxter 2}, their Wronskians
\begin{eqnarray}
W_{\om_1}[q_1,q_2](\la) &=& q_1(\la) q_2(\la+\i\om_1)-q_2(\la)q_1(\la+\i\om_1) \; , \\
W_{\om_2}[q_1,q_2](\la)&=& q_1(\la) q_2(\la+\i\om_2)-q_2(\la)q_1(\la+\i\om_2) \; ,
\end{eqnarray}
 satisfy to the finite difference equations
\beq
W_{\om_1}[q_1,q_2](\la+\i\om_1) = \sg^2 \varkappa^{\om_1} W_{\om_1}[q_1,q_2](\la) \; , \; 
W_{\om_2}[q_1,q_2](\la + \i\om_2) = \sg^2 \varkappa^{\om_2} W_{\om_2}[q_1,q_2](\la) \; .
\label{ecriture eqns differences finie q Wronsikens}
\enq

Furthermore, with any solution  $u$ of the set of dual Baxter equations \eqref{ecriture ensemble autodual eqns Baxter 1}-\eqref{ecriture ensemble autodual eqns Baxter 2} one can associated a self-dual Wronskian
\beq
\mc{W}[u](\la) =  u(\la) u(\la+\i\Om) - \sg^{2} u(\la+\i\om_1) u(\la+\i\om_2) \;, 
\label{definition Wronskien self dual}
\enq
where $\sg$ is as defined in eq. \eqref{ecriture parametres eqn Baxter qToda} or eq. \eqref{ecriture parametres eqn Baxter Toda2}, depending on the model of interest. 
The self-dual Wronskian satisfies to the difference equations
\beq
\mc{W}[u](\la+\i\om_1) \, =\, \varkappa^{\om_1} \mc{W}[u](\la) \quad  \e{and}  \quad
\mc{W}[u](\la+\i\om_2) \, = \, \varkappa^{\om_2} \mc{W}[u](\la) \;.
\label{ecriture eqns diff finies pour Wronskien Self dual}
\enq

We now compute the Wronskians of the functions $q_{\pm}$ introduced in \eqref{definition fcts q pm solutions de TQ self dual}, or equivalently, \eqref{definition fcts q pm base des solutions de TQ}. 
Their explicit value will be of use in the course of the analysis. However, at this stage, one should consider the objects below
without having any \textit{a priori} connection with a $t-q$ equation. 
It holds that 
\beq
W_{\om_1}[q_+,q_-](\la) \; = \; \f{ \check{W}_{\om_1}[Q_+,Q_-](\la)  }{ \th_{ \bs{\de}}(\la)  \th_{ -\bs{\de}}(-\la-\i\om_1) }
\quad \e{and} \quad 
W_{\om_2}[q_+,q_-](\la) \; = \; \f{ \check{W}_{\om_2}[Q_+,Q_-](\la)  }{ \th_{ \bs{\de}}(\la)  \th_{ -\bs{\de}}(-\la) }
\label{ecriture formule close Wronskiens q pm}
\enq
where 
\begin{eqnarray}
 \check{W}_{\om_1}[Q_+,Q_-](\la) & = & Q_+(\la) Q_{-}(\la+\i\om_1)\, - \, q^{2N} Q_+(\la+\i\om_1) Q_{-}(\la)   \vspace{2mm} \\
 & = & 
\varkappa^{-\i\la} g^{-N\om_1} \wt{\psi}_{+}(\la)  \wt{\psi}_{-}(\la)   f_{p_0}^{(+)}(\la)    f_{p_0}^{(-)}(\la+\i\om_1)   \th_{ \bs{\de}}(\la)
\label{ecriture check W om1}
\end{eqnarray}
and 
\begin{eqnarray}
 \check{W}_{\om_2}[Q_+,Q_-](\la) & = & Q_+(\la) Q_{-}(\la+\i\om_2)\, - \,  Q_+(\la+\i\om_2) Q_{-}(\la) \vspace{2mm} \\
& = &
\varkappa^{-\i\la} g^{-N\om_2} \psi_{+}(\la)  \psi_{-}(\la)   f_{p_0}^{(+)}(\la)    f_{p_0}^{(-)}(\la+\i\om_2)   \wt{\th}_{ \tilde{\bs{\de}} }(\la) \;.
\label{ecriture check W om2}
\end{eqnarray}
Those results follow from direct calculations based on the finite difference equations satisfied by the $q$-infinite products, the ones satisfied by the functions $f_{p_0}^{(\pm)}$:
\beq
f_{p_0}^{(+)}(\la+\i\om_1) \; = \;   \ex{  \om_1  p_0 } \; f_{p_0}^{(+)}(\la) \;, \qquad
f_{p_0}^{(-)}(\la+\i\om_1) \; = \; (-1)^N \ex{ - \f{2\pi}{\om_2} N \la } f_{p_0}^{(-)}(\la) \;, 
\label{eqn diff finie f pm qToda}
\enq
for the Toda$_2$ chain, and 
\beq
f_{p_0}^{(+)}(\la+\i\om_1)  =      (-\i)^N \ex{-\f{\pi N }{\om_2}(\la+\i\om_1)  }  \ex{ \f{\om_1}{2}  p_0 }  \, f_{p_0}^{(+)}(\la) \;, \;
f_{p_0}^{(-)}(\la+\i\om_1)  =   (\i)^N \ex{-\f{\pi N }{\om_2}\la  } \ex{ \f{\om_1}{2}  p_0 }   \, f_{p_0}^{(-)}(\la) \;, 
\label{eqn diff finie f pm Toda2}
\enq
for the $q$-Toda chain. Note that, in order to set the first order difference equations in this form, we made use of the constraints \eqref{ecriture liste contraintes sur les deltas}. 
Finally, in the course of the computations, one also uses  the Wronskian relations satisfied by $v_{\ua/\da}$, eq. \eqref{ecriture relation Wronskien}. 
One can check by direct inspection that the two Wronskians satisfy to the finite difference equations eq. \eqref{ecriture eqns differences finie q Wronsikens}.

Finally, the self-dual Wronskian satisfies
\beq
\mc{W}[q_+](\la) \, = \, \mc{W}[q_-](\la) \, = \, 0 
\enq
while, given two elliptic functions $\mc{P}_+, \mc{P}_-$,  it holds
\beq
\mc{W}[\mc{P}_+ q_++ \mc{P}_- q_-](\la) \, = \, \mc{P}_+(\la) \mc{P}_-(\la)\varkappa^{-\i\la} g^{-N \Om} C_{ \tilde{\bs{\de}} }  \cdot \f{ \th_{\tilde{\bs{\de}}}(\la) }{ \th_{\bs{\de}}(\la) }   \;. 
\label{calcul explicite du Wronskien}
\enq
$C_{\tilde{\bs{\de}} }  $ is a constant depending on the model. 
\beq
C_{\tilde{\bs{\de}} } \, = \, 
\exp\bigg\{ -\f{\i\pi }{ \om_1 \om_2 } \sul{a=1}{N} \tilde{\de}_a^2 \, + \, \f{\i \pi N }{ 6 \om_1 \om_2} \big( \om_1^2+3\om_1\om_2+\om_2^2 \big) \bigg\}
\label{definition cste C tilde delta qToda}
\enq
for the $q$-Toda chain, while, in the Toda$_2$ case, it rather reads
\beq
C_{\tilde{\bs{\de}} } \, = \, 
\exp\bigg\{ -\f{\i\pi }{ \om_1 \om_2 } \sul{a=1}{N} \tilde{\de}_a^2 \,- \, \f{\pi N \Om }{ \om_1 \om_2 } \sul{a=1}{N} \tilde{\de}_a    \,  + \, \f{\i \pi N }{ 6 \om_1 \om_2} \big( \om_1^2+3\om_1\om_2+\om_2^2 \big) \bigg\} \;. 
\label{definition cste C tilde delta Toda2}
\enq
Here, again, the results follow from a direct computation and the use of Wronskian relations eq. \eqref{ecriture relation Wronskien}.

\subsection{General form of the solutions}
\label{Appendix Subsection forme generale solutions tq}

 We have now introduced enough notations so as to characterise the form of any solution to the set of dual Baxter equations \eqref{ecriture ensemble autodual eqns Baxter 1}-\eqref{ecriture ensemble autodual eqns Baxter 2}. 

 Let $q$ be any meromorphic on $\Cx$ solution to the set of two dual Baxter equations \eqref{ecriture ensemble autodual eqns Baxter 1}-\eqref{ecriture ensemble autodual eqns Baxter 2}
and let $\mf{s}_{\pm}$ be any two linearly independent meromorphic on $\Cx$ solutions to \eqref{ecriture ensemble autodual eqns Baxter 1}-\eqref{ecriture ensemble autodual eqns Baxter 2}. 
Then, there exist two elliptic functions $\mc{P}_{\pm}(\la)$ of periods $\i\om_1$, $\i\om_2$ such that
\beq
q(\la)=\mc{P}_+(\la) \mf{s}_+(\la)+\mc{P}_-(\la)  \mf{s}_-(\la)  \;.
\label{forme generale solutions equation TQ et dual}
\enq

To establish this property, first suppose that formula \eqref{forme generale solutions equation TQ et dual} holds. Then one can reconstruct the function $\mc{P}_{\pm}(\la)$ by computing the Wronskians $W_{\om_1}[q,\mf{s}_{\pm}]$. Indeed,
\beq
W_{\om_1}[q,\mf{s}_{\pm}](\la) \, = \, \mc{P}_{+}(\la) W_{\om_1}[\mf{s}_+,\mf{s}_{\pm}](\la)+
\mc{P}_{-}(\la) W_{\om_1}[\mf{s}_-,\mf{s}_{\pm}](\la) \, = \,  \mc{P}_{\mp}(\la) W_{\om_1}[\mf{s}_{\mp},\mf{s}_{\pm}](\la) \;.
\enq
Thus
\beq
\mc{P}_{\pm}(\la)= \pm \f{ W_{\om_1}[q,\mf{s}_{\mp}](\la)  }{ W_{\om_1}[\mf{s}_+,\mf{s}_-](\la)  } \;.
\enq
Note that this representation is also valid if one replaces $\om_1$ by $\om_2$, what ensures the consistence with $\mc{P}_{\pm}$ being elliptic. 

Now suppose that $q$ is any solution to the set of two dual Baxter $t-q$ equations \eqref{ecriture ensemble autodual eqns Baxter 1}-\eqref{ecriture ensemble autodual eqns Baxter 2}. Then define
\beq
q_{\e{red}}(\la)=q(\la)-\f{ W_{\om_1}[q,\mf{s}_{-}](\la)  }{ W_{\om_1}[\mf{s}_+,\mf{s}_-](\la)  } \mf{s}_+(\la)
+ \f{ W_{\om_1}[q,\mf{s}_{+}](\la)  }{ W_{\om_1}[\mf{s}_+,\mf{s}_-](\la)  } \mf{s}_-(\la) \; .
\enq

As the ratio of any two $W_{\om_1}$ is $\i\om_1$ periodic, one gets that, by construction
\beq
W_{\om_1}[q_{\e{red}},\mf{s}_+](\la)=W_{\om_1}[q_{\e{red}},\mf{s}_-](\la)=0 \;.
\enq
This leads to the system of equations for $q_{\e{red}}(\la)$:
\bem
\left( \ba{cc}  q_{\e{red}}(\la+\i\om_1) \mf{s}_+(\la)-q_{\e{red}}(\la) \mf{s}_+(\la+\i\om_1) \\
              q_{\e{red}}(\la+\i\om_1) \mf{s}_-(\la)-q_{\e{red}}(\la) \mf{s}_-(\la+\i\om_1)   \ea \right) \\
= \left( \ba{cc} \mf{s}_+(\la) &  \mf{s}_+(\la+\i\om_1) \\   \mf{s}_-(\la) &  \mf{s}_-(\la+\i\om_1)      \ea \right)
\left( \ba{cc} q_{\e{red}}(\la+\i\om_1) \\ -q_{\e{red}}(\la) \ea \right) = 0 \; . 
\end{multline}
There can exist non-trivial solutions only if the determinant  
\beq
\det \left( \ba{cc} \mf{s}_+(\la) &  \mf{s}_+(\la+\i\om_1) \\   \mf{s}_-(\la) &  \mf{s}_-(\la+\i\om_1)      \ea \right) \, = \, W_{\om_1}[\mf{s}_+,\mf{s}_-](\la)  
\enq
vanishes. However, $W_{\om_1}[\mf{s}_+,\mf{s}_-](\la)$ is a meromorphic function on $\Cx$ that is non-identically zero since the solutions $\mf{s}_{\pm}$ are linearly independent. Therefore
it can only vanish at isolated points. Hence, we get that $q_{\e{red}}(\la) \not= 0$ only for $\la$'s belonging to an
locally finite set. $q_{\e{red}}(\la)$ being meromorphic  on $\Cx$, we infer that $q_{\e{red}}=0$. 
A similar reasoning can be carried out by considering the $\om_2$ Wronskian $W_{\om_2}$. One thus gets that there exists $ \i\om_1$ periodic functions $\ga_{\om_1}(\la)$ and $\rho_{\om_1}(\la)$, 
as well as $\i\om_2$ periodic functions $\ga_{\om_2}(\la)$ and $\rho_{\om_2}(\la)$ such that
\beq
q(\la) \, = \,  \ga_{\om_1}(\la) \mf{s}_+(\la) + \rho_{\om_1}(\la) \mf{s}_-(\la) \, = \, 
\ga_{\om_2}(\la) \mf{s}_+(\la) \, + \, \rho_{\om_2}(\la) \mf{s}_-(\la) \;.
\enq
By using the the Wronskian reconstitution formulae once again, we get that 
\beq
\ga_{\om_1}(\la)=\ga_{\om_2}(\la) \quad \e{and} \quad   \rho_{\om_1}(\la)=\rho_{\om_2}(\la) \; .
\enq
In other words, there exists two meromorphic, $ \i\om_1$ and $ \i\om_2$ periodic, functions $\mc{P}_{\pm}(\la)$  such that
\beq
q(\la)= \mc{P}_{+}(\la) \mf{s}_+(\la) + \mc{P}_-(\la) \mf{s}_-(\la) \;.
\enq

\section{Special functions}
\label{Appendix Special functions}

\subsection{q products}
\label{Appendix q products}

Given $|p|<1$ one denotes
\beq
( z ; p)\, = \, \pl{k\geq0}{}(1-zp^k) \qquad \e{and} \qquad 
\big( w(\la); p \big)_{\bs{\mu}} \, = \,\pl{a=1}{N}  \pl{k\geq 0}{} \big(1-w(\la-\mu_a)\cdot p^k \big) \; , 
\enq
for any function $w$ and a collection of $N$ parameters $\bs{\mu}=(\mu_1,\dots, \mu_N)$. 
In particular, the $\th$ function and its dual take the form 
\beq
\th(\la)\, = \, \big( \ex{-\frac{2\pi}{\om_2}\la}; q^{2} \big)\cdot \big( q^2 \ex{\frac{2\pi}{\om_2}\la}; q^{2} \big) \qquad \e{and} \qquad 
\tilde{\th}(\la)\, = \, \big( \ex{\frac{2\pi}{\om_1}\la}; \tilde{q}^{\, -2} \big)\cdot \big( \, \tilde{q}^{\,-2} \ex{-\frac{2\pi}{\om_1}\la}; \tilde{q}^{\,-2} \big) \;. 
\label{definition des fonction theta et theta duale}
\enq
They satisfy to the first order finite difference equations
\beq
\th(\la-\i\om_1)\, = \, - \ex{ \frac{2\pi}{\om_2}\la}\th(\la) \quad \e{and} \quad 
\tilde{\th}(\la-\i\om_2)\, = \, - \ex{ - \frac{2\pi}{\om_1}\la} \tilde{\th}(\la)  
\enq
and enjoy the reflection relation
\beq
\th(-\la-\i\om_1)\, = \, \th(\la) \quad \e{and} \quad 
\tilde{\th}(-\la-\i\om_2)\, = \, \tilde{\th}(\la)   \;. 
\enq

Note that, up to a constant and an exponential prefactor, $\th(\la)$ coincides with the usual theta function $\th_1(\la\mid \tau)$. 
The modular transformation formula for $\th_3(\la\mid \tau)$ translates into
\beq
\th(\la) \, = \, \tilde{\th}(\la) \ex{ \i B(\la) }  
\label{ecriture transfo modulare fct theta}
\enq
in which 
\beq
B(z) \, = \,  \f{\pi  }{\om_1 \om_2 } z^2 +  \i \f{\pi \Om}{\om_1 \om_2 }z - \f{\pi}{6 \om_1 \om_2}\Big(\om_1^2+3\om_1\om_2+\om_2^2 \Big) \,  
\enq
and  we remind that
\[
\Omega= \omega_1+\omega_2 \;.
\]
Given a collection of $N$ parameters $\bs{\mu}=(\mu_1,\dots, \mu_N)$ it also appears convenient to denote 
\beq
\th_{\bs{\mu}}(\la) \, = \, \pl{a=1}{N}\th(\la-\mu_a) \qquad \e{and} \qquad \tilde{\th}_{ \bs{\mu} }(\la) \, = \, \pl{a=1}{N}\tilde{\th}(\la-\mu_a) \;. 
\enq

\subsection{The double sine function}
\label{Appendix Double Sine}

The double sine function $\mc{S}$ is defined by the integral representation
\beq
\ln \mc{S}(z) \, = \,  \int\limits_{\R+ \i 0^+}^{} \f{\dd t}{t}  \f{ \ex{ \i z t} }{  \big(\ex{\om_1 t }-1 \big) \big( \ex{\om_2 t} - 1 \big) } \;.
\label{definition Double Sine}
\enq
$\mc{S}$ can be represented as a convergent infinite product in the case where $\Im\big(\tfrac{\om_1}{\om_2} \big)>0$,
\textit{i.e.} $|q| < 1$ and $|\tilde{q}|>1$:
\beq
\mc{S}(\la)=  \f{ \Big( \ex{-\f{2\pi}{\om_2}\la } ; q^2  \Big) }{ \Big( \wt{q}^{\, -2} \ex{-\f{2\pi}{\om_1}\la } ; \wt{q}^{\, -2}   \Big) }
\, = \, \ex{ \i  B(\la) } \cdot \f{ \Big( \ex{\f{2\pi}{\om_1}\la } ; \wt{q}^{\, -2}   \Big) }{ \Big( q^2 \ex{\f{2\pi}{\om_2}\la } ; q^2  \Big) } \;. 
\label{fonction S: formules produit infini}
\enq
The equivalence of these two representation is a consequence of the modular transformation relation for theta functions \eqref{ecriture transfo modulare fct theta}.

The double sine function satisfies the quasi-periodicity relations
\beq
\f{ \mc{S}(z - \i \om_1) }{\mc{S}(z)} = \f{1}{ 1-\ex{-\f{2\pi}{\om_2}z}} \quad ,  \quad
\f{ \mc{S}(z - \i \om_2) }{\mc{S}(z)} = \f{1}{ 1-\ex{-\f{2\pi}{\om_1}z}} 
\enq
and enjoys a reflection property
\beq
\mc{S}(\la)\,  \mc{S}(-\la - \i \Om ) \, =\,   \ex{\i  B(\la) } \;. 
\enq

The zeroes and poles of $\mathcal{S}(z)$ are all simple and located on the lattices 
\begin{empheq}{align}
   \i m \omega_1 + \i n \omega_2\; , &\qquad m, n, \geq 0  &   \mathrm{zeroes} \\
   \i m \omega_1 +  \i n \omega_2 \; , &\qquad m,  n, \leq -1  &   \mathrm{poles} \;.
\end{empheq}

Finally, $\mc{S}$ has the $\la \rightarrow \infty$ asymptotics 
\beq
\mc{S}(\la)\sim
\left\{ \ba{cc }
1  &  \e{arg}(\om_1)-\tfrac{\pi}{2} < \e{arg}(\la) < \e{arg}(\om_2) +\tfrac{\pi}{2}   \vspace{2mm} \\
\ex{ \i B(\la) }  &  \e{arg}(\om_1)-3\tfrac{\pi}{2} < \e{arg}(\la) < \e{arg}(\om_2) -\tfrac{\pi}{2}    \vspace{2mm}  \\
\ex{ \i B(\la) } \Big( q^{2}\ex{\f{2\pi}{\om_2}\la } ; q^2  \Big)^{-1} &  \e{arg}(\om_2)-\tfrac{\pi}{2} < \e{arg}(\la) < \e{arg}(\om_1) -\tfrac{\pi}{2} \vspace{2mm}  \\
 \Big( q^2 \ex{-\f{2\pi}{\om_2}\la } ; q^2  \Big) &  \e{arg}(\om_2)+\tfrac{\pi}{2} < \e{arg}(\la) < \e{arg}(\om_1) +\tfrac{\pi}{2}   
 \ea \right. \;.
\label{Appendix Asymptotiques double Sine}
\enq

\subsection{The quantum dilogarithm}
\label{Appendix Section quantum dilog}

The quantum dilogarithm \cite{FaKa93} $\varpi$ is a  meromorphic function that  is directly related to the double sine function $\mc{S}$:
\beq
\varpi\Big( \la \, + \, \i\f{\Om}{2} \Big) \, =\, \ex{ - \i \f{B(\la)}{2}} \mc{S}(\la) \;.
\label{definition dilogarithme}
\enq
It when $\om_1=\overline{\om}_2$, it satisfies to the relations
\beq
 \overline{\varpi(z)} \, = \, \varpi(-\overline{z}) \;.
\label{equation conj complx dilogarithme}
\enq

\end{appendix}

\end{document}